\documentclass[aps,a4paper, preprint, superscriptaddress,preprintnumbers,floatfix,nofootinbib,amsmath,amssymb,table]{revtex4-1}
\usepackage{hyperref}
\usepackage{cleveref}
\usepackage{subfig}
\usepackage{braket}

\usepackage{amstext,amssymb}
\usepackage[section]{placeins}
\usepackage{amsmath}
\usepackage{graphicx}
\usepackage{xspace}
\usepackage{color}
\usepackage{float}
\usepackage{comment}
\usepackage{amstext,amssymb}
\usepackage{amsmath}
\usepackage[section]{placeins}
\usepackage{xspace}
\usepackage{units}
\usepackage{array}
\usepackage{amsmath,bm}
\usepackage{amssymb}
\usepackage{gensymb}
\usepackage{soul}
\usepackage{textcomp}
\usepackage{slashed} 
\usepackage{multirow}
\usepackage{hyperref}
\usepackage{mathrsfs}
\usepackage{enumitem}
\usepackage{graphicx}
\usepackage{comment}
\usepackage{epstopdf}
\usepackage{float}
\usepackage{units}
\usepackage{feynmp}
\usepackage{slashed}
\usepackage{amsmath,bm}
\usepackage[table]{xcolor}
\usepackage[normalem]{ulem}
\definecolor{antiquewhite}{rgb}{0.98, 0.92, 0.84}
\definecolor{aqua}{rgb}{0.0, 1.0, 1.0}
\definecolor{amethyst}{rgb}{0.6, 0.4, 0.8}
\definecolor{applegreen}{rgb}{0.55, 0.71, 0.0}
\definecolor{byzantine}{rgb}{0.74, 0.2, 0.64}
\definecolor{cadetgrey}{rgb}{0.57, 0.64, 0.69}
\definecolor{candypink}{rgb}{0.89, 0.44, 0.48}
\usepackage{lipsum}
\usepackage{cals}
\usepackage{slashed} 
\newcommand{\be}{\begin{equation}}
\newcommand{\ee}{\end{equation}}
\newcommand{\bea}{\begin{eqnarray}}
\newcommand{\eea}{\end{eqnarray}}

\def\s1{\hat s}

\usepackage{xcolor}

\makeatletter

    \def\CT@@do@color{%
      \global\let\CT@do@color\relax
            \@tempdima\wd\z@
            \advance\@tempdima\@tempdimb
            \advance\@tempdima\@tempdimc
    \advance\@tempdimb\tabcolsep
    \advance\@tempdimc\tabcolsep
    \advance\@tempdima2\tabcolsep
            \kern-\@tempdimb
            \leaders\vrule
                    \hskip\@tempdima\@plus  1fill
            \kern-\@tempdimc
            \hskip-\wd\z@ \@plus -1fill }
    \makeatother

\usepackage{slashed}
\definecolor{ashgrey}{rgb}{0.7, 0.75, 0.71}
\definecolor{aureolin}{rgb}{0.99, 0.93, 0.0}
\definecolor{babypink}{rgb}{0.96, 0.76, 0.76}
\definecolor{buff}{rgb}{0.94, 0.86, 0.51}
\definecolor{chamoisee}{rgb}{0.63, 0.47, 0.35}
\definecolor{chartreuse(web)}{rgb}{0.5, 1.0, 0.0}
\definecolor{citrine}{rgb}{0.89, 0.82, 0.04}
\definecolor{emerald}{rgb}{0.31, 0.78, 0.47}
\definecolor{fawn}{rgb}{0.9, 0.67, 0.44}
\definecolor{fulvous}{rgb}{0.86, 0.52, 0.0}

\newcommand{\nua}[1]{\ensuremath{\rlap{\kern-2.5pt\ensuremath{\overset{\scriptscriptstyle(-)}{\phantom{\nu}}}}{\ensuremath{{\nu}_{#1}}}}\xspace}

\begin{document}
\title{Effect of Off-diagonal NSI Parameters on Entanglement Measurements in Neutrino Oscillations}

\author{Lekhashri Konwar}
\email{konwar.3@iitj.ac.in (Corresponding author)}
\affiliation{Indian Institute of Technology Jodhpur, Jodhpur - 342030, India}
\author{Papia Panda}
\email{ppapia93@gmail.com}
\affiliation{School of Physics,  University of Hyderabad, Hyderabad - 500046,  India}
\author{Rukmani Mohanta}
\email{rmsp@uohyd.ac.in}
\affiliation{School of Physics,  University of Hyderabad, Hyderabad - 500046,  India}

\begin{abstract}

In this work, we explore the influence of off-diagonal non-standard interaction (NSI) parameters on quantum entanglement within the three-flavor neutrino oscillation framework. By expressing three key entanglement measures: Entanglement of Formation (EOF), Concurrence, and Negativity in terms of oscillation probabilities, we analyze how these quantum correlations are affected by the NSI parameters $\epsilon_{e\mu}$, $\epsilon_{e\tau}$, and $\epsilon_{\mu\tau}$, including their complex phases. The quantum correlation measures considered in this work cannot be extracted directly from event rates, but solely in terms of oscillation probabilities. Using the DUNE experiment as a reference point, our analysis shows that NSI effects are most pronounced at lower energies, while Negativity continuing to dominate even at higher energies. It is observed that $\epsilon_{e \mu}$ and $\epsilon_{e \tau}$ affect entanglement measures mainly through the appearance channel, while the impact of $\epsilon_{\mu \tau}$ on EOF, Concurrence, and Negativity is predominantly linked to the disappearance channel.  Further, our results show that Negativity is more sensitive than EOF and Concurrence in the [Energy ($E$) – $\delta_{CP}$] plane under the influence of off-diagonal NSI scenarios, displaying a clear dependence of the CP-violating phase, $\delta_{CP}$ on specific energy ranges, particularly in the lower energy regime.
\end{abstract}

\maketitle
\flushbottom

\section{Introduction }
Neutrino oscillation \cite{Fukuda_1998,Ahmad_2002} is a quantum mechanical phenomenon that arises from the mixing between neutrino flavor and mass eigenstates. Since neutrino flavor states are quantum superpositions of mass eigenstates, a neutrino originated with a specific flavor can be detected as a different flavor during propagation. The probabilities of such oscillations are governed by six key parameters: three mixing angles $(\theta_{12}, \theta_{13}, \theta_{23})$, two mass-squared differences $(\Delta m^{2}_{21}, \Delta m^{2}_{31})$, and a CP-violating phase, $\delta_{CP}$. Over the past few decades, a broad range of solar, atmospheric, reactor, and accelerator neutrino experiments have measured most of these parameters with remarkable precision. Despite this progress, several fundamental questions remain unresolved, most notably the ordering of neutrino masses, commonly known as the mass hierarchy problem, the exact value of the CP-violating phase $\delta_{CP}$, and whether the atmospheric mixing angle $\theta_{23}$ lies in the lower or upper octant. Addressing these unknowns is a primary goal of upcoming and ongoing high-precision neutrino oscillation experiments \cite{Abe_2023,Acero_2022,Abi:2020wmh,JUNO:2021vlw}, which are designed not only to rigorously test the three-flavor paradigm but also to search for signatures of physics Beyond the Standard Model (BSM) \cite{Abi_2021,Majhi_2023,mohanta:2025,Panda:2024avc,Panda:2024qsh,pusty2024,bera2025,singha2024}.\\
 Neutrino Physics has now entered the precision era, and at this point, it is crucial to study the sub-leading eﬀects in neutrino oscillations, commonly termed as  Non-Standard neutrino Interactions (NSIs), which introduce new coupling parameters to the theory, and thus play a pivotal role on the determination of unknowns in the neutrino sector. 
A model-independent framework for exploring NSI effects in neutrino oscillations is discussed in Refs.\cite{Ohlsson:2012kf,Miranda:2015dra,farzan2018}. Among the various possible new physics effects in the neutrino sector,  NSIs have received significant attention and effectively modify neutrino production, detection, and most significantly propagation of neutrinos through matter \cite{Biggio_2009}. These interactions can be described by dimension-six four-fermion operators, which introduce additional terms into the neutrino evolution Hamiltonian. In this work, we focus on propagation NSI, as constraints on production and detection NSI are typically more stringent. The matter NSI terms are parametrized by dimensionless coefficients $\epsilon_{\alpha\beta}$ ($\alpha, \beta= e, \mu, \tau$) that modify the effective matter potential experienced by neutrinos in propagation. These include both diagonal (flavor-conserving) and off-diagonal (flavor-violating) components, where the off-diagonal terms can be complex and introduce new CP-violating effects. Since NSIs affect the coherent forward scattering of neutrinos with electrons and quarks in matter, they can lead to significant modifications in oscillation probabilities, especially in long-baseline experiments with specific beamlines and detector setups \cite{deepthi2015,fukasawa2017,masud2016probing,de2016non,Deepthi_2018,Blennow_2017,Agarwalla_2016,Masud_2016,Coloma_2017,Liao_2016,Coloma_2016,Denton:2022pxt}.\\
Recent developments in quantum information theory have opened a novel and exciting perspective in the neutrino system \cite{Alok_2016q,Blasone_2009Entanglement,Blasone_2008,banerjee2015quantum,formaggio2016violation,fu2017testing,naikoo2019leggett,naikoo2020quantum,shafaq2021enhanced,sarkar2021effects,blasone2023leggett,chattopadhyay2023quantum,konwar2024violation,dixit2024quantum,Bouri:2024kcl,banerjee2024analysis,konwar2025steering,konwar2025neutrino,Li:2022mus,banerjee2024analysis,Blasone_2010,Li:2022mus,Blasone_2021,KumarJha:2020pke,Konwar_2024}. Specifically, the concept of quantum entanglement, once considered relevant primarily in systems like atomic and optical systems, is now being investigated in the context of fundamental particles, including neutrinos. 
Several studies have explored the role of quantum entanglement in neutrino oscillations, revealing the presence of nonclassical correlations in flavor evolution. Blasone et al. \cite{Blasone_2009Entanglement} pioneered this approach by treating different neutrino flavor modes as subsystems, showing that flavor oscillations naturally generate bipartite entanglement. Extending this framework, Jha et al. \cite{KumarJha:2020pke} investigated genuine tripartite entanglement in three-flavor oscillations, highlighting how oscillation dynamics encode multipartite quantum correlations. Li et al. \cite{li2021characterizing} examined the interplay between entanglement and measurement uncertainty, demonstrating that uncertainty relations can serve as indicators of quantum correlations in neutrino systems. Subsequent studies by Li et al. \cite{Li:2022mus} and Banerjee et al. \cite{banerjee2015quantum} further quantified genuine tripartite entanglement and applied quantum information theoretic tools to elucidate the connection between neutrino oscillations. More recently, Konwar and Yadav \cite{Konwar_2024} analyzed the impact of real NSI for all NSI parameters at a time, on tripartite entanglement, showing that NSI parameters can significantly modify quantum correlations, providing new insights into physics beyond the SM. Together, these works establish neutrino oscillations as a promising platform to study quantum correlations and their effect on both standard and non-standard effects. Entanglement represents a core feature of quantum mechanics, characterising non-classical correlations between different components of a system \cite{Horodecki_2009}. In neutrino physics, the coherent superposition of mass and flavor eigenstates naturally leads to a form of single-particle mode entanglement \cite{Blasone_2008,Blasone_2009Entanglement,blasone2014field}, where the different flavor modes of a single neutrino become quantum-mechanically correlated. The key insight is that these entanglement measures can be quantified using neutrino oscillation probabilities, which are directly measurable in neutrino oscillation experiments. To quantify entanglement in neutrino systems, several well-established measures from quantum information theory are employed. In this work, we focus on three such measures: Entanglement of Formation (EOF) \cite{Bennett:1996gf,Guo2020}, Concurrence \cite{Hill:1997pfa,Wootters:1997id,Guo2019} and Negativity \cite{Peres:1996dw,Vidal:2002zz, Sabin2008}. \\
Among the various entanglement measures, the EOF plays a prominent role and captures the minimum average entanglement over all pure state decompositions of a mixed state.  Concurrence, initially formulated for bipartite qubit systems, has been successfully extended to tripartite scenarios and provides a robust way to evaluate the degree of entanglement through the reduced density matrices. Another widely used measure is Negativity, which is derived from the Peres-Horodecki criterion \cite{Peres:1996dw}. It quantifies entanglement based on the partial transpose of the density matrix, acquiring negative eigenvalues, which serves as a clear signature of non-separability and, consequently, the presence of entanglement in the system. All three measures can be effectively expressed in terms of flavor transition probabilities, providing a framework to study quantum correlations in neutrino oscillations. In this study, we explore the behavior of three key entanglement measures within the framework of three-flavor neutrino oscillations by formulating them in terms of oscillation probabilities. We further examine for first time, how these measures are influenced by the presence of three complex off-diagonal NSI scenarios.\\
The paper is organized as follows. Section \ref{sec2} presents the theoretical formalism underlying neutrino oscillation with NSI, along with a description of mode entanglement in the three-flavor neutrino system. In Section \ref{sec3}, we define the three entanglement measures used in our analysis: EOF, Concurrence and Negativity, and express them in terms of neutrino oscillation probabilities. Section \ref{sec4} discusses the experimental description of the DUNE experiment.
Section \ref{sec5} emphasis the major results coming out from our work. 
Finally, Section \ref{sec6} summarizes the conclusions of the study.

\section{Formalism}
\label{sec2}
Neutrino oscillations arise as a consequence of both neutrino mixing and non-degenerate neutrino masses. The neutrino mass eigenstates $\ket{\nu_{j}}, j=1,2,3$, associated with definite masses $m_{j}$ and energies $E_{j}$, evolve in time of the form:
\begin{equation}\label{1}
     \ket{\nu_{j}(t)}=e^{-\iota  E_{j}t}\ket{\nu_{j }}.
\end{equation}
The flavor eigenstates $\ket{\nu_{\alpha}}$, $\alpha= e, \mu, \tau$ are linear superpositions of the mass eigenstates is defined as,
\begin{equation}\label{2}
     \ket{\nu_{\alpha}}= \sum_{j}U_{\alpha j} \ket{\nu_{j }}\end{equation}
     and, vice versa, we can also write
\begin{equation}\label{3}
    \ket{\nu_{j}}= \sum_{\alpha}U^{\ast }_{\alpha j}\ket{\nu _{\alpha}}.
\end{equation}
     The mixing matrix U, known as the Pontecorvo–Maki–Nakagawa–Sakata (PMNS) matrix, can be parametrized in terms of three mixing angles $\theta _{12}$, $\theta _{23}$ and $\theta _{13}$  and a CP-violating phase $\delta$. The elements of the PMNS matrix are given by
\begin{equation}\label{4}
\begin{pmatrix}
c_{12}c_{13} & s_{12}c_{13}&s_{13}e^{-\iota\delta_{CP}}\\ 
-s_{12}c_{23}-c_{12}s_{13}s_{23}e^{\iota\delta_{CP}}& c_{12}c_{23}-s_{12}s_{13}s_{23}e^{\iota\delta_{CP} }&c_{13}s_{23}\\ 
s_{12}s_{23}-c_{12}s_{13}c_{23}e^{\iota\delta_{CP} }& -c_{12}s_{23}-s_{12}s_{13}c_{23}e^{\iota\delta_{CP}}&c_{13}c_{23}
\end{pmatrix}.
 \end{equation}
Here $s_{ij} = \sin \theta _{ij}$ and $c_{ij} = \cos \theta _{ij}$ with i, j= 1, 2, 3.
The time evolution of the flavor state can be written as
\begin{align}\label{5}
    \ket{\nu_{\alpha}(t)} 
        &= \sum_{j}^{} U_{\alpha j}\, e^{-\iota  E_{j}t}\ket{\nu_{j}} \nonumber\\
    &= \sum_{j,\beta}^{} U_{\alpha j}\, e^{-\iota  E_{j}t}\, U^{\ast}_{\beta j}\ket{\nu_{\beta}} \nonumber\\
    &= U^{f}_{\alpha \beta}(t)\ket{\nu_\beta}
\end{align}
where $U^f_{\alpha \beta}(t)= U e^{-\iota H_{vac}t}U^{-1}$ is the time-evolution operator with $\mathcal{H}_{vac} = \text{diag}(E_{1}, E_{2}, E_{3})$. The flavor state $\ket{\nu_{\alpha}(t)}$ at time $t$ of Eq. \ref{5} can then be expressed in terms of the flavor basis as:
\begin{eqnarray}\label{6}
    \ket{\nu _{\alpha }(t)}=U^f_{\alpha e}(t)\ket{\nu}_{e}+U^f_{\alpha \mu}(t) \ket{\nu}_{\mu}
    +U^f_{\alpha \tau}(t) \ket{\nu}_{\tau}.
\end{eqnarray}
The transition probability for a neutrino of flavor $\alpha$ to be detected as a neutrino of flavor $\beta$ at time $t$ is given by \cite{Giunti:2007ry}:
\begin{equation}
P_{(\nu_{\alpha}\rightarrow\nu_{\beta})}(t)= \left|\left< \nu _{\beta}|\nu _{\alpha }(t)\right>\right|^{2}= \left|U^f_{ \alpha \beta }(t)\right|^{2}
\end{equation}
This probability depends on the energy differences $\Delta E_{jk} = E_{j}-E_{k} $ for ($j, k=1,2,3$) and the parameters of the mixing matrix. In the ultra-relativistic limit, we approximate as, $\Delta E_{jk}\simeq  ({\Delta m^{2}_{jk}}/{2E})$ with $\Delta m^{2}_{jk}=m^{2}_{j}-m^{2}_{k} $ and $E$ is the neutrino energy, assuming all mass eigenstates share the same momentum $p$.

In the scenarios beyond the SM, neutrino propagation can be affected by NSI with matter, which introduce additional effective potentials modifying the flavor evolution Hamiltonian. The NSI formalism extends the standard oscillation framework by incorporating these new interactions as perturbations in the matter potential, typically parameterized by dimensionless coefficients 
$\epsilon_{\alpha\beta}$ with $\alpha, \beta= e, \mu, \tau$. This approach allows for a systematic study of how BSM effects can influence neutrino oscillation probabilities. The widely studied dimension six four-fermion operators responsible for NSIs can be written as \cite{Ohlsson:2012kf,Miranda:2015dra,Liao:2016orc}
\begin{eqnarray}\label{nsi1}
\mathcal{L}_{NSI}^{NC}=2\sqrt{2}G_F\sum\limits_{\alpha, \beta, C} \epsilon_{\alpha\beta}^{f,C}(\bar{\nu}_\alpha \gamma^{\mu} P_{L} \nu_\beta)(\bar{f} \gamma_\mu P_{C} f),
\end{eqnarray}
where $G_{F}$ is the Fermi constant, $\epsilon_{\alpha\beta}^{f, C}$ is the parameter which describes the strength of the NSI, $f$ is a first generation
SM fermion ($e, u,$ or $d$)), $P_{C}$ denotes the chiral projector with $C= L, R$ corresponding to the left and right-handed chiral projections defined as $P_{L}=(1- \gamma^{5})/2$, and $P_{R}=(1 + \gamma^{5})/2$. The indices $\alpha$ and $\beta$ denote the neutrino flavors with $e, \mu, \text{and} ~ \tau.$\\
The total Hamiltonian in the presence of matter with NSI can be written as
\begin{eqnarray}
\mathcal{H}_{tot}&=&
\mathcal{H}_{vac}+\mathcal{H}_{mat}+\mathcal{H}_{NSI}
\end{eqnarray}
\begin{eqnarray}\label{nsi2}
   \mathcal{H}_{tot}&=&
\begin{pmatrix}
E_{1} & 0 & 0 \\ 
0 & E_{2}& 0 \\
0 & 0 & E_{3}
\end{pmatrix}+U^{\dagger } A\begin{pmatrix}
1+\epsilon_{ee} &\epsilon_{e\mu}  &\epsilon_{e\tau}\\ 
\epsilon_{\mu e} & \epsilon_{\mu \mu} & \epsilon_{\mu \tau}\\
 \epsilon_{\tau e} & \epsilon_{\tau \mu} & \epsilon_{\tau \tau}
\end{pmatrix} U.
\end{eqnarray}\\
Where A is the matter potential arising from the coherent forward scattering of electron neutrinos with electrons in matter, and is given by 
$A=\pm \sqrt{2}G_{F}N_{e}$
with $G_{F}$ is the Fermi constant and $N_{e}$ is the electron number density. The sign of A is positive for neutrinos and negative for antineutrinos. In our analysis, we consider the matter potential as $A$= $1.01  \times  10^{-13}$ eV, which is associated with Earth’s matter density of 
$\rho= 2.8$ gm/cc. And the NSI parameters $\epsilon _{\alpha \beta }$ are given by
\begin{equation}\label{nsi3}
    \epsilon _{\alpha \beta }=\sum_{C,f}\frac{N_{f} }{N_{e}} \epsilon _{\alpha \beta }^{f,C},
\end{equation}
$\epsilon _{\alpha \beta }^{f, C}$ are the NSI parameters, which describe the interaction strengths between neutrinos of flavors $\alpha$ and $\beta$ with fermions of type $f$, and $N_{f}$ is the number density of $f$ type fermions.

In the relativistic limit, neutrino flavor states can be treated as distinguishable modes within a single-particle quantum field framework \cite{blasone2014field}. When two or more of these modes exhibit non-factorizable quantum correlations, the system is said to possess mode entanglement. Specifically, neutrino mode entanglement refers to the entanglement among different flavor modes, such as electron, muon, and tau, within a single neutrino field. This type of entanglement arises naturally from the time evolution of flavor states due to neutrino mixing. As a result, a time-evolved neutrino flavor state can be interpreted as an entangled superposition across the flavor modes. For a three-flavor neutrino system, each flavor state can be formally represented in the occupation number basis as a three-qubit configuration \cite{Blasone_2009Entanglement}:  $\ket{\nu _{e}}\equiv \ket{1}_{e} \ket{0}_{\mu }\ket{0}_{\tau }$, $ \ket{\nu _{\mu}}\equiv \ket{0}_{e} \ket{1}_{\mu }\ket{0}_{\tau }$ and  $\ket{\nu _{\tau}}\equiv \ket{0}_{e} \ket{0}_{\mu }\ket{1}_{\tau }$. Here, $\ket{0}_{\alpha}$ and $\ket{1}_{\alpha}$ denote the absence and presence, respectively, of a neutrino in the flavor mode $\nu_{\alpha}$. The general time-evolved evolution matrix of Eq. \eqref{6} takes the form: 
\begin{eqnarray}\label{12}
    \ket{\nu _{\alpha }(t)}={U}^{f}_{\alpha e}(t)\ket{1}_{e} \ket{0}_{ \mu}\ket{0}_{\tau } +{U}^{f}_{\alpha \mu}(t) \ket{0}_{e} \ket{1}_{\mu }\ket{0}_{\tau }
    +{U}^{f}_{\alpha \tau}(t) \ket{0}_{e} \ket{0}_{\mu }\ket{1}_{\tau }.
\end{eqnarray}
Here, each term in this expression represents the occupation of one specific flavor mode (e, $\mu$, or $\tau$) while the others remain unoccupied. To characterize entanglement of neutrino states, one constructs the density matrix of the system and evaluates the reduced density matrices after tracing over one or more modes. If the reduced states are mixed, the original state is entangled. The degree of entanglement can be quantified using entanglement measures; EOF, Concurrence, Negativity, etc. In principle, both flavor and mass bases can be used for this characterization, but in this work, we employ the flavor basis.
To analyze this entanglement quantitatively, the density matrix of the system at time $t$ is constructed as,
$\rho _{ABC}^{\alpha}(t) = \ket{\nu _{\alpha }(t)}\bra{\nu _{\alpha }(t)}$ where $A$, $B$, and $C$ are the three subsystems of a tripartite system. For the neutrino, these subsystems are chosen as the three flavor modes ($\nu_{e}, \nu_{\mu}, \nu_{\tau}$) and the density matrix takes the form \cite{banerjee2024analysis}:
\begin{equation}{\label{rho1}}
    \rho _{e\mu\tau}^{\alpha}(t)=\begin{pmatrix}
0 & 0 & 0 & 0 & 0 & 0 & 0 & 0\\ 
0 & \rho _{22}^{\alpha } & \rho _{23}^{\alpha } & 0 & \rho _{25}^{\alpha } & 0 & 0 & 0\\ 
0 & \rho _{32}^{\alpha } & \rho _{33}^{\alpha } & 0 & \rho _{35}^{\alpha } & 0 & 0 & 0\\ 
0 & 0 & 0 & 0 & 0 & 0 & 0 & 0\\ 
0 & \rho _{52}^{\alpha } & \rho _{53}^{\alpha } & 0 & \rho _{55}^{\alpha } & 0 & 0 & 0\\ 
0 & 0 & 0 & 0 & 0 &0  & 0 & 0\\ 
0 & 0 & 0 & 0 & 0 &0  & 0 & 0\\
0 & 0 & 0 & 0 & 0 &0  & 0 & 0 
\end{pmatrix},
\end{equation}
where the elements of this matrix in the presence of matter and NSI can be represented as. 
\begin{eqnarray}{\label{rhop}}
  &&  \rho _{22}^{\alpha}= \left |\Bar{U}^{f}_{\alpha \tau }(t) \right |^{2};  ~~~~~~~~\rho _{23}^{\alpha}= \Bar{U}^{f}_{\alpha \tau }(t)\Bar{U}_{\alpha \mu }^{f\ast }(t); ~~~~\rho _{25}^{\alpha}= \Bar{U}^{f}_{\alpha \tau }(t)\Bar{U}_{\alpha e }^{f\ast }(t);\notag\\&&
    \rho _{32}^{\alpha}= \Bar{U}^{f}_{\alpha \mu }(t)\Bar{U}_{\alpha \tau }^{f\ast }(t);~~~
    \rho _{33}^{\alpha}= \left |\Bar{U}^{f}_{\alpha \mu }(t) \right |^{2}; ~~~~~~~~~
   \rho _{35}^{\alpha}= \Bar{U}_{\alpha \mu }(t)\Bar{U}_{\alpha e }^{f\ast }(t);\notag\\
   &&  \rho _{52}^{\alpha}= \Bar{U}^{f}_{\alpha e }(t)\Bar{U}_{\alpha \tau }^{f\ast }(t); ~~~~\rho _{53}^{\alpha}= \Bar{U}^{f}_{\alpha e }(t)\Bar{U}_{\alpha \mu }^{f\ast }(t); ~~~~
     \rho _{55}^{\alpha}= \left |\Bar{U}^{f}_{\alpha e }(t) \right |^{2}.
\end{eqnarray}\\
Here $\Bar U^f_{\alpha \beta}(t)= U e^{-\iota H_{tot}t}U^{-1}$ is the time-evolution operator in the presence of matter and NSI, where $H_{tot}$ is defined in Eq. \eqref{nsi2}. The probabilities are as follows: $P_{\alpha e}(t)= \left |\Bar{U}^{f}_{\alpha e}(t) \right |^{2}$, $P_{\alpha \mu}(t)= \left |\Bar{U}^{f}_{\alpha \mu }(t) \right |^{2}$,  and $P_{\alpha \tau}(t)= \left |\Bar{U}^{f}_{\alpha \tau }(t) \right |^{2}$. In the next section, we will discuss the mathematical expressions of EOF, Concurrence and Negativity in three-flavor neutrino system.
\section{Entanglement measures}
\label{sec3}
Entanglement is one of the most fundamental features of quantum mechanics, reflecting the existence of non-classical correlations between different parts of a composite system. A state is said to be entangled if it cannot be expressed as a simple product of the states of its individual subsystems, meaning that the description of each part is inseparable from the whole. Quantifying entanglement is essential not only for understanding fundamental aspects of quantum theory but also for exploring its role as a resource in quantum communication, computation, and information processing. To capture its different facets, several entanglement measures have been introduced, each offering a unique perspective on the strength and nature of quantum correlations. In this work, we focus on three widely employed measures: Entanglement of Formation, Concurrence, and Negativity and derive these measures in the context of neutrino oscillation probabilities.
\subsection{Entanglement of Formation}
In quantum mechanics, a pure state is said to be entangled if it cannot be written as a product of states of its subsystems. That is, when the total state $\gamma$ cannot be expressed as $\gamma_A \otimes \gamma_B$, then the state is entangled. A convenient way to measure this entanglement is through the entropy of entanglement, which is defined as the von Neumann entropy of either of the reduced density matrices: \cite{Bennett:1996gf}
$$E\left ( \gamma  \right ) = S\left ( \rho _{A} \right )= S\left ( \rho _{B} \right ),$$
where $S(\rho) = -\text{Tr}(\rho \log_2 \rho)$ is the von Neumann entropy. The reduced density matrix $\rho_A$ is obtained by tracing out subsystem $B$, $\rho_A = \text{Tr}_B\left (\ket{\gamma}\bra{\gamma}\right )$, and similar for $\rho_B$.\\

When dealing with mixed states, entanglement can still be quantified using the EOF. For a mixed state, EOF is defined as the minimum average entanglement over all possible pure-state decompositions that realize $M$ \cite{Bennett:1996gf}:
 $$ EOF(M)= \min \sum_{i}^{}p_{i}E\left ( \gamma _{i} \right ),$$
where $M = \sum_{i}^{}p_i \ket{\gamma_i}\bra{\gamma_i}$ and the minimization is over all possible sets of pure states $\ket{\gamma_i}$ and corresponding probabilities $p_i$ that reproduce the mixed state $M$. \\
For a given pure state $\gamma$, EOF is computed to the entropy of entanglement:
\begin{equation}
    EOF(\gamma)=-Tr(\rho _{A} \log_{2}\rho _{A})=-Tr(\rho _{B} \log_{2}\rho _{B}),
\end{equation}
where $\rho_{A}$ and $\rho_{B}$ are the reduced density matrices of the two subsystems. This definition can also be extended to tripartite systems. For a pure tripartite system, the density matrix is denoted as $\rho_{ABC}(t)$ and the EOF is given by \cite{Guo2020}:
\begin{equation}\label{EOF3}
   EOF(\rho _{ABC}(t))=\frac{1}{2}[S(\rho _{A})+S(\rho _{B})+S(\rho _{C})],
\end{equation}
where $S(\rho _{A})$, $ S(\rho _{B})$ and $ S(\rho _{C})$ are von Neumann entropies defined as $S(\rho _{A})=-Tr(\rho _{A}\log\rho _{A})$ and same with $S(\rho _{B})$ and $S(\rho _{C})$. $\rho_{A}$, $\rho_{B}$ and $\rho_{C}$ are reduced density matrices which have the expressions \cite{Guo2020}, $\rho _{A}=Tr_{BC}\left (\rho _{ABC}(t)\right )$, $\rho _{B}=Tr_{AC}\left (\rho _{ABC}(t)\right )$ and $\rho _{C}=Tr_{AB}\left (\rho _{ABC}(t)\right )$. \\

The EOF for a neutrino initial flavor state, $\mu$ can be expressed in terms of the flavor transition probabilities as follows \cite{li2021characterizing}:
\begin{eqnarray}\label{EOF}
  &&  EOF^{\mu}=-\frac{1}{2}[P_{\mu e} \log_{2}P_{\mu e}+P_{\mu \mu} \log_{2}P_{\mu \mu}+P_{\mu \tau} \log_{2}P_{\mu \tau}
    +(P_{\mu \mu}+P_{\mu \tau}) \log_{2}(P_{\mu \mu}+P_{\mu \tau})\nonumber\\
  &&~~~~~~~~~~~~  +(P_{\mu e}+P_{\mu \tau}) \log_{2}(P_{\mu e}+P_{\mu \tau}) +(P_{\mu \mu}+P_{\mu e}) \log_{2}(P_{\mu \mu}+P_{\mu e})],
\end{eqnarray}
where $P_{\alpha \beta}$ are the oscillation probabilities from $\nu_{\alpha}$ to $\nu_{\beta}$ transition. The detailed derivation of the above expression of the Entanglement of Formation (EOF) in terms of oscillation probabilities is provided in Appendix \ref{sec:analytic}.

\subsection{Concurrence}
For bipartite systems, particularly two-qubit systems, one of the most widely recognized and analytically tractable measures of entanglement is Concurrence. Originally introduced by Hill and Wootters \cite{Hill:1997pfa} and later generalized by Wootters \cite{Wootters:1997id}, Concurrence provides a quantification of the entanglement of formation for arbitrary mixed states of two qubits. However, a recent study has established that the framework of Concurrence can be generalized to entanglement in three-qubit states, thereby extending its applicability beyond bipartite systems. This development constitutes a significant contribution to the theoretical understanding of entanglement in multipartite quantum systems, \cite{Guo2019}
\begin{equation}\label{con}
    C(\rho _{ABC})=[3-Tr(\rho _{A})^{2}-Tr(\rho _{B})^{2}-Tr(\rho _{C})^{2}]^{\frac{1}{2}},
\end{equation}
where $\rho _{A}$, $\rho _{B}$, and  $\rho _{C}$ are the reduced density matrices. 

For oscillating neutrinos, the Concurrence for $\mu$ as initial flavor, $C^{\mu}$ can be expressed in terms of oscillation and survival probabilities as \cite{li2021characterizing}
\begin{equation}\label{C} 
 C^{\mu}=\sqrt{3-3P_{S}-2P_{\mu \mu}P_{\mu \tau}-2P_{\mu e}(P_{\mu \mu}+P_{\mu \tau})},
\end{equation}
where $P_{S}$ is the sum of the square of probabilities, and defined as $(P_{\mu e}^{2}+P_{\mu \mu}^{2}+P_{\mu \tau}^{2})$. The derivation of the Concurrence expression in terms of neutrino oscillation probabilities can be found in Appendix \ref{sec:analytic}.

\subsection{Negativity}
A necessary condition for separability is provided by the Peres-Horodecki criterion \cite{Peres:1996dw}, which states that the partial transpose of a separable state must have only non-negative eigenvalues. Violation of this condition implies that the state is entangled.
Negativity is a computable entanglement measure based on this criterion. It is defined using the partial transpose $\rho^{T_{A}}$ of the bipartite density matrix $\rho$ with respect to subsystem A. Negativity quantifies how much $\rho^{T_{A}}$ fails to be positive and is given by \cite{Vidal:2002zz}:
$$\emph{N}\left ( \rho  \right )=  -2\sum _{i} \sigma _{i} (\rho ^{T_{A}}),$$
where $\sigma _{i} (\rho ^{T_{A}})$ are the negative eigenvalues of the partial transpose $\rho ^{T_{A}}$ of the total state $\rho$ with respect to the subsystem A. The Negativity vanishes for separable states and provides a quantitative measure of the degree of entanglement present in composite quantum systems.\\ 
The tripartite Negativity of a state $\rho_{ABC}$ is defined as \cite{Sabin2008}
$$N=(N_{A-BC}N_{B-CA}N_{C-AB})^{\frac{1}{3}},$$
with $N_{I-JK} = -2 \sum_{i} \sigma_{i} (\rho^{T_{I}})$ being the negative eigenvalues of $\rho^{T_{I}}$, the partial transpose of $\rho$ with respect to subsystem $I$, with $I = A, B, C,$ and $JK = BC, CA, AB$, respectively.\\

In terms of survival and oscillation probabilities of neutrino oscillation, Negativity for $\mu$ as initial flavor, $N^{\mu}$ is given as \cite{li2021characterizing} :
\begin{equation}\label{N}
    N^{\mu}=[\sqrt{P_{\mu e}}\sqrt{P_{\mu \mu}+P_{\mu \tau}}\sqrt{P_{\mu e}+P_{\mu \tau}}\sqrt{P_{\mu \mu}}\sqrt{P_{\mu e}+P_{\mu \mu}}\sqrt{P_{\mu \tau}}]^{\frac{1}{3}}.
\end{equation}
The explicit derivation of the Negativity formula, expressed through oscillation and survival probabilities, is presented in Appendix \ref{sec:analytic}.

\section{Experimental Description}
\label{sec4}
In this analysis, we consider the upcoming long-baseline neutrino experiment Deep Underground Neutrino Experiment (DUNE).  DUNE is an ambitious forthcoming project featuring a baseline of 1300 km, extending from the Fermi National Accelerator Laboratory (FNAL) to the Sanford Underground Research Facility (SURF). For simulation purposes, we utilize the official configuration files based on the technical design report \cite{DUNE:2021cuw}. The experimental setup includes four liquid argon time-projection chamber (LArTPC) detectors, each with a mass of 10 kilotons, and operates with a 1.2 MW proton beam.  The files represent an
exposure of 624 kt-MW-years which corresponds to 6.5 years of run each in neutrino (FHC) and
antineutrino (RHC) modes, using a 40 kt fiducial mass liquid argon time-projection chamber
(LArTPC) far detector. In this work, we restrict our results to the normal mass ordering scenario, i.e., $\Delta m_{31}^2>0.$

\begin{table}
    \centering
    \begin{tabular}{|c|c|c|c|c|c|c|c|c|}
        \hline
        
        $~\Delta m^2_{21} ({\rm eV}^{2})$ ~&~ $\Delta m_{31}^2 ({\rm eV}^{2})$ ~&~  $\sin^2{\theta_{12}}$ ~&~ $\sin^2{\theta_{13}}$~&~ $\sin^2{\theta_{23}}$~&~ $\delta_{CP}$~\\
       \hline
         $7.49 \times 10^{-5}$ &$2.513 \times 10^{-3}$ & 0.308  & 0.02215 & 0.470&  212$^{\circ}$ \\
        \hline
    \end{tabular}
    \caption{ Best fit values of standard three-neutrino oscillation parameters used in our analysis \cite{Esteban_2024}.}
    \label{del1}
\end{table}

\section{Results}
\label{sec5}
In this section, we will illustrate the main results of our simulation. In the first subsection, we show how the appearance and disappearance probabilities are modified in the presence of off-diagonal NSI parameters, incorporating one parameter at a time. We then extend this study to the event level in the following subsection. Then, in the next subsection, we discuss the effect of off-diagonal NSI parameters in the entanglement measures: EOF, Concurrence, and Negativity. The final subsection shows the variation of entanglement measures with neutrino energy and CP-violating phase, $\delta_{CP}$.

\begin{table}[]
    \centering
    \begin{tabular}{|c|c|}
    \hline
       NSI parameters  & Best fit value \\
       \hline
        $|\epsilon_{e \mu}|$ & 0.15\\
        \hline
        $|\epsilon_{e \tau}|$ & 0.27 \\
        \hline
        $|\epsilon_{\mu \tau}|$  &  0.35 \\
        \hline
        \hline
        $\phi_{e \mu}$ & $1.38 \pi$ rad \\
        \hline
        $\phi_{e \tau}$ & $1.62 \pi$ rad\\
        \hline
        $\phi_{\mu \tau}$ & $0.6 \pi$ rad\\
        \hline
    \end{tabular}
    \caption{The best fit value of the NSI parameters used in the analysis \cite{Chatterjee:2020kkm,Denton:2020uda}.}
    \label{nsi}
\end{table}
\subsection{Analysis in probability level}
In this subsection, we discuss how non-standard interaction (NSI) parameters influence the appearance and disappearance probabilities within the framework of the DUNE experiment. At the source, a beam of muon neutrinos ($\nu_\mu$) is generated and travels through matter over a baseline of 1300 km to reach the far detector. Therefore, our primary focus is on the oscillation ($P_{\mu e}$) and survival ($P_{\mu \mu}$) probabilities. The presence of NSI modifies the standard probability expressions. In this analysis, we concentrate specifically on the off-diagonal NSI parameters: $\epsilon_{e\mu}= \left |\epsilon_{e\mu} \right | e^{-i\phi_{e\mu}}$, $\epsilon_{e\tau}= \left |\epsilon_{e\tau} \right | e^{-i\phi_{e\tau}}$, and $\epsilon_{\mu\tau}= \left |\epsilon_{\mu\tau} \right | e^{-i\phi_{\mu\tau}}.$ Since these are off-diagonal elements, they can possess both real and imaginary components, each associated with a corresponding phase: $\phi_{e \mu}$, $\phi_{e \tau}$, and $\phi_{\mu \tau}$. The appearance probability $P_{\mu e}$, modified by the presence of off-diagonal NSI parameters, can be expressed as follows  \footnote{In Ref. \cite{Liao:2016orc}, the expression for $P_{\mu e}$ also includes contributions from diagonal NSI terms (such as $\epsilon_{ee}$). However, since our analysis focuses exclusively on the off-diagonal parameters, Eq. \ref{app} is formulated solely in terms of these off-diagonal contributions, keeping $\epsilon_{ee}=0$.} \cite{Liao:2016orc,Liao:2016hsa,Kopp:2007ne,Kikuchi:2008vq}:
\begin{align}
\label{app}
    P (\nu_\mu \to \nu_e) &= x^2 f^2 + 2xyfg \cos(\Delta + \delta_{CP}) + y^2 g^2 \nonumber \\
    &\quad + 4 \hat{A} \left |\epsilon_{e\mu} \right | \Big\{ x f [s_{23}^2 f \cos(\phi_{e\mu} + \delta_{CP}) + c_{23}^2 g \cos(\Delta + \delta_{CP} + \phi_{e\mu})] \nonumber \\
    &\quad + y g [c_{23}^2 g \cos\phi_{e\mu} + s_{23}^2 f \cos(\Delta - \phi_{e\mu})] \Big\} \nonumber \\
    &\quad + 4 \hat{A} \left |\epsilon_{e\tau} \right | s_{23} c_{23} \Big\{ x f [f \cos(\phi_{e\tau} + \delta_{CP}) - g \cos(\Delta + \delta_{CP} + \phi_{e\tau})] \nonumber \\
    &\quad - y g [g \cos\phi_{e\tau} - f \cos(\Delta - \phi_{e\tau})] \Big\} \nonumber \\
    &\quad + 4 \hat{A}^2 \Big( g^2 c_{23}^2 |c_{23}\epsilon_{e\mu}  - s_{23} \epsilon_{e\tau}|^2 + f^2 s_{23}^2 |s_{23} \epsilon_{e\mu} + c_{23} \epsilon_{e\tau}|^2 \Big) \nonumber \\
    &\quad + 8 \hat{A}^2 fg s_{23} c_{23} \Big\{ c_{23} \cos\Delta [s_{23} (\left |\epsilon_{e\mu} \right |^2 - \left |\epsilon_{e\tau} \right |^2) + 2 c_{23} \left |\epsilon_{e\mu} \right | \left |\epsilon_{e\tau} \right | \cos(\phi_{e\mu} - \phi_{e\tau})] \nonumber \\
    &\quad - \left |\epsilon_{e\mu} \right | \left |\epsilon_{e\tau}  \right |\cos(\Delta - \phi_{e\mu} + \phi_{e\tau}) \Big\} + \mathcal{O}(s_{13} \epsilon, s_{13} \epsilon^2, \epsilon^3),
\end{align}
where,
\begin{align}
    x &\equiv 2 s_{13} s_{23}, \quad y \equiv 2 \alpha s_{12} c_{12} c_{23}, \quad \alpha = \frac{\Delta m^2_{21}}{\Delta m^2_{31}} , \nonumber \\
    f &\equiv \frac{\sin[\Delta(1 - \hat{A})]}{(1 - \hat{A})}, ~~~~~~~~~~~~~\quad g \equiv \frac{\sin(\hat{A} \Delta)}{\hat{A}}, \nonumber \\
    ~~~\Delta &=  \frac{\Delta m^2_{31} L}{4E} , ~~~~~~~~~~~~~~~~~~~~~\quad \hat{A} =  \frac{A}{\Delta m^2_{31}},
\end{align}
for normal ordering. For the inverted mass ordering, the corresponding substitutions are: $\Delta \rightarrow -\Delta, y \rightarrow -y$ and $\hat{A} \rightarrow -\hat{A}$. In the case of antineutrinos, Eq.~\ref{app} is modified by the following transformations: $\hat{A} \rightarrow - \hat{A}, \delta_{CP} \rightarrow - \delta_{CP}$ and $\phi_{\alpha \beta} \rightarrow - \phi_{\alpha \beta}$. 
Here $c_{ij}= \cos\theta _{ij}$, $s_{ij}= \sin\theta _{ij}$ with $i,j=1,2,3$. From Eq.~\ref{app}, it is evident that $P_{\mu e}$ largely depends on the off-diagonal NSI parameters $\left |\epsilon_{e \mu}\right |$, $\left |\epsilon_{e \tau}\right |$ and their associated phases, while it remains nearly independent of $ \epsilon_{\mu \tau}$ (however, at subleading order, a dependence on $\epsilon_{\mu \tau}$ does appear). \\
The disappearance probability in presence of off-diagonal NSI parameters can be written as \footnote{Similar to \ref{app}, in the Ref. \cite{Liao:2016orc}, the expression for $P_{\mu \mu}$ also includes contributions from diagonal NSI terms (such as $\epsilon_{\mu \mu}, \epsilon_{\tau \tau}$). However, Eq. \ref{dis} is formulated solely in terms of off-diagonal contributions, keeping $\epsilon_{\mu \mu}= \epsilon_{\tau \tau}= 0$.} \cite{Liao:2016orc,Liao:2016hsa,Kopp:2007ne,Kikuchi:2008vq}:
\begin{align}
\label{dis}
    P (\nu_\mu \to \nu_\mu) &= 1 - \sin^2 2\theta_{23} \sin^2 \Delta + \alpha c_{12}^2 \sin^2 2\theta_{23} \Delta \sin2 \Delta - \frac{4 s_{23}^4 s_{13}^2 \sin^2 (1-\hat{A})\Delta}{(1-\hat{A})^2} \nonumber \\ & -\frac{\sin^2 2 \theta_{23} s_{13}^2}{(1- \hat{A})^2} \left[ \hat{A}(1-\hat{A}) \Delta \sin 2 \Delta + \sin(1-\hat{A}) \Delta \sin (1+ \hat{A}) \Delta \right] \nonumber \\ & - 2 \hat{A} \left |\epsilon_{\mu\tau} \right | \cos\phi_{\mu\tau} \Big[ \sin^3 2\theta_{23} \Delta \sin 2\Delta + 2 \sin 2\theta_{23} \cos^2 2\theta_{23} \sin^2 \Delta \Big] \nonumber \\
    &  - 2 \hat{A}^2 \sin^2 2\theta_{23}  \left |\epsilon_{\mu\tau} \right |^2 \Big[ 2 \sin^2 2\theta_{23} \cos^2\phi_{\mu\tau} \Delta^2 \cos 2\Delta + \sin^2\phi_{\mu\tau} \Delta \sin 2\Delta \Big] .
\end{align}

From Eq. \ref{dis}, it is evident that $P_{\mu \mu}$ depends mainly on the NSI parameter $\epsilon_{\mu \tau}$, and is unaffected by $\epsilon_{e \mu}$ and $\epsilon_{e \tau}$. Hence, to investigate the influence of $\epsilon_{e \mu}$ and $\epsilon_{e \tau}$, one should primarily focus on the appearance probability $P_{\mu e}$, while the impact of $\epsilon_{\mu \tau}$ is most effectively studied through the disappearance probability $P_{\mu \mu}$.

\begin{figure}[htb]
   \hspace{-0.2cm}
   \includegraphics[height=50mm, width=54mm]{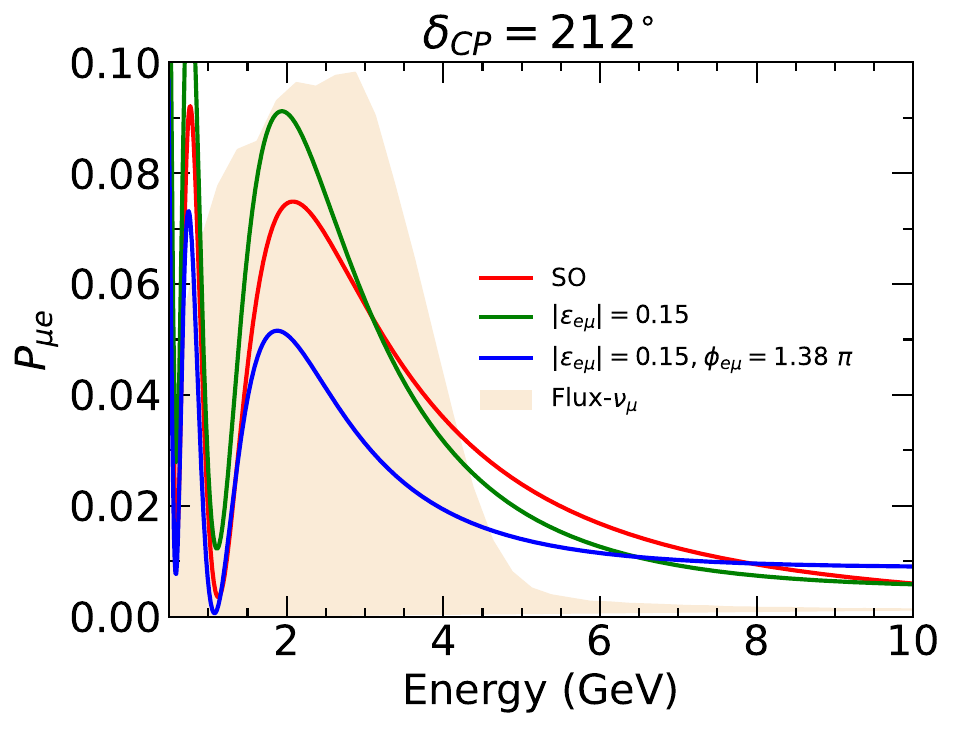}
\includegraphics[height=50mm, width=54mm]{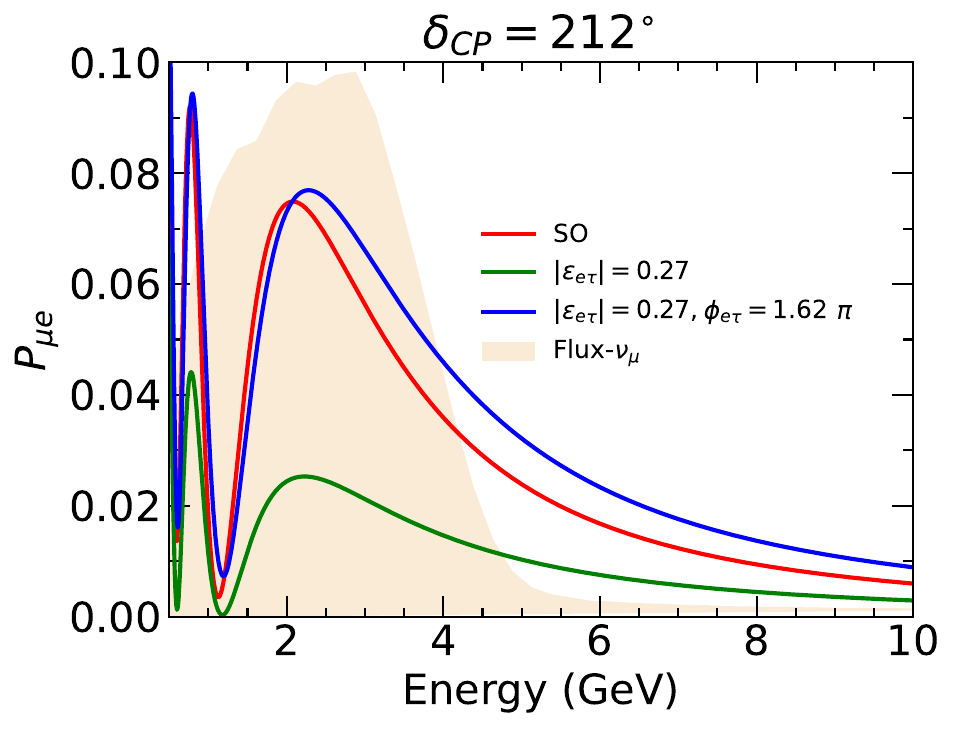}   \includegraphics[height=50mm, width=54mm]{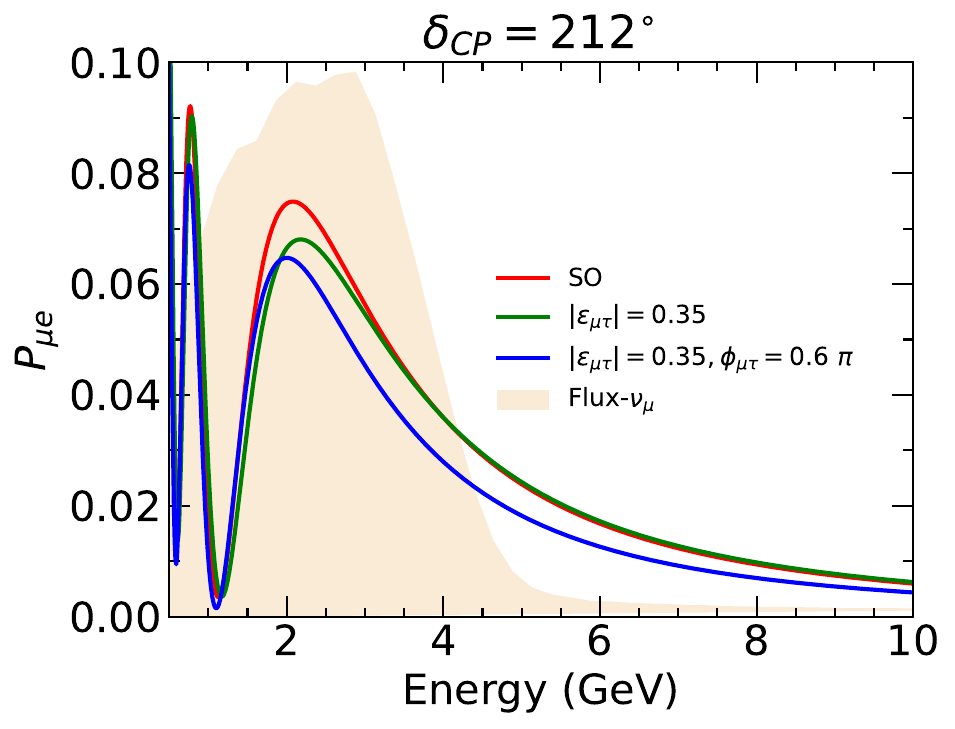}\\
 \includegraphics[height=50mm, width=54mm]{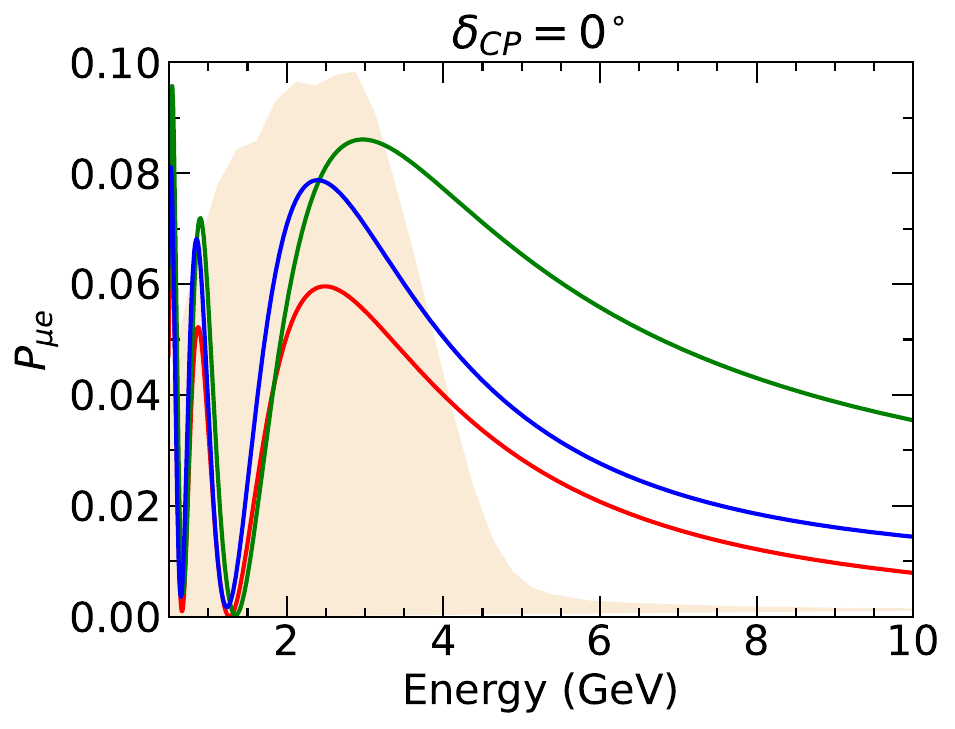}
 \includegraphics[height=50mm, width=54mm]{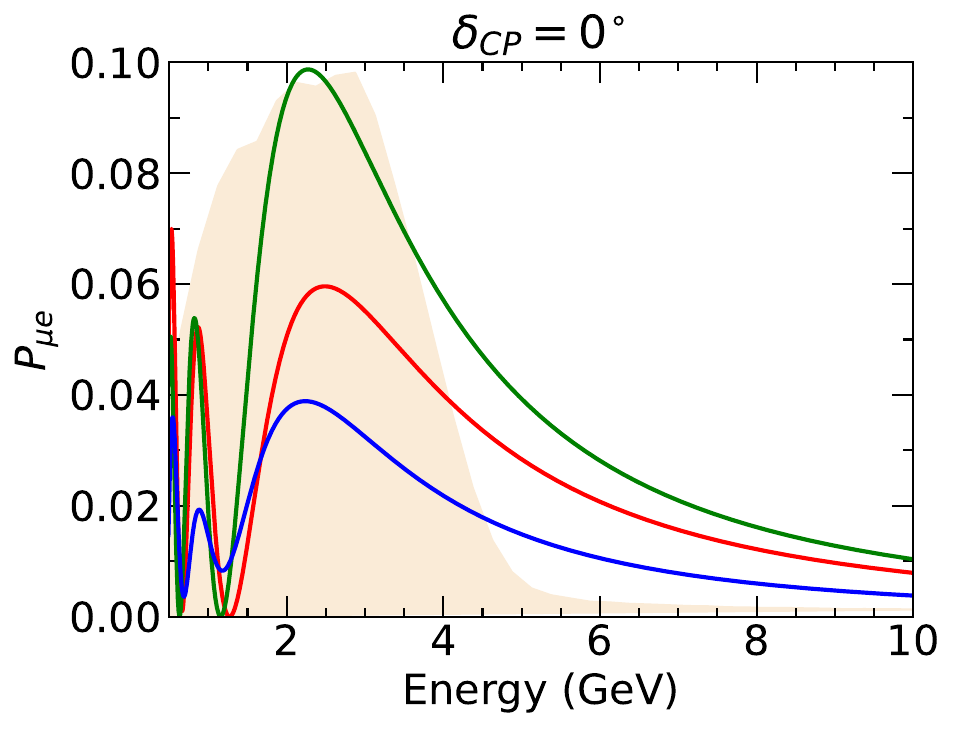}
 \includegraphics[height=50mm, width=54mm]{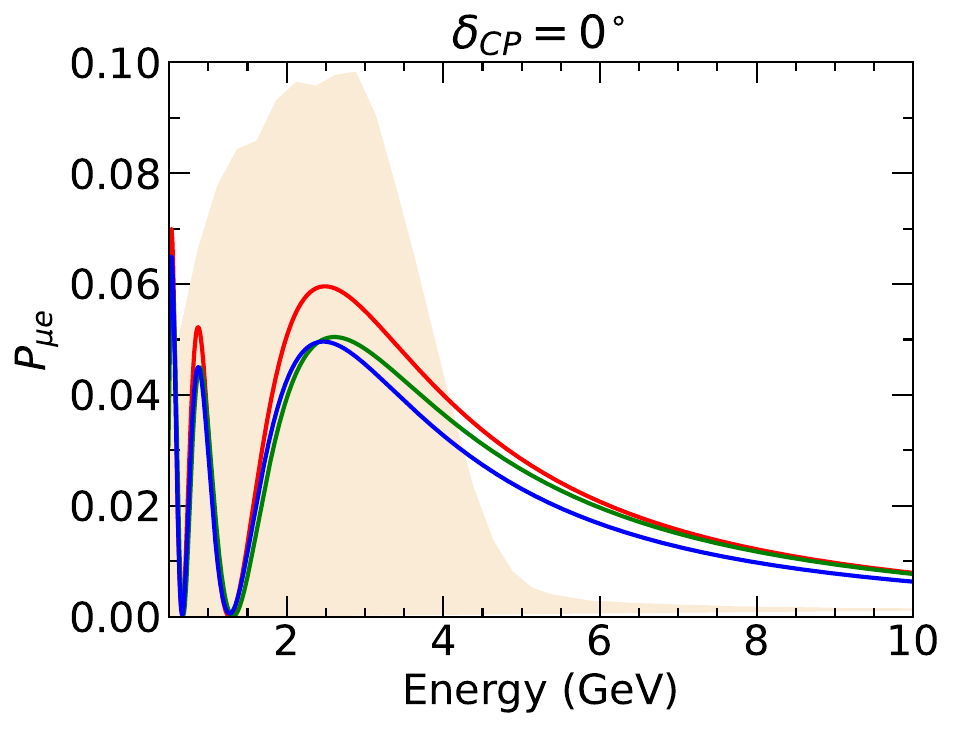}
    \caption{Upper (lower) row shows the appearance probability as a function of neutrino energy for DUNE experiment with $\delta_{CP} = 212^{\circ} (0^{\circ})$. Left (middle) [right] panel represent the curves with one off-diagonal parameter at a time, $\left| \epsilon_{e \mu}\right|$ ($\left|\epsilon_{e \tau}\right|$) [$\left|\epsilon_{\mu \tau}\right|$] and their associated phases. Color codes are given in the legend.}
    \label{fig1}
\end{figure}

\begin{figure}[htb]
   \hspace{-0.2cm}
\includegraphics[height=65mm, width=80mm]{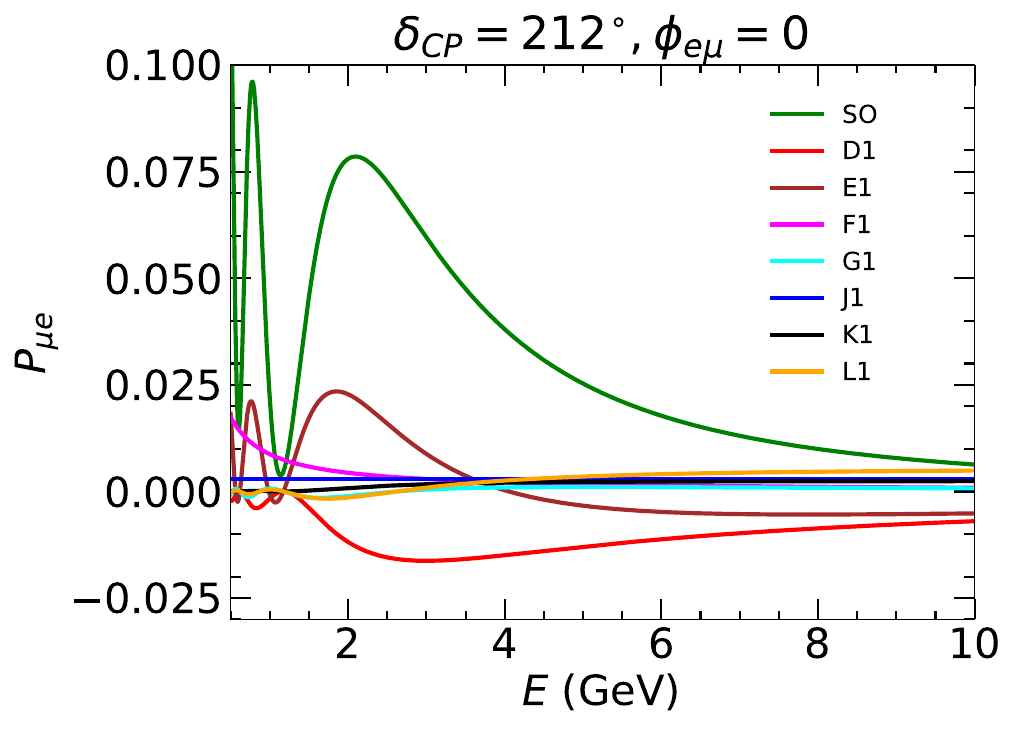}   \includegraphics[height=65mm, width=80mm]{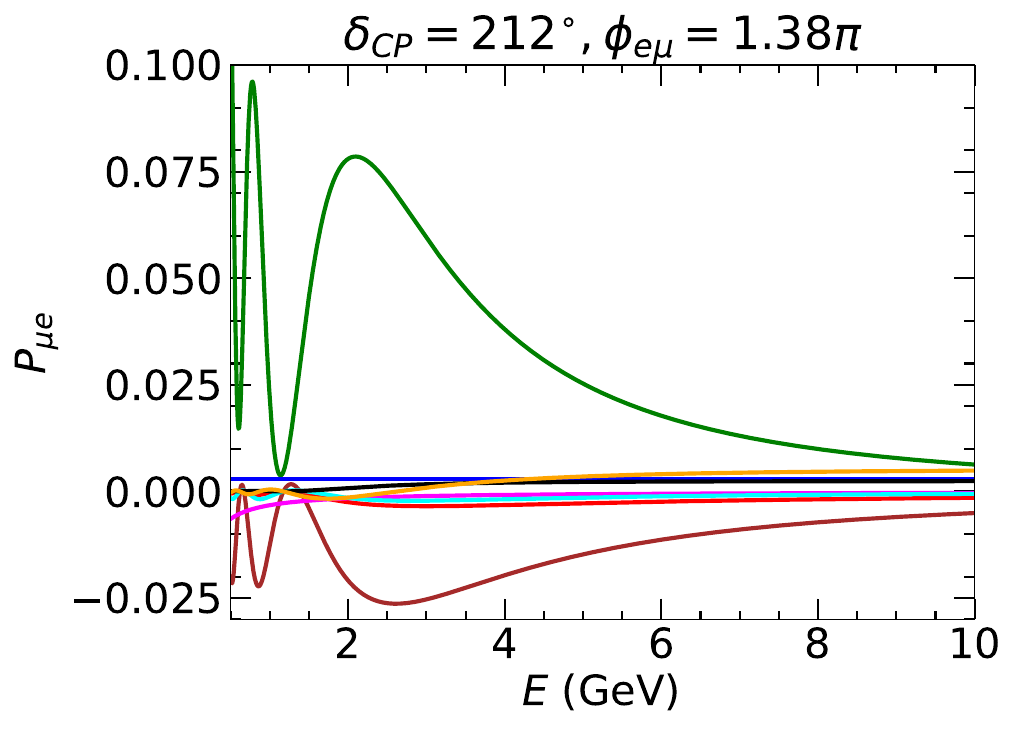}\\
\includegraphics[height=65mm, width=80mm]{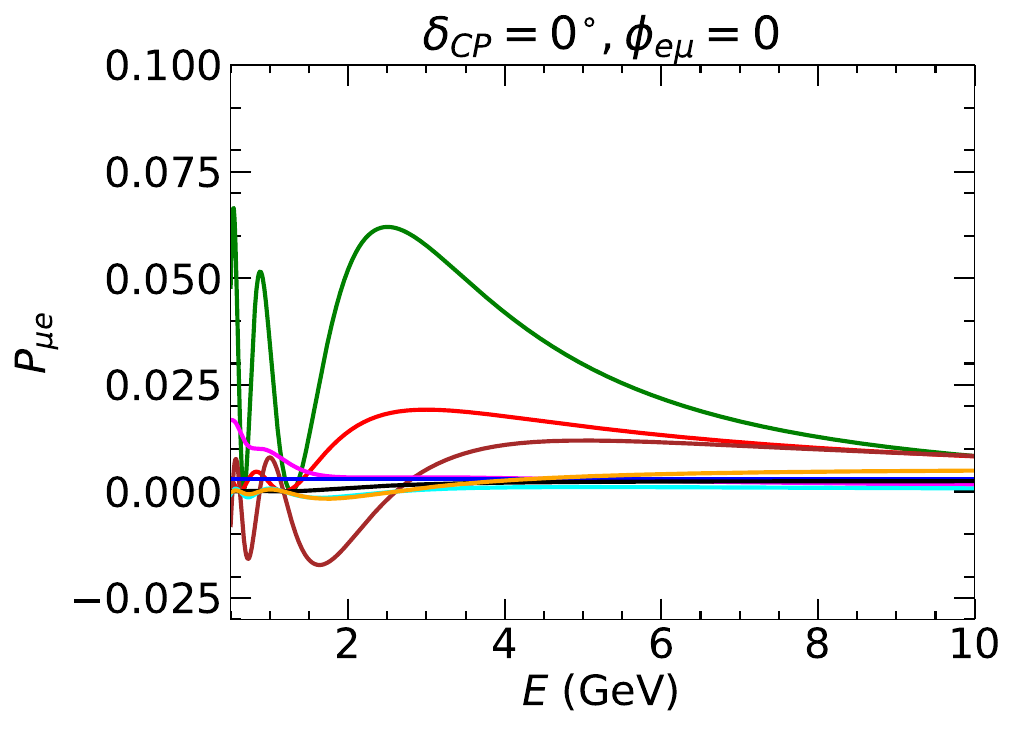}   \includegraphics[height=65mm, width=80mm]{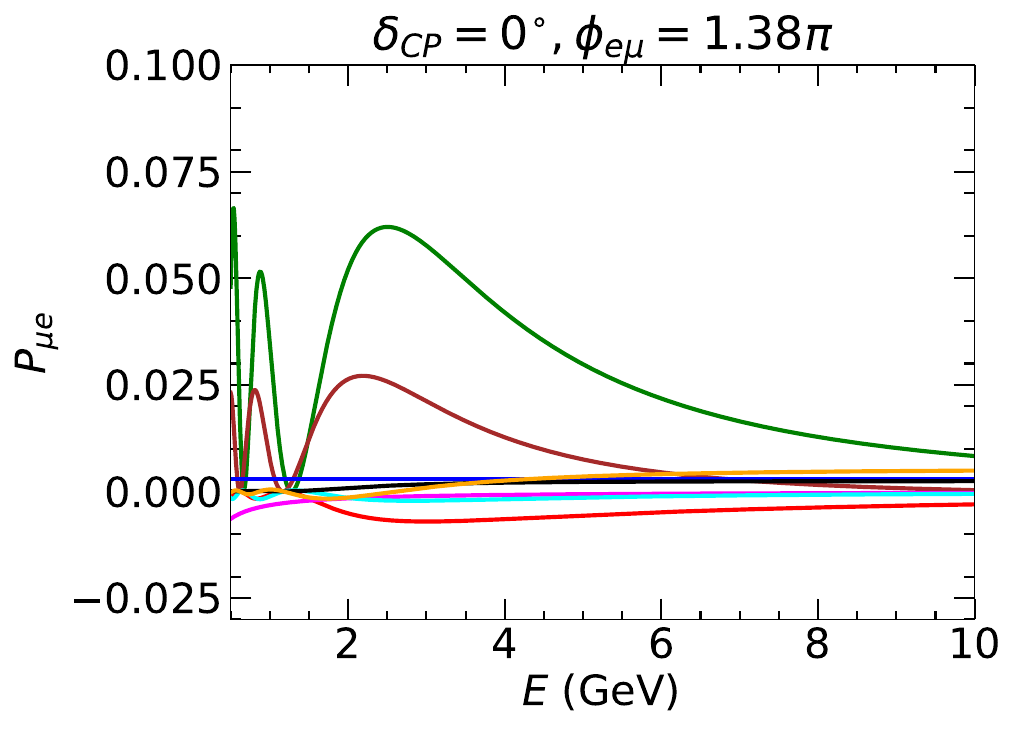}
    \caption{Upper (lower) row shows the effect of each term containing $\left |\epsilon_{e \mu} \right |$ and $\phi_{e \mu}$ in appearance probability as a function of neutrino energy with $\delta_{CP} = 212^{\circ} ~(0^{\circ})$. Left (right) panel shows the results with $\phi_{e \mu} = 0 ~(1.38 \pi$). Color codes are given in the legend. Each term is thoroughly explained within the text.}
    \label{fig21}
\end{figure}

\begin{figure}[htb]
   \hspace{-0.2cm}
\includegraphics[height=65mm, width=80mm]{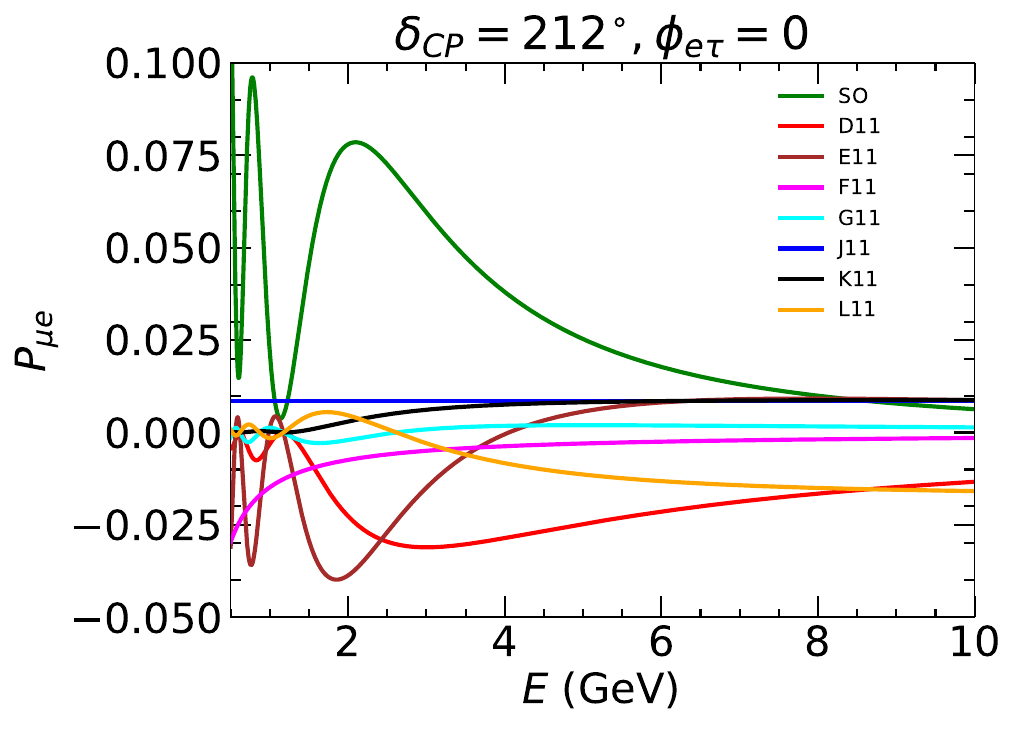}   \includegraphics[height=65mm, width=80mm]{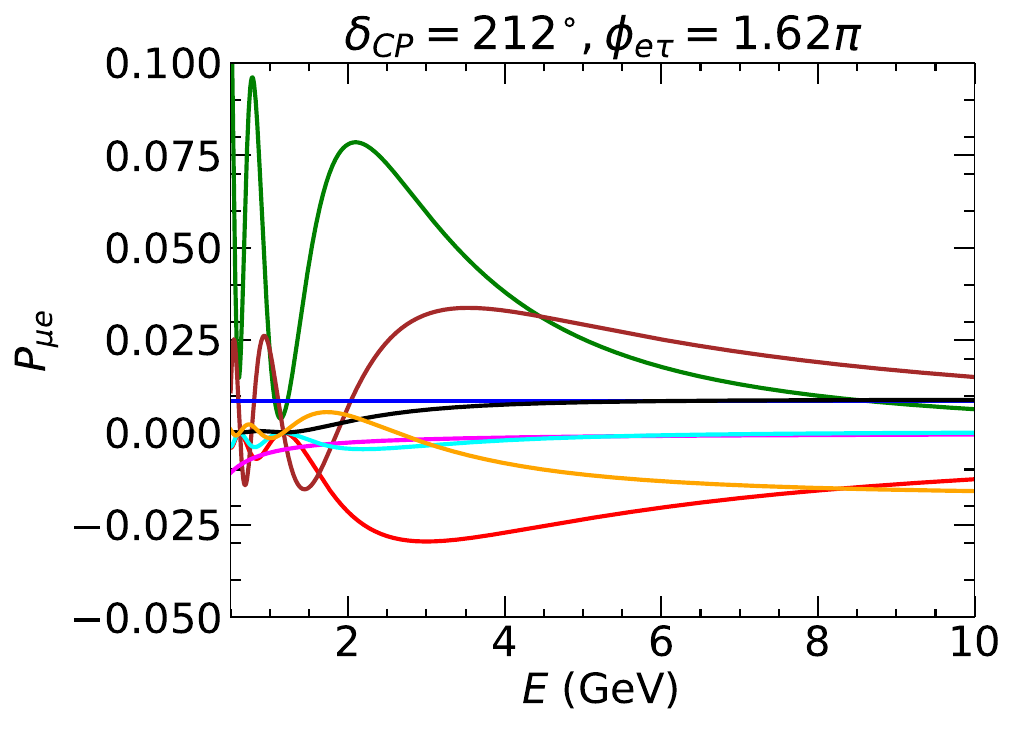}\\
\includegraphics[height=65mm, width=80mm]{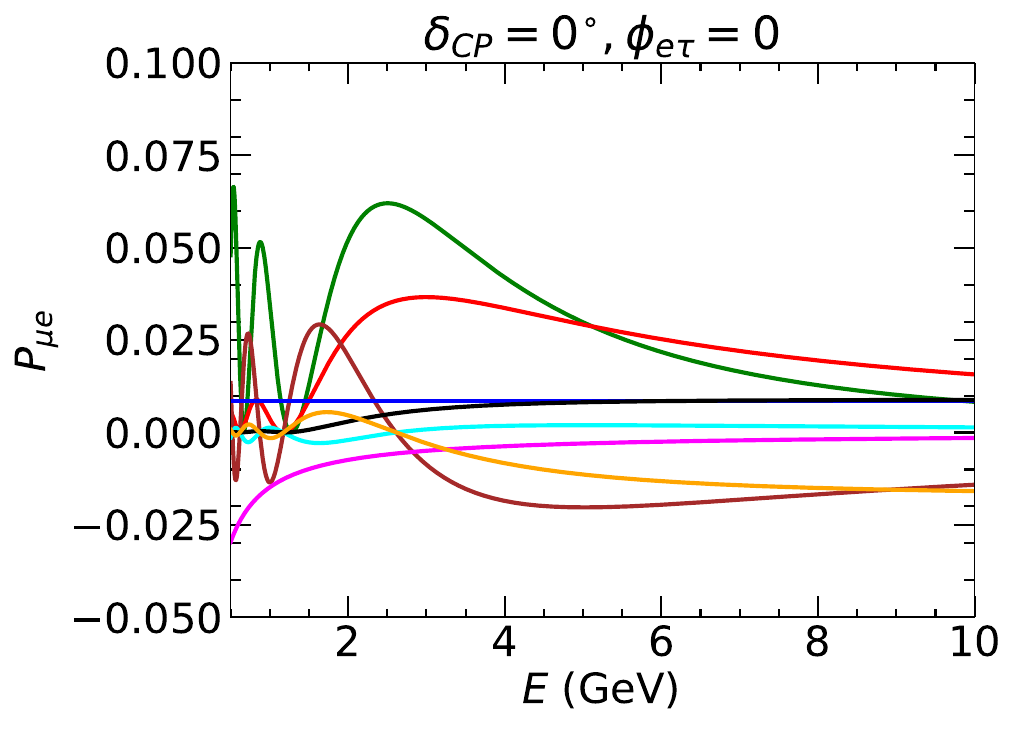}   \includegraphics[height=65mm, width=80mm]{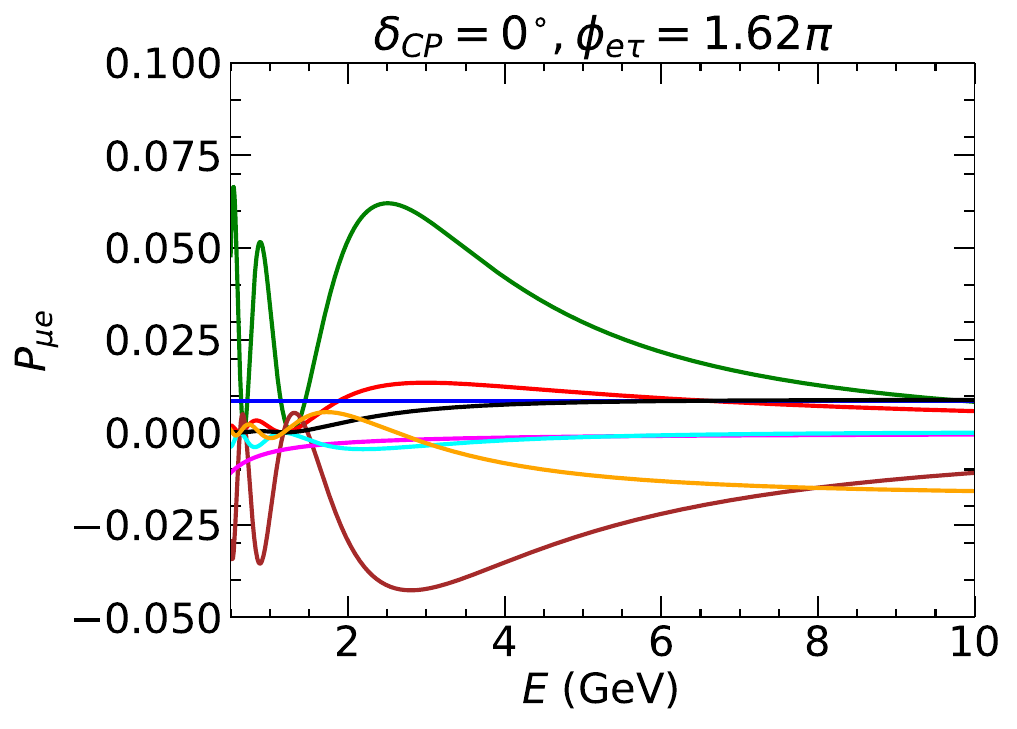}
    \caption{Upper (lower) row shows the effect of each term containing $\epsilon_{e \tau}$ and $\phi_{e \tau}$ in appearance probability as a function of neutrino energy with $\delta_{CP} = 212^{\circ} ~(0^{\circ})$. Left (right) panel shows the results with $\phi_{e \tau} = 0 ~(1.62 \pi$ rad). Color codes are given in the legend. For more about each term, see the text.}
    \label{fig31}
\end{figure}

Let's first focus on the effect of NSI parameters on the appearance channel. Fig. \ref{fig1} shows the appearance probability with respect to neutrino energy for different scenarios. Left (middle) panel shows the dependency of $\epsilon_{e \mu} ~(\epsilon_{e \tau})$ on $P_{\mu e}$ whereas right panel depicts the effect of $\epsilon_{\mu \tau}$. In generating these plots, we use the values of standard oscillation parameters from Table \ref{del1}. For NSI parameters, we take the values from Table \ref{nsi}. 
The specific values of the off-diagonal NSI parameters $ \left | \epsilon_{\alpha \beta}\right |$ and their associated phases $\phi_{\alpha \beta}$ considered in Table \ref{nsi} are motivated by their potential to resolve the observed tension between the NO$\nu$A and T2K experimental results in the measurement of the leptonic CP-violating phase $\delta_{\rm CP}$ and the atmospheric mixing angle $\theta_{23}$ \cite{Chatterjee:2020kkm,Denton:2020uda}.\\
In this analysis, we focus solely on the off-diagonal NSI parameters $\left |\epsilon_{\alpha \beta} \right |$ ($\alpha  \neq \beta$), incorporating their associated complex phases $\phi_{\alpha\beta}$. To simplify the discussion, we consider one NSI parameter at a time. This allows us to examine the distinct effects of each off-diagonal NSI term on the oscillation probabilities. We explore three cases for each off diagonal NSI parameters $\epsilon_{\alpha \beta}$: (i) the standard oscillation scenario (SO) (ii) NSI with a vanishing complex phase $(\phi_{\alpha\beta}=0)$, and (iii) NSI with a non-zero complex phase $(\phi_{\alpha\beta} \neq 0)$. In each panel of Fig.~\ref{fig1}, the red curve represents the standard oscillation (SO) scenario, while the green curve corresponds to the case with $\left |\epsilon_{\alpha \beta} \right |$ included. The blue curve illustrates the effect when both $\left |\epsilon_{\alpha \beta} \right |$ and its associated phase $\phi_{\alpha\beta}$ are taken into account. The shaded region corresponds to the DUNE $\nu_{\mu}$ neutrino flux in arbitrary units.  \\
The left-most panel of the upper row in Fig. \ref{fig1} displays the effect of the NSI parameter $\epsilon_{e\mu}$ on the appearance probability when we take $\delta_{CP}=212^{\circ}$. From the plot, it is observed that the green curve exhibits a higher amplitude than the red curve. In contrast, the blue curve shows a reduced amplitude compared to the red one.
This variation in amplitudes can be understood by examining the modified expression for $P_{\mu e}$ in the presence of NSI. Considering only $\left|\epsilon_{e \mu}\right|$ and $\phi_{e \mu}$ (with $\left|\epsilon_{e \tau}\right|,$ and $\phi_{e \tau}=0$), the appearance probability can be expanded, and the individual contributions are explicitly displayed in Fig. \ref{fig21} for a clearer understanding of the results in Fig.~\ref{fig1}. Similarly, to further interpret the features of Fig.~\ref{fig1}, we analyze the case with only $\left|\epsilon_{e \tau}\right|$ and $\phi_{e \tau}$ (with $\left|\epsilon_{e \mu}\right|,$ and $\phi_{e \mu}=0$), which is presented in Fig. \ref{fig31}.

To understand this behavior of Fig. \ref{fig1}, we refer to the modified expression for $P_{\mu e}$ in the presence of NSI, considering only $\left |\epsilon_{e \mu} \right |$ and $\phi_{e \mu}$ (i.e., setting $\left |\epsilon_{e \tau} \right |, \phi_{e\tau}= 0$). The corresponding appearance probability can be expanded as:
\begin{align}
\label{app-split}
P(\nu_\mu \to \nu_e)|_{\left |\epsilon_{e \tau} \right |}, \phi_{e \tau} = 0 &= SO + D1 + E1 + F1 + G1 + J1 + K1 + L1,
\end{align}
where 
\begin{eqnarray}
&&SO = x^2 f^2 + 2xyfg \cos(\Delta + \delta_{CP}) + y^2 g^2, ~~~~~
 D1 = 4 \hat{A} \left |\epsilon_{e \mu} \right | x f^2 s_{23}^2 \cos(\phi_{e \mu} + \delta_{CP}),\nonumber\\
 &&E1 = 4 \hat{A} \left |\epsilon_{e \mu} \right | x f g c_{23}^2 \cos(\phi_{e \mu} + \Delta + \delta_{CP}), ~~~~~~~~
 F1 = 4 \hat{A} \left |\epsilon_{e \mu} \right | y g^2 c_{23}^2 \cos(\phi_{e \mu}),\nonumber\\
  &&G1 = 4 \hat{A} \left |\epsilon_{e \mu} \right | y g f s_{23}^2 \cos(\Delta - \phi_{e \mu}),~~~~~~~~~~~~~~~~~
 J1 = 4 \hat{A}^2 g^2 \left |\epsilon_{e \mu} \right |^2 c_{23}^2,\nonumber\\
 &&K1 = 4 \hat{A}^2 f^2 \left |\epsilon_{e \mu} \right |^2 s_{23}^4,~~~~~~~
 L1 = 8 \hat{A}^2 f g \left |\epsilon_{e \mu} \right |^2 \sin \theta_{23} \cos \theta_{23} (\cos \theta_{23} \cos \Delta \sin \theta_{23}).    
\end{eqnarray}

The upper row of Fig. \ref{fig21} shows $P_{\mu e}$ as a function of neutrino energy for $\left |\epsilon_{e \mu} \right |$ = 0.15 with $\phi_{e \mu} = 0$ and $1.38\pi$. Individual contributions from Eq. \ref{app-split} reveal that D1 and E1 dominate. For $\phi_{e\mu} = 0$, D1 is negative while E1 enhances the amplitude up to around 4 GeV, then decreases. Despite opposing signs, their combined effect raises the overall amplitude without shifting the oscillation peak, as their extrema occur at the same energy.\\
For $\phi_{e \mu} = 1.38\pi$, the behavior changes notably. The E1 term now contributes negatively across the entire energy range, and the D1 term becomes nearly negligible. This arises from the cosine dependence in E1, $\cos(\phi_{e\mu} + \Delta + \delta_{CP})$, which is positive near the oscillation peak for $\phi_{e\mu}=0$ and $\delta_{CP} = 212^\circ$ but becomes entirely negative at $\phi_{e\mu}=1.38\pi$. Similarly, the D1 term, proportional to $\cos(\phi_{e\mu} + \delta_{CP})$, contributes minimally at $1.38\pi$. Consistently, in Fig. \ref{fig1} (upper-left panel), the green curve exceeds the red curve only up to $\sim$4 GeV, after which the red dominates, matching the trend in Fig. \ref{fig21} where the E1 term turns negative beyond 4 GeV for $\phi_{e \mu} = 0$. Unlike the green curve, the blue curve consistently shows a lower amplitude than the red curve for SO across the full energy range up to 8 GeV. This suppression is directly attributable to the negative E1 contribution shown in the right panel of the upper row of Fig. \ref{fig21}, which dominates when $\phi_{e \mu} = 1.38\pi$.

To investigate the combined effect of NSI and the CP-violating phase $\delta_{CP}$ on the appearance probability, we consider a CP-conserving scenario with $\delta_{CP} = 0^\circ$, as shown in the lower row in Fig. \ref{fig1}. The color scheme in this panel follows the same convention as the upper row: the red curve corresponds to standard oscillation, the green curve represents the case with only $\left |\epsilon_{e \mu} \right |$, and the blue curve includes both $\left |\epsilon_{e \mu} \right |$ and the phase $\phi_{e \mu}$. From the leftmost plot of the lower row of Fig. \ref{fig1}, it is evident that the red curve lies below both the green and blue curves across the entire energy range. Between the green and blue curves, the green curve shows a higher amplitude than the blue one. This behavior is further clarified in the lower panel of Fig. \ref{fig21}, which decomposes the individual contributions to $P_{\mu e}$ under the same parameter settings as the upper panel but with $\delta_{CP}=0^{\circ}$. The D1 and E1 terms are found to contribute most significantly to the overall modification of the appearance probability. The overall amplitude for the configuration with $\left |\epsilon_{e \mu} \right | = 0.15$ and $\phi_{e \mu} = 0$  is larger than that of the setup with $ \left | \epsilon_{e \mu} \right |= 0.15$ and $\phi_{e \mu} = 1.38\pi$; this trend is clearly visible in the lower panel of Fig. \ref{fig21}.

Now we analyze the impact of the NSI parameter $\epsilon_{e \tau} $ on the appearance probability. The middle panel of Fig. \ref{fig1} illustrates this effect, where the upper row corresponds to $\delta_{CP} = 212^\circ$ and the lower row to $\delta_{CP} = 0^\circ$. The red curve represents the standard oscillation scenario, the green curve shows the probability in the presence of non-zero $\left |\epsilon_{e \tau} \right |$, and the blue curve corresponds to the case with $\left |\epsilon_{e \tau} \right |$ = 0.27 and $\phi_{e \tau} = 1.62\pi$. For $\delta_{CP} = 212^\circ$, the green curve has a significantly lower amplitude than the red curve, indicating a suppression in the appearance probability due to the presence of $\epsilon_{e\tau}$. The blue curve, on the other hand, closely follows the red curve, suggesting only a minor effect from the inclusion of the NSI phase in this case. However, when $\delta_{CP} = 0^\circ$ in Fig. \ref{fig1}, the overall pattern changes. The green curve now shows a much higher amplitude than the red curve, while the blue curve exhibits the lowest amplitude among the three. This change highlights the sensitivity of the appearance probability to both the magnitude and phase of the NSI parameter $\epsilon_{e \tau}$, as well as its strong interplay with the CP-violating phase $\delta_{CP}$.

To understand the behavior of the effect of $\left |\epsilon_{e \tau} \right |$ on $P_{\mu e}$, we split each term of the appearance probability given in Eq. \ref{app} as, by taking $\epsilon_{e \tau}$ and $\phi_{e \tau}$ only, keeping $\epsilon_{e \mu} = 0$,
\begin{align}
\label{app-split-etau}
P(\nu_\mu \to \nu_e)|_{\left |\epsilon_{e \mu} \right |, \phi_{e \mu} = 0} &= SO + D11 + E11 + F11 + G11 + J11 + K11 + L11,
\end{align}
 where, the expression of $SO$ is given in eq. \ref{app-split} and the other terms are given as
 \begin{eqnarray}
&&D11= 4 \hat{A} \left |\epsilon_{e \tau} \right |s_{23} c_{23} x f^2 \cos(\phi_{e \tau} + \delta_{CP}),~~~~ E11= - 4 \hat{A} \left |\epsilon_{e \tau} \right | s_{23} c_{23} x f g \cos(\Delta + \phi_{e \tau} + \delta_{CP}),\nonumber\\ &&F11= -4 \hat{A} \left |\epsilon_{e \tau} \right | s_{23} c_{23} y g^2 \cos \phi_{e \tau}, ~~~~~~~~~~~~ G11= 4 \hat{A} \left |\epsilon_{e \tau} \right | s_{23} c_{23} y g f \cos(\Delta - \phi_{e \tau}),\nonumber\\ &&J11= 4 \hat{A}^2 g^2  \left |\epsilon_{e \tau} \right |^2 s^2_{23} c^2_{23},~~~~K11= 4 \hat{A}^2 f^2  \left |\epsilon_{e \tau} \right |^2 s^2_{23} c^2_{23},~~~~L11= -8 \hat{A}^2 f g  \left |\epsilon_{e \tau} \right |^2 s^2_{23} c^2_{23} \cos \Delta. \nonumber\\   
 \end{eqnarray}

Figure \ref{fig31} presents the energy dependence of the individual terms contributing to the appearance probability. The upper and lower panels correspond to $\delta_{CP} = 212^\circ$ and $\delta_{CP} = 0^\circ$, respectively, while the left and right columns show results for $\phi_{e \tau} = 0$ and $\phi_{e \tau} = 1.62\pi$.
In the case of $\delta_{CP} = 212^\circ$, it is evident that the terms D11, E11, and L11 dominate the behavior of $P_{\mu e}$, with the remaining terms contributing negligibly. For $\phi_{e\tau} = 0$, all three dominant terms, D11, E11, and L11 exhibit negative amplitudes throughout the energy range, leading to an overall suppression of the appearance probability compared to the standard oscillation scenario. This observation aligns with the green curve in the middle panel of the upper row in Fig. \ref{fig1}. When $\phi_{e \tau} = 1.62\pi$, the E11 term becomes positive, while D11 and L11 remain negative. This results in a slight enhancement of the overall amplitude compared to the case with $\phi_{e \tau} = 0$, although the appearance probability is still close to that of standard oscillations, as shown by the blue curve in the corresponding panel of Fig. \ref{fig1}.

For the CP-conserving case with $\delta_{CP} = 0^\circ$, the behavior changes considerably. The lower panel of Fig. \ref{fig31} reveals that when $\phi_{e\tau} = 0$, D11 and E11 contribute positively, while L11 remains negative. The combined effect of these terms leads to a significantly higher appearance probability than that of standard oscillations, which is consistent with the green curve in the middle column of the lower row of Fig. \ref{fig1}.
However, for $\phi_{e\tau} = 1.62\pi$, E11 and L11 contribute large negative amplitudes, while D11 shows positive amplitude. This results in a strong suppression of the overall appearance probability, reflected in the blue curve in the corresponding panel of Fig. \ref{fig1}.

Finally, we examine the influence of $\epsilon_{\mu \tau}$ on the appearance probability. The right-most panels of the upper and lower rows of Fig. \ref{fig1} illustrate this effect for $\delta_{CP} = 212^\circ$ and $\delta_{CP} = 0^\circ$, respectively. The color scheme remains consistent with the previous panels: the red curve corresponds to standard oscillations, the green curve represents the scenario with non-zero $\left |\epsilon_{\mu \tau} \right |$ and $\phi_{\mu \tau} = 0$, and the blue curve shows the result for $\left |\epsilon_{\mu \tau} \right |$ = 0.35 and $\phi_{\mu \tau} = 0.6\pi$. As observed in both CP-violating and CP-conserving cases, the green and blue curves exhibit only slight deviations from the red curve. This indicates that the presence of $\epsilon_{\mu\tau}$ leads to relatively minor modifications in the appearance probability.
This behavior is consistent with the analytical expression of $P_{\mu e}$ given in Eq. \ref{app}, which shows that $\epsilon_{\mu \tau}$ does not contribute to the probability up to $\mathcal{O}(\epsilon^2)$. However, if the expression is expanded to higher orders, specifically beyond $\mathcal{O}(\epsilon^2)$, subleading terms involving $\epsilon_{\mu\tau}$ may appear. These higher-order contributions can account for the small deviations observed in the rightmost column of Fig. \ref{fig1}.

\begin{figure}[htb]
    \hspace{-0.2cm}
    \includegraphics[height=50mm, width=54mm]{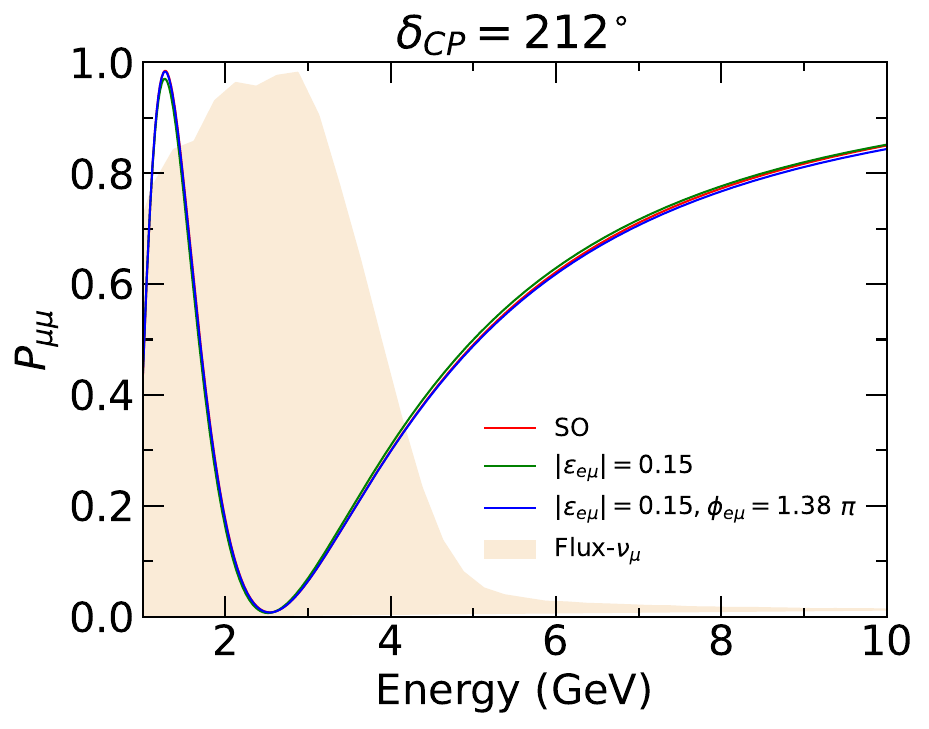}
\includegraphics[height=50mm, width=54mm]{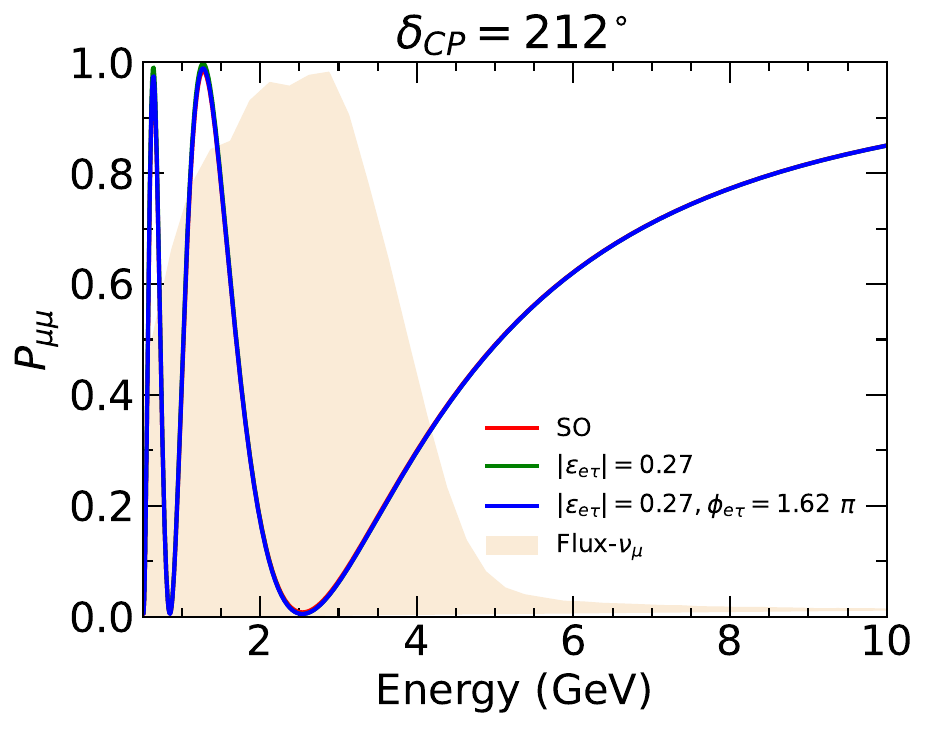} 
\includegraphics[height=50mm, width=54mm]{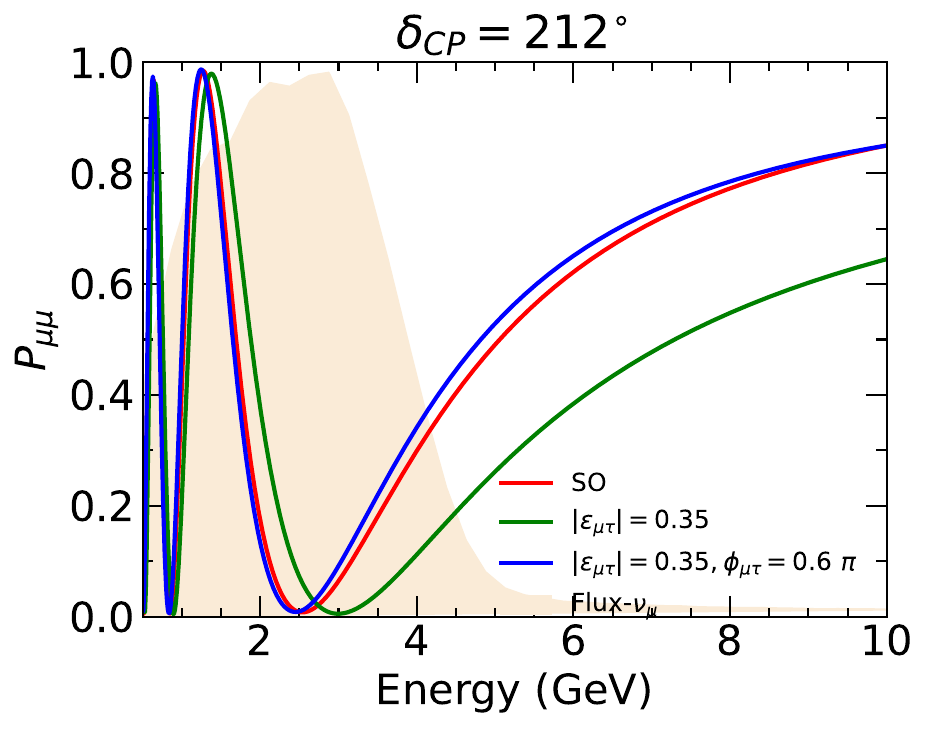}
    \caption{The disappearance probability as a function of neutrino energy for the DUNE experiment with $\delta_{CP} = 212^{\circ}$. Left (middle) [right] panel represent the curves with one off-diagonal parameter one at a time, $\epsilon_{e \mu}$ ($\epsilon_{e \tau}$) [$\epsilon_{\mu \tau}$] and their associated phases. Color codes are given in the legend.}
    \label{fig-pmumu}
\end{figure}

Now, we explore the impact of NSI parameters on the disappearance probability. According to Eq. \ref{dis}, the disappearance channel $P_{\mu \mu}$ is primarily influenced by $\left |\epsilon_{\mu \tau} \right |$ and its associated phase $\phi_{\mu \tau}$. Figure \ref{fig-pmumu} illustrates the effects of three different NSI parameters on the disappearance probability separately. The left and middle columns show the effects of $\epsilon_{e \mu}$ and $\epsilon_{e \tau}$, respectively, while the rightmost column presents the impact of $\epsilon_{\mu \tau}$.
The color scheme remains consistent throughout: the red curve represents the standard oscillation scenario, the green curve shows the result when $\left |\epsilon_{\alpha \beta} \right |$ is non-zero with the corresponding phase set to zero, and the blue curve corresponds to the case where both $\left |\epsilon_{\alpha \beta} \right |$ and $\phi_{\alpha \beta}$ are non-zero. The shaded area represents the $\nu_{\mu}$ flux for the DUNE experiment in arbitrary units.

\begin{figure}[htb]
   \hspace{-0.2cm}
\includegraphics[height=65mm, width=80mm]{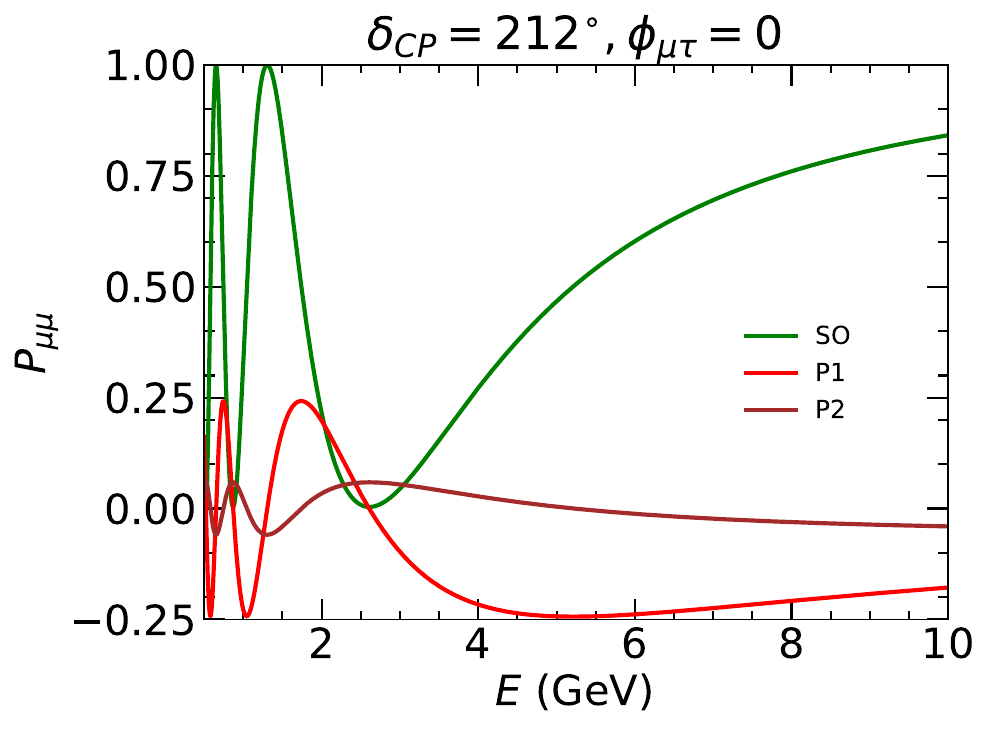}   \includegraphics[height=65mm, width=80mm]{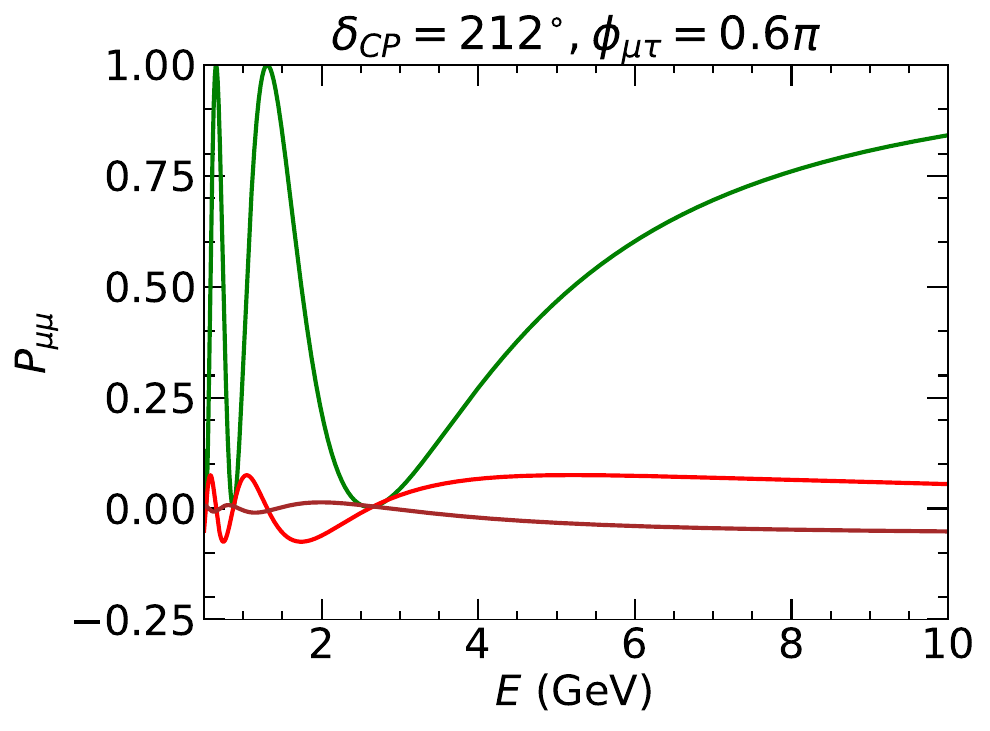}
    \caption{The effect of each term containing $\left |\epsilon_{\mu \tau} \right |$ and $\phi_{\mu \tau}$ in disappearance probability as a function of neutrino energy with $\delta_{CP} = 212^{\circ}$. Left (right) panel shows the results with $\phi_{\mu \tau} = 0 ~(0.6\pi$ rad). Color codes are given in the legend. For more about each term, see the text.}
    \label{fig41}
\end{figure}

From the figure (Fig. \ref{fig-pmumu}), it is evident that when either $\epsilon_{e \mu}$ or $\epsilon_{e\tau}$ is present, the green and blue curves show negligible deviations from the standard oscillation (red curve). This observation is consistent with the analytical form of Eq. \ref{dis}, which shows independent behaviour on these parameters in the disappearance channel.
However, the presence of $\left |\epsilon_{\mu \tau} \right |$ and its phase leads to a noticeable change in the disappearance probability. The rightmost column clearly shows a significant deviation of the green curve from the red curve, indicating the strong effect of $\left |\epsilon_{\mu \tau} \right |$ when the phase is set to zero. Interestingly, the blue curve, which includes both $\left |\epsilon_{\mu \tau} \right |$ and $\phi_{\mu \tau}$, mostly overlaps with the red curve. Furthermore, there is an observable shift in the oscillation minimum for the green curve relative to the red and blue curves, highlighting the behavior of $P_{\mu \mu}$ to $\left |\epsilon_{\mu \tau} \right |$ in the absence of a complex phase.
To understand the nature, we need to split the modified expression of $P_{\mu \mu}$, 
\begin{align}
\label{dis-split}
    P (\nu_\mu \to \nu_\mu) &= SO  + P1 + P2,
\end{align}
where $SO$ is the first five terms of Eq. \ref{dis} and 
\begin{eqnarray}
&&P1=-2 \hat{A} \left |\epsilon_{\mu \tau} \right | \cos\phi_{\mu\tau} \Big[ \sin^3 2\theta_{23} \Delta \sin 2\Delta + 2 \sin 2\theta_{23} \cos^2 2\theta_{23} \sin^2 \Delta \Big]\nonumber\\
&&P2=- 2 \hat{A}^2 \sin^2 2\theta_{23} \left |\epsilon_{\mu \tau} \right |^2 \Big[ 2 \sin^2 2\theta_{23} \cos^2\phi_{\mu\tau} \Delta^2 \cos 2\Delta + \sin^2\phi_{\mu\tau} \Delta \sin 2\Delta \Big] . 
\end{eqnarray}

Figure \ref{fig41} illustrates the contribution of each term to the disappearance probability. Let us first consider the scenario where $\left |\epsilon_{\mu \tau} \right |$ = 0.35 and $\phi_{\mu \tau} = 0$. In this case, the left panel of Fig. \ref{fig41} shows that the term P1 causes a noticeable shift in the oscillation minimum compared to the standard oscillation. Additionally, beyond the oscillation minimum, the P1 curve exhibits a negative amplitude, leading to an overall suppression of the probability relative to the standard case.

However, when the phase is set to $\phi_{\mu\tau} = 0.6\pi$, both P1 and P2 have a negligible impact on the standard oscillation behavior. As a result, the corresponding blue curve in the rightmost column of Fig. \ref{fig-pmumu} nearly overlaps with the standard oscillation curve, indicating minimal deviation.

\subsection{Analysis in event level}

 To further complement the probability level study, we simulate the expected event rates at DUNE using the General Long-Baseline Experiment Simulator (GLoBES) framework \cite{Huber:2004ka, Huber:2007ji}, where the probability engine has been modified to include the effects of off-diagonal NSI parameters. While probabilities provide the underlying theoretical description of oscillations, experiments measure the corresponding event rates. However, these two quantities are not independent, the event rates are derived from oscillation probabilities by incorporating the source flux, neutrino interaction cross sections, and detector response. Consequently, the trends observed at the probability level are directly reflected in the event rates.

\begin{figure}[htb]
   \hspace{-0.2cm}
   \includegraphics[height=50mm, width=54mm]{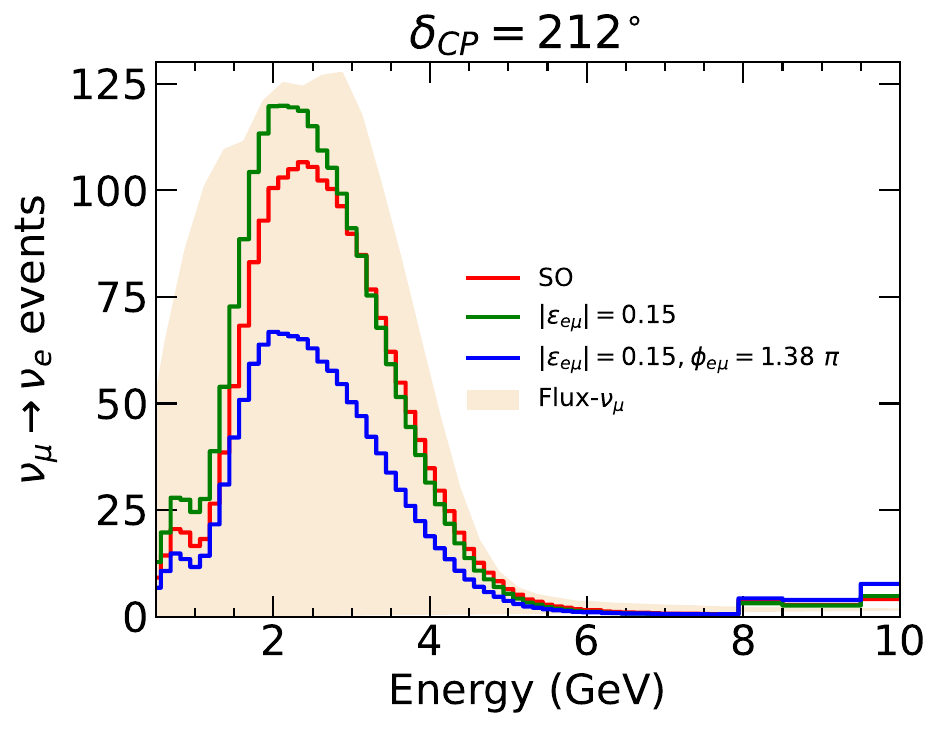}
\includegraphics[height=50mm, width=54mm]{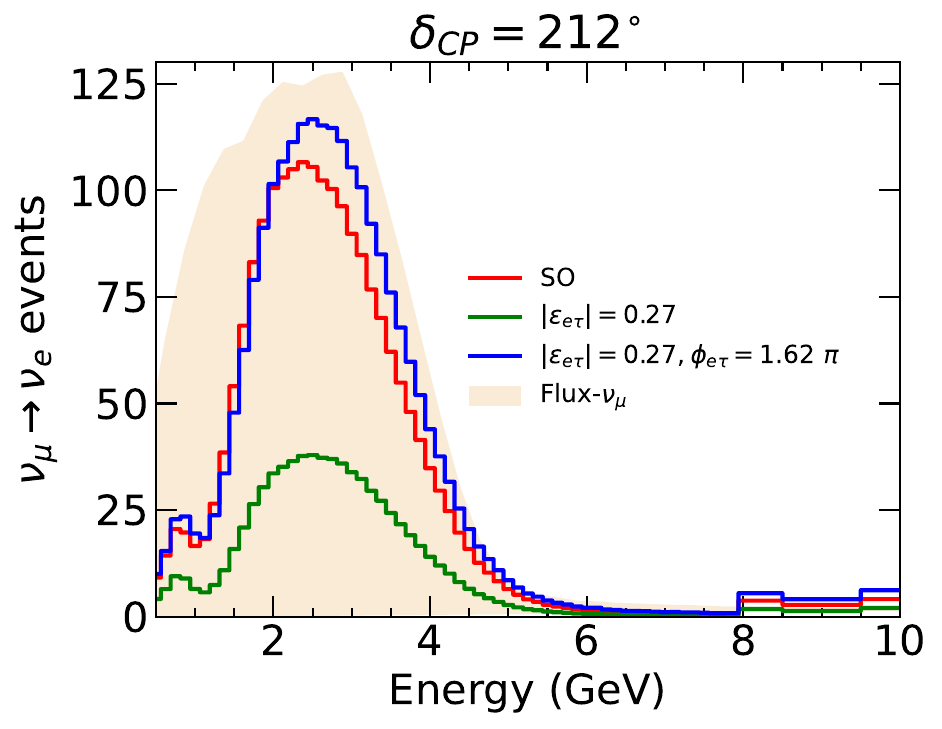}   \includegraphics[height=50mm, width=54mm]{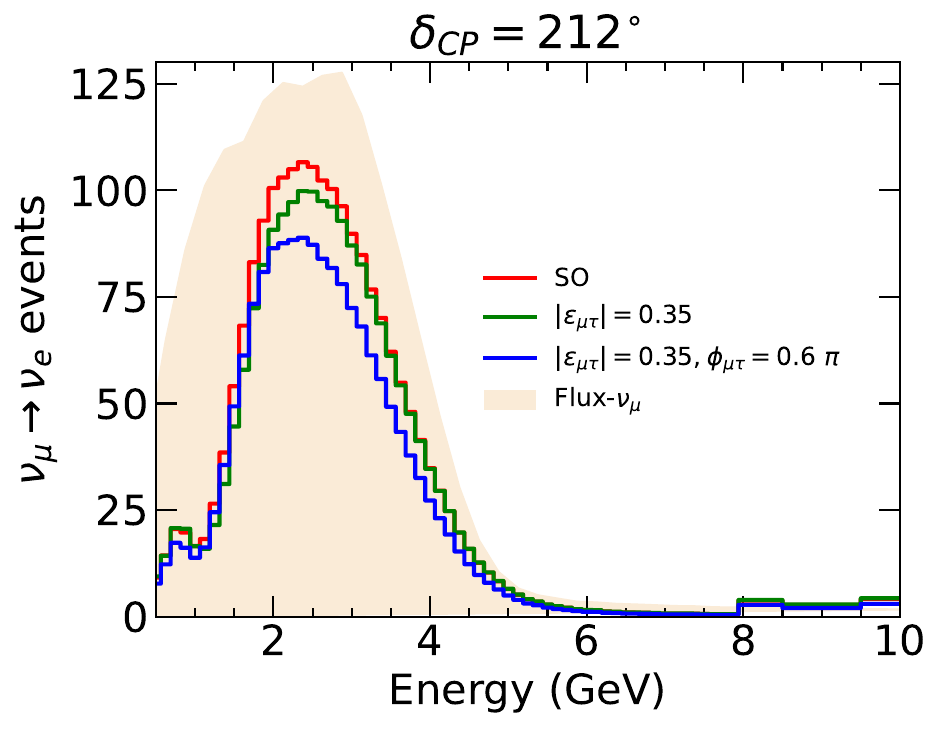}
\\
\includegraphics[height=50mm, width=54mm]{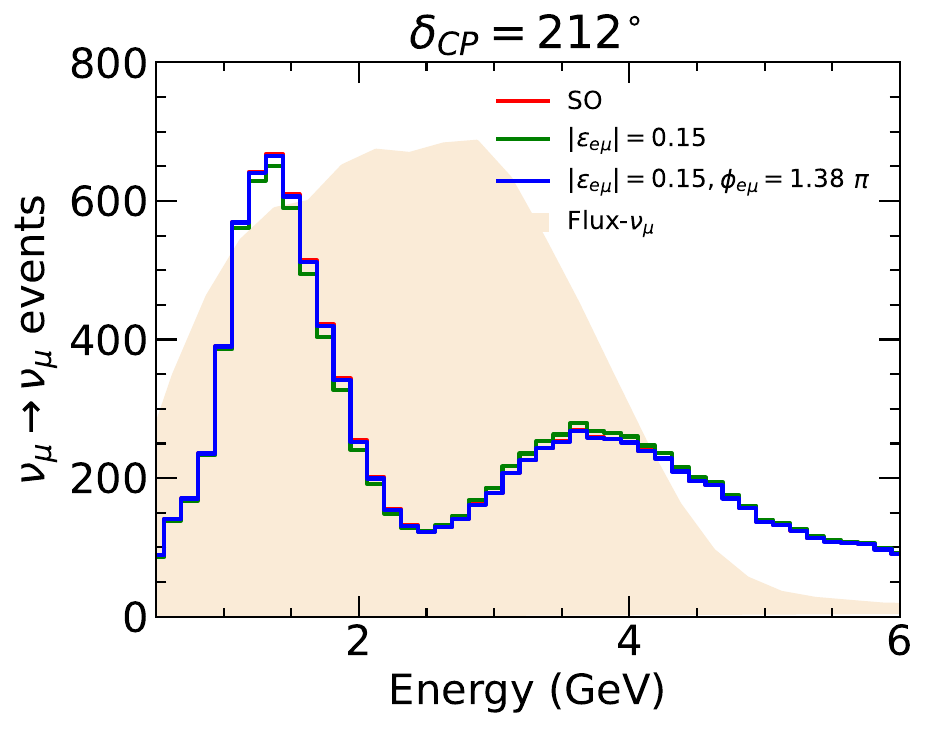}
\includegraphics[height=50mm, width=54mm]{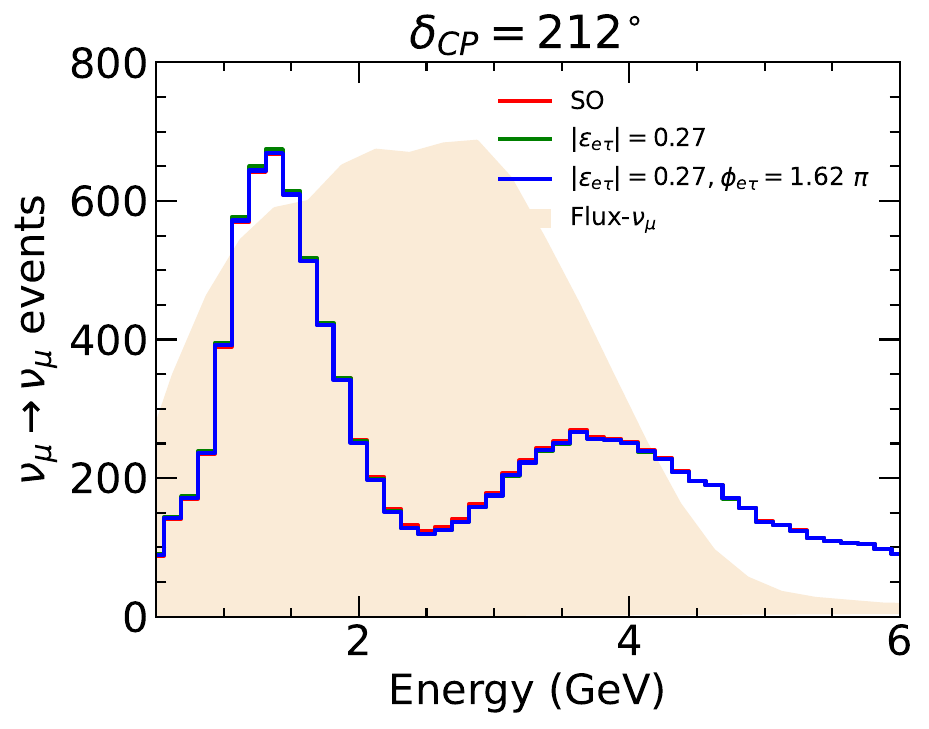}   \includegraphics[height=50mm, width=54mm]{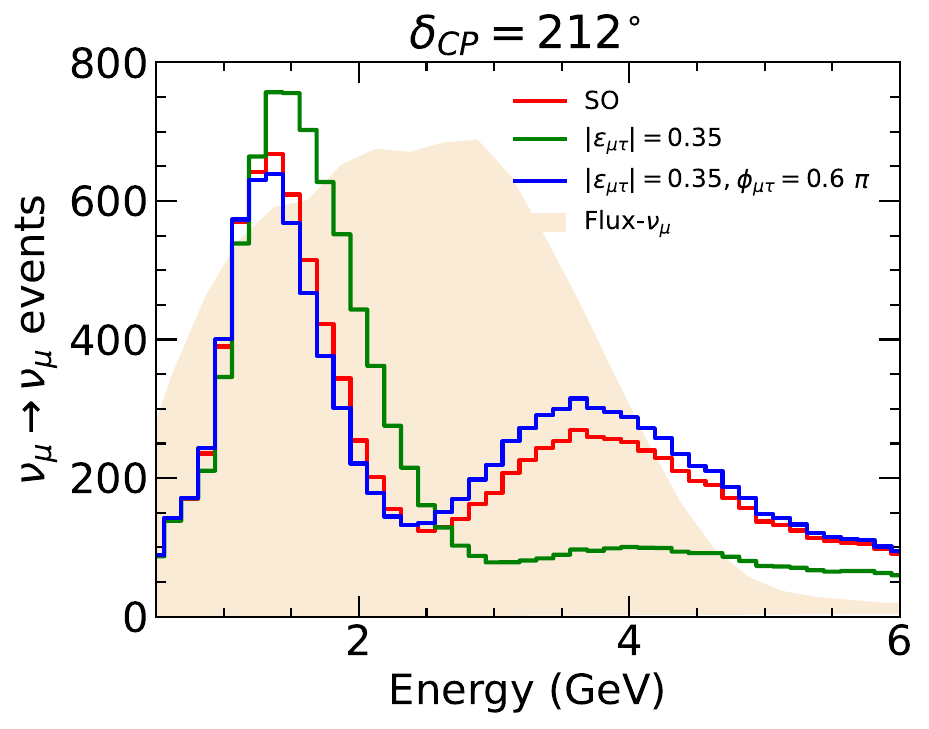}
    \caption{Figure shows the event rates for appearance (upper) and disappearance (lower) channels as a function of neutrino energy for the DUNE experiment with $\delta_{CP} = 212^{\circ}$. Left (middle) [right] panel represent the curves with one off-diagonal parameter at a time, $\epsilon_{e \mu}$ ($\epsilon_{e \tau}$) [$\epsilon_{\mu \tau}$] and their associated phases. Color codes are given in the legend.}
    \label{fig1-event}
\end{figure}

The general expression for the differential event rate in a given interaction channel can be written as \cite{Huber:2004ka, Huber:2007ji}
\begin{equation} \label{eventr}
\frac{dn^{\mathrm{IT}}_{\beta}}{dE'} =
N \int_{0}^{\infty} dE \int_{0}^{\infty} d\hat{E} \,
\underbrace{\Phi_{\alpha}(E)}_{\text{Source}} \,
\underbrace{\frac{1}{L^{2}} P(\alpha \rightarrow \beta)(E,L,\rho;\theta_{12},\theta_{13},\theta_{23},\Delta m^{2}_{31},\Delta m^{2}_{21},\delta_{\mathrm{CP}})}_{\text{Propagation}} ,
\end{equation}
\[
\times \underbrace{\sigma^{\mathrm{IT}}_{f}(E)\, k^{\mathrm{IT}}_{f}(E-\hat{E})}_{\text{Interaction}}
\times \underbrace{T_{f}(\hat{E})\, V_{f}(\hat{E}-E')}_{\text{Detection}} .
\]
Here, $\alpha$ denotes the initial neutrino flavor and $\beta$ the detected flavor. The quantity $\Phi_{\alpha}(E)$ corresponds to the neutrino flux at the source of the initial flavor $\alpha$, $L$ is the baseline length, $N$ represents an overall normalization factor, and $\rho$ is the matter density along the path.  
The different energies in this formula have different meanings, such as $E$ is the true energy of the incoming neutrino, which cannot be directly observed, $\hat{E}$ is the energy of the produced secondary particle and $E'$ denotes the reconstructed neutrino energy, i.e., the experimentally measured value.\\
Fig. \ref{fig1-event} displays the simulated event rates at DUNE as functions of neutrino energy, considering the influence of the off-diagonal NSI parameters $\epsilon_{e\mu}$, $\epsilon_{e\tau}$ and $\epsilon_{\mu\tau}$. Upper panel for the appearance channel and lower panel for the disappearance channel, with no background contributions included. In each panel, the red curve corresponds to the standard oscillation scenario, the green curve to the case with a non-zero real part of the NSI parameter, and the blue curve to the full complex parameter with its associated phase. The shaded regions denote the neutrino flux distribution.\\
A clear correspondence with the probability level plots of Fig. \ref{fig1} and Fig. \ref{fig-pmumu} is observed. For the appearance channel ($P_{\mu e}$), deviations in the event rate in Fig. \ref{fig1-event} closely follow those seen in the probability plots in Fig. \ref{fig1}. The suppression or enhancement of the blue and green curves relative to the red standard oscillation case is mirrored in the number of electron neutrino events. Around the flux peak ($ \sim$2.5 GeV), these deviations are most pronounced.\\
For the disappearance channel, the event rate spectra shown in the lower panels of Fig. \ref{fig1-event} mirror the behavior of the survival probability $P_{\mu\mu}$ in Fig. \ref{fig-pmumu}. As in the disappearance probability analysis, the event distributions are largely insensitive to 
$\epsilon_{e\mu}$ and $\epsilon_{e\tau}$, with the corresponding curves almost overlapping with the standard oscillation case. In contrast, the parameter $\epsilon_{\mu\tau}$ produces a visible shift. Thus, the disappearance event rates clearly follow the same trend as Fig. \ref{fig-pmumu}.\\
Overall, the event-level results confirm that the modifications due to NSI parameters are not independent from those seen in the oscillation probabilities. Instead, the event rates naturally follow the same trends, with the DUNE probability depending on the energy range. Thus, the analysis at the probability level provides the fundamental explanation for the event rate patterns observed in the simulations. For this reason, we only present the event rates to demonstrate their similarity with the corresponding probabilities. We have not performed any sensitivity study or incorporated systematic effects such as flux normalization, cross-section, or detector efficiency uncertainties. Since the main emphasis of this paper is on probabilities rather than on experimental sensitivity, systematic uncertainties are also not considered here.

\subsection{Analysis based on quantum entanglement measures}

In this subsection, we discuss how the off-diagonal NSI parameters influence the entanglement measures: Entanglement of Formation (EOF), Concurrence, and Negativity. Since all three entanglement measures are functions of neutrino oscillation and survival probabilities, the impact of NSI parameters on these quantities is essentially determined by their influence on the oscillation probabilities. The event rates and the corresponding oscillation probabilities are correlated, as shown in Eq. \eqref{eventr}, the overall structure and trends of the event rates follow those of the underlying probabilities. The quantum correlation measures considered in this work cannot be extracted directly from event rates but are expressed solely as functions of these oscillation probabilities. These measures are defined in terms of the neutrino density matrix, which itself depends on the oscillation probabilities. Since the probabilities can be inferred from the measured event rates, the corresponding quantum correlation measures can be accessed indirectly through experimental data, even though they are not directly observable quantities.
\\
Quantum correlation measures have been widely investigated in neutrino systems, consistently revealing nonclassical features in oscillation phenomena. A key motivation behind such studies is to examine whether neutrino oscillations being inherently quantum mechanical could be mimicked or altered by alternative formulations of quantum theory. In this work, we focus on the role of non-standard interactions (NSI) and explore how they affect various quantum correlation measures, with particular attention to entanglement.
\\
The quantities plotted in our work are standard entanglement measures, Entanglement of Formation, Concurrence and Negativity constructed from the neutrino oscillation probabilities. By definition these measures are zero only when the underlying neutrino state is unentangled and become larger as quantum correlations increase; hence the value of the plotted function directly indicates the degree of entanglement between flavor states.\\
Entanglement provides a complementary lens for studying quantum correlations in neutrino oscillations, as it can be expressed directly in terms of oscillation probabilities. Since our study emphasizes the impact of NSI on entanglement, any modifications to oscillation probabilities induced by NSI are reflected in the entanglement measure. Therefore, identifying NSI effects through probability measurements can be complemented by observing their imprint on entanglement. This dual perspective enriches our understanding of NSI and improves our ability to assess their impact on neutrino dynamics.
\begin{figure}[htb]
\hspace{-0.2cm}
\includegraphics[height=50mm, width=54mm]{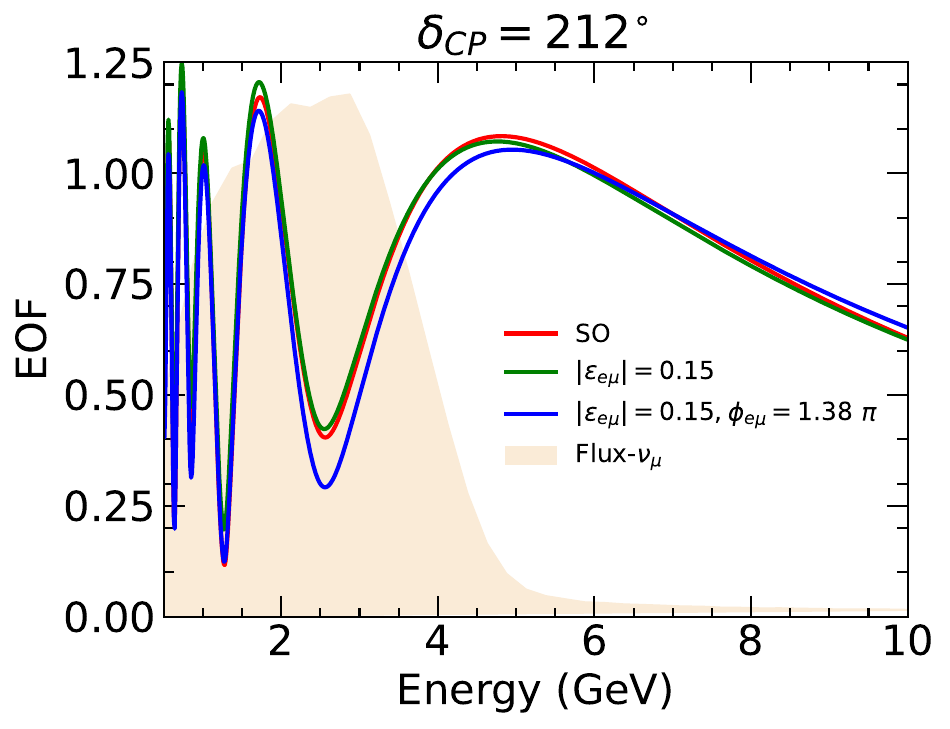}
\includegraphics[height=50mm, width=54mm]{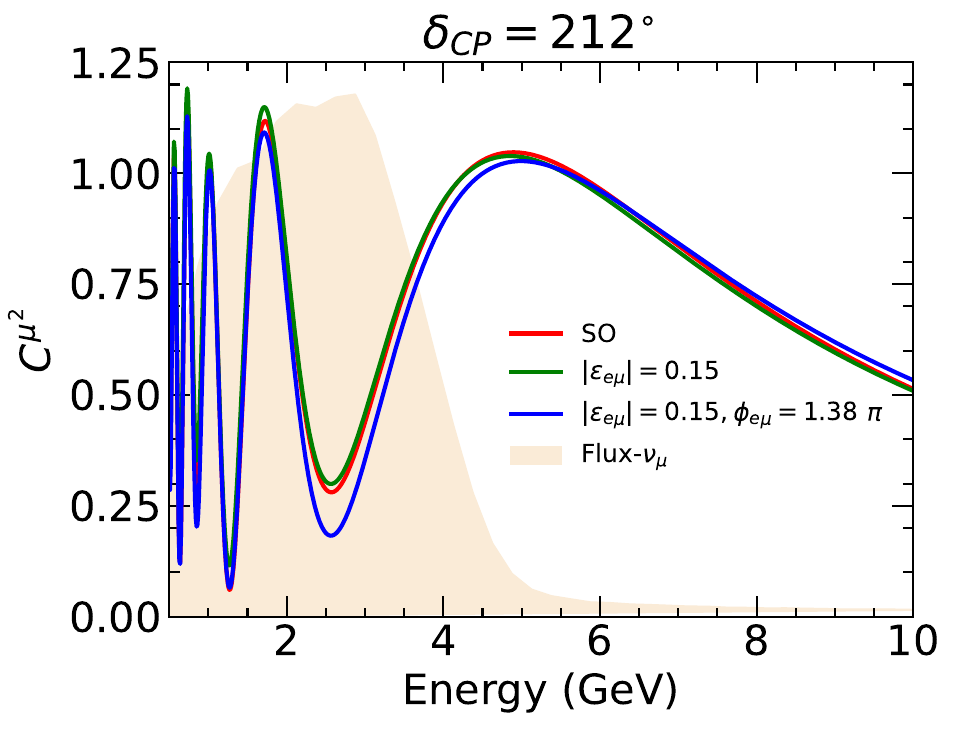}   \includegraphics[height=50mm, width=54mm]{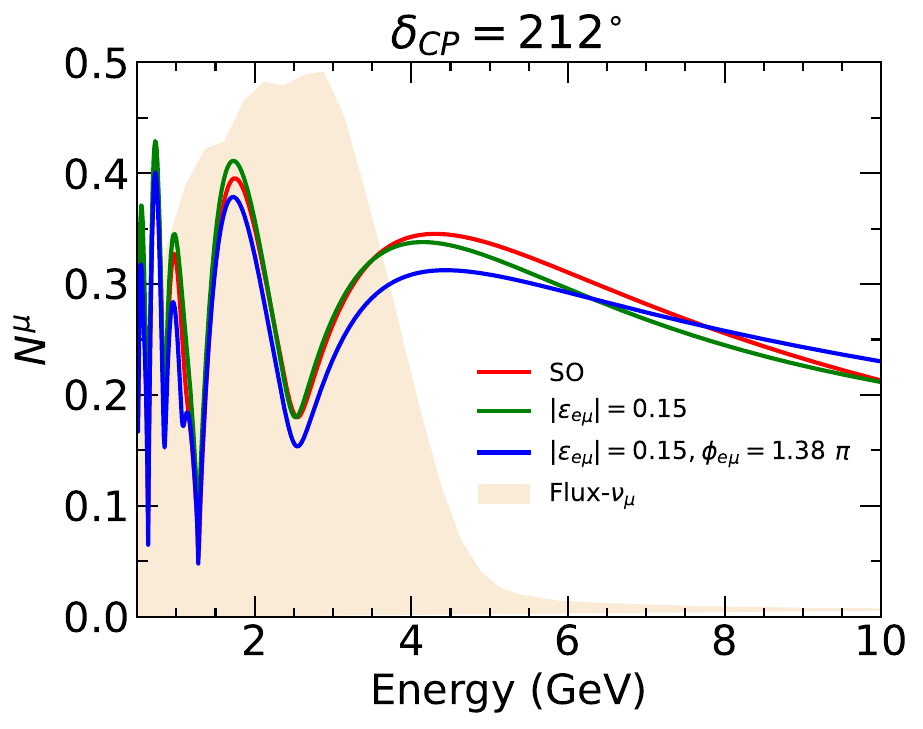}\\
\includegraphics[height=50mm, width=54mm]{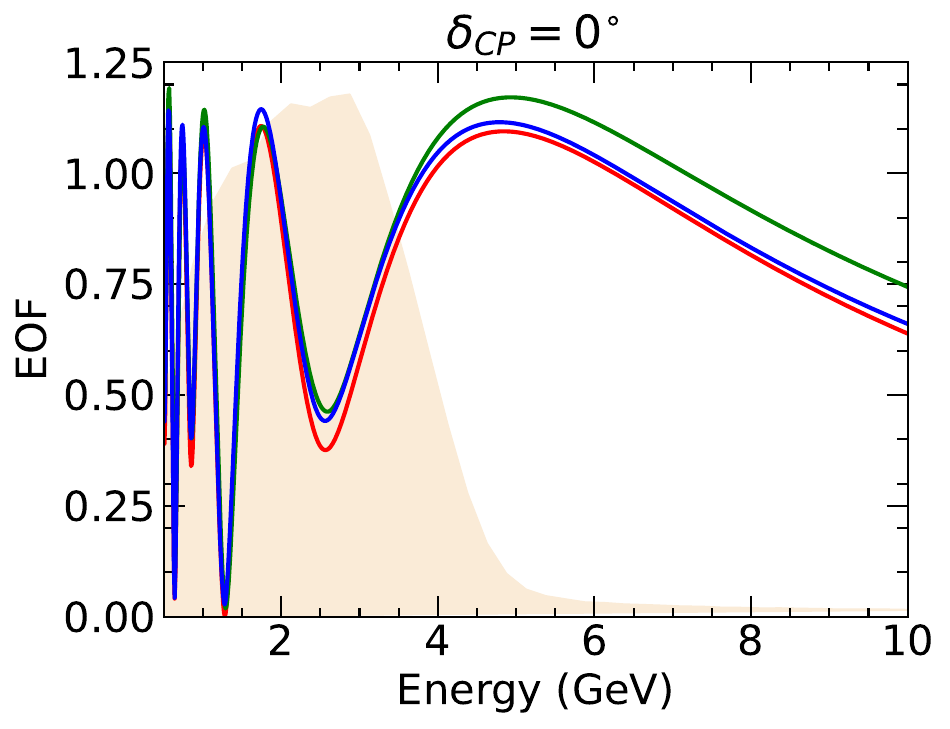}
\includegraphics[height=50mm, width=54mm]{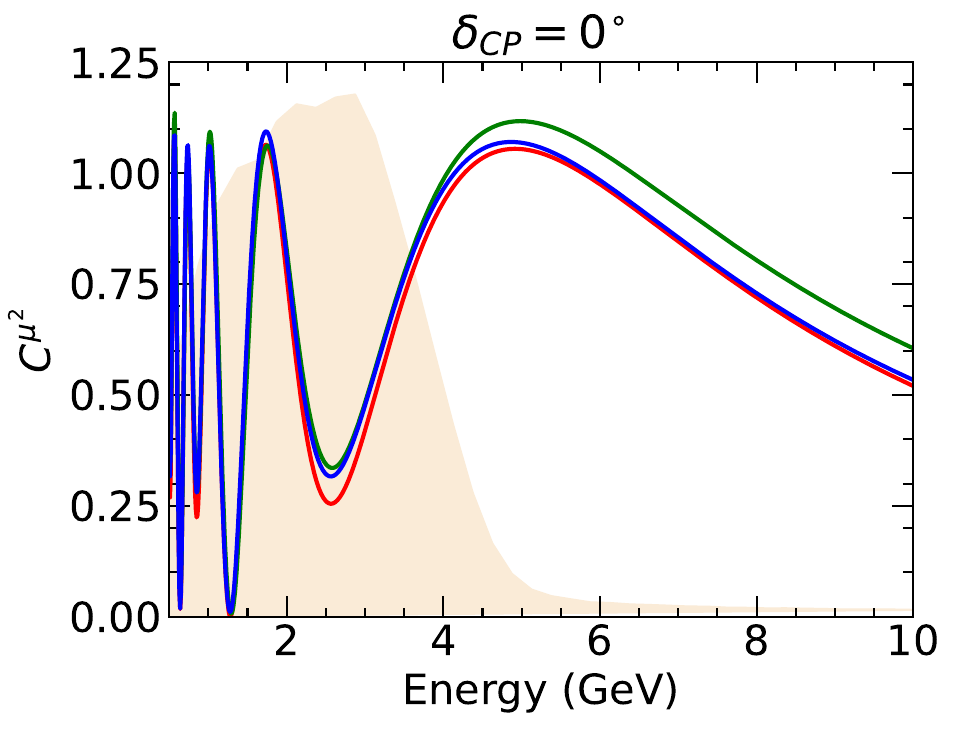}   \includegraphics[height=50mm, width=54mm]{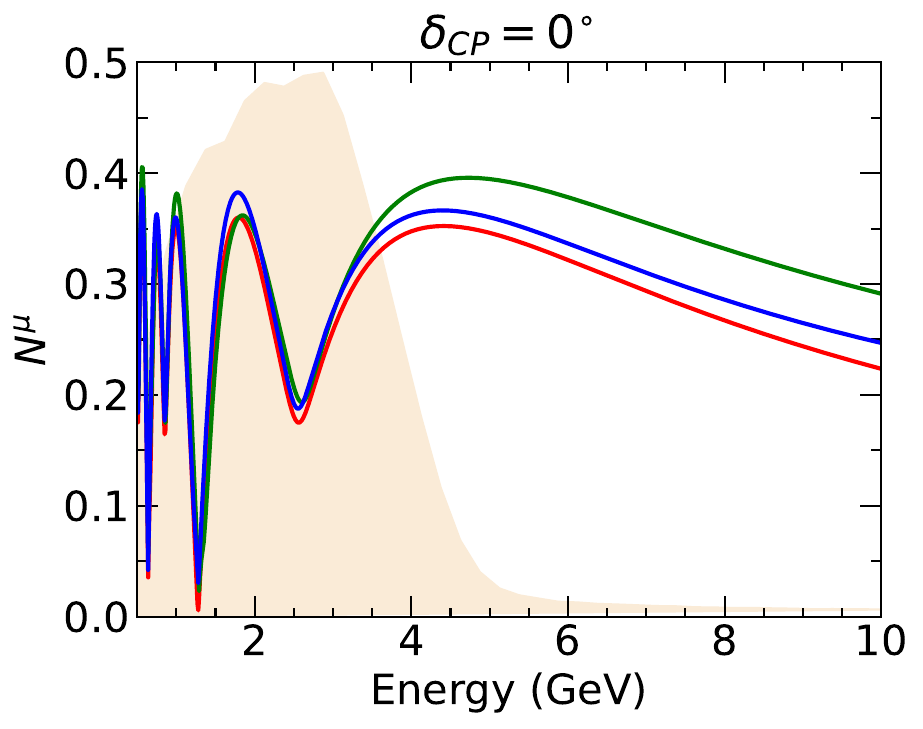}
    \caption{Upper (lower) row shows the NSI parameter ($\epsilon_{e\mu}$) dependency on EOF (left), Concurrence (middle) and Negativity (right) for DUNE experiment with $\delta_{CP} = 212^{\circ} ~(0^{\circ})$.}
    \label{fig2}
\end{figure}
\\
Figure \ref{fig2} illustrates how the NSI parameter, $\epsilon_{e\mu}$ affects EOF, Concurrence, and Negativity. In each panel, the red curve corresponds to the standard oscillation scenario with no NSI present. The green curve shows the effect of a non-zero real part, specifically $\left |\epsilon_{e\mu } \right |$ = 0.15 with the associated phase $\phi_{e\mu} = 0$. The blue curve includes both the real and imaginary components of $\epsilon_{e\mu}$, i.e., the full complex parameter. The shaded region represents the $\nu_{\mu}$ flux in arbitrary units. The left, middle, and right panels of Fig. \ref{fig2} respectively depict the variation of EOF, Concurrence, and Negativity due to the presence of $\epsilon_{e\mu}$. For these plots, the standard oscillation parameters are taken from Table \ref{del1}, while the NSI values are from Table \ref{nsi}. Upper row of Fig. \ref{fig2} shows the result when $\delta_{CP} = 212^{\circ}$ whereas lower row represents the result with $\delta_{CP} = 0^{\circ}$.

To understand the influence of the $\epsilon_{e\mu}$ parameter on the Entanglement of Formation (EOF), we refer to the leftmost panel of the upper row in Fig. \ref{fig2}. A clear deviation of the blue curve from the standard oscillation curve (red) is observed near the DUNE flux oscillation peak approximately at 2.5 GeV, corresponding to an approximately 28\% deviation from the SO scenario for $\delta_{CP}= 212^{\circ}$.
When both the real and imaginary components of $\epsilon_{e\mu}$ are included, the EOF value decreases at this oscillation peak. In contrast, if only the real part of $\epsilon_{e\mu}$ is considered, there is a slight enhancement in the EOF amplitude at the same energy compared to the standard case. This behavior continues consistently up to around 5 GeV, beyond which all three curves converge and exhibit negligible differences. In the lower row of the leftmost panel, with $\delta_{CP} = 0^{\circ}$, the EOF displays a different trend. Here, the green curve diverges significantly from both the red and blue curves, particularly beyond 4 GeV, where the difference between the green and red curves becomes more pronounced, reaching approximately 18\%. A similar qualitative behavior is observed for Concurrence. For $\delta_{CP} = 212^{\circ}$, the red curve deviates from the blue curve by approximately 38\% at $E \approx$ 2.5 GeV. In contrast, for $\delta_{CP} = 0^{\circ}$, the green curve deviates from the red curve by about 6\% at energies above 4 GeV. For Negativity, in the upper-left panel, the blue curve can be clearly distinguished from the red curve at $E$ approx 2.5 GeV, with a deviation of approximately 16\%. However, for $\delta_{CP} = 0^{\circ}$, at energies above 4 GeV, all three curves (red, blue, and green) become clearly distinguishable, with the blue curve deviating from the red by about 9\% and the green curve by approximately 31\%.

To explain the nature of both the panels, we need to see the numerical expression.
The expression of EOF in terms of oscillation probabilities is given in Eq. \ref{EOF}. In this study, we mainly focus on the accelerator-based long-baseline experiment DUNE. For the DUNE experimental setup, the experiment is sensitive to measure the appearance and disappearance probabilities of muon electron neutrinos. Thus, we need to express Eq. \ref{EOF} in terms of $P_{\mu e}$ and $P_{\mu \mu}$.  
The modified form of the Eq.  \ref{EOF} (considering $\nu_{\mu}$ as initial neutrino flavor) in terms of $P_{\mu e}$ and $P_{\mu \mu}$ is given by:
\begin{eqnarray}\label{EOF1}
    &&EOF=-\frac{1}{2}\Big[P_{\mu e} \log_{2}P_{\mu e}+P_{\mu \mu} \log_{2}P_{\mu \mu}+ (1-P_{\mu e}- P_{\mu \mu}) \log_{2}{(1-P_{\mu e}- P_{\mu \mu})}+(1-P_{\mu e})  \nonumber\\ 
    &&~~~~~~~~~~\times\log_{2}(1-P_{\mu e})
    +(1-P_{\mu \mu}) \log_{2}(1-P_{\mu \mu}) +(P_{\mu \mu}+P_{\mu e}) \log_{2}(P_{\mu \mu}+P_{\mu e})\Big].
\end{eqnarray}
From Eq. \ref{EOF1}, it is evident that the EOF  depends equally on both the appearance probability $P_{\mu e}$ and the disappearance probability  $P_{\mu \mu}$. However, as shown in Eq. \ref{dis} and Fig. \ref{fig-pmumu}, the influence of the NSI parameter $\epsilon_{e \mu}$ on the disappearance channel is minimal. Therefore, the behavior of EOF is primarily governed by the appearance probability in the presence of the NSI parameter $\epsilon_{e\mu}$, as described in Eq. \ref{app} and illustrated in Fig. \ref{fig1}.

Both the analytical expression and the plots indicate that the behavior of $P_{\mu e}$ in the presence of $\epsilon_{e \mu}$ closely mirrors the pattern observed in the leftmost column of Fig. \ref{fig2}. For instance, at $\delta_{CP} = 212^\circ$, near the oscillation peak, the blue curve lies below the red (standard) curve, while the green curve exhibits a higher amplitude, matching the pattern seen in the left panel of the upper row of Fig. \ref{fig1}. Similarly, for $\delta_{CP} = 0^\circ$, the behavior in the lower row of the leftmost panel in Fig. \ref{fig2} closely resembles that of the corresponding panel in Fig. \ref{fig1}. From this analysis, it becomes clear that the impact of $\epsilon_{e\mu}$ on the EOF is predominantly driven by its effect on the appearance channel.

Next, we turn our attention to understanding how $\epsilon_{e\mu}$ influences the other two entanglement measures: Concurrence and Negativity. The simplified form of Concurrence (Eq. \ref{C}) in terms of electron neutrino appearance and muon neutrino disappearance probabilities as,
\begin{eqnarray}
     C^{\mu}= 2\sqrt{P_{\mu \mu}(1- P_{\mu \mu} - P_{\mu e}) + P_{\mu e}(1-P_{\mu e})}.
\end{eqnarray}
\begin{figure}[htb]
    \hspace{-0.2cm}
    \includegraphics[height=50mm, width=54mm]{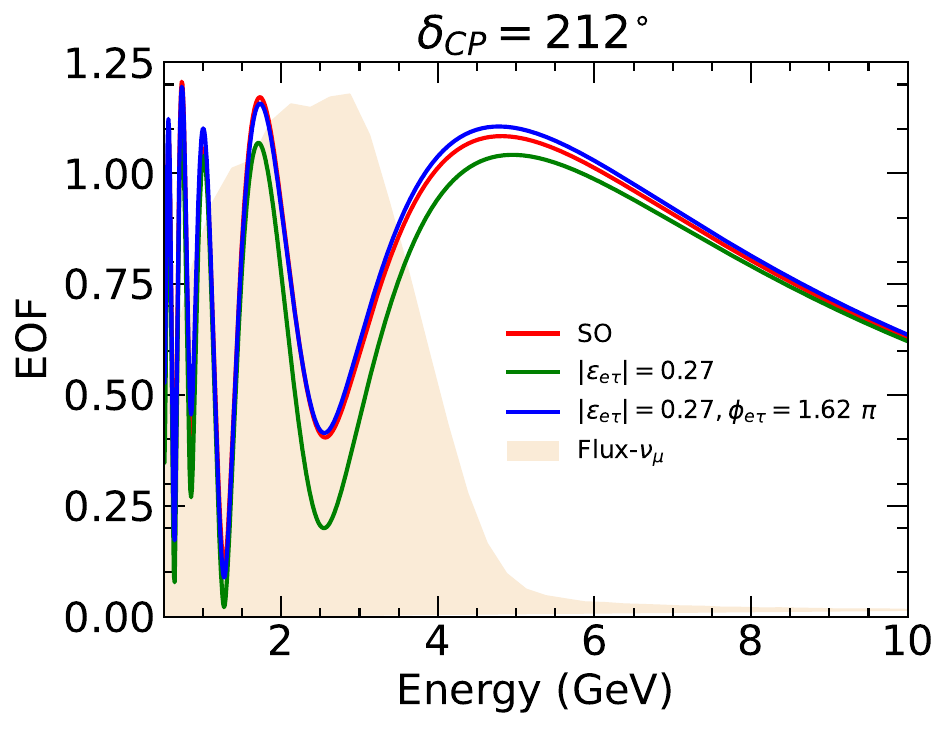}
    \includegraphics[height=50mm, width=54mm]{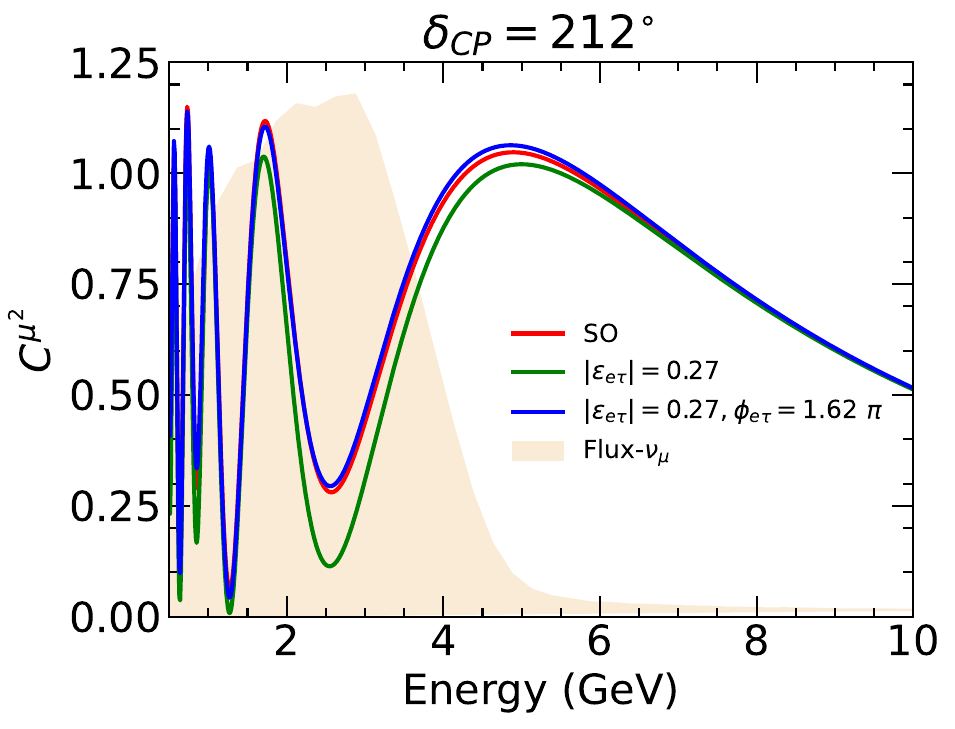}
    \includegraphics[height=50mm, width=54mm]{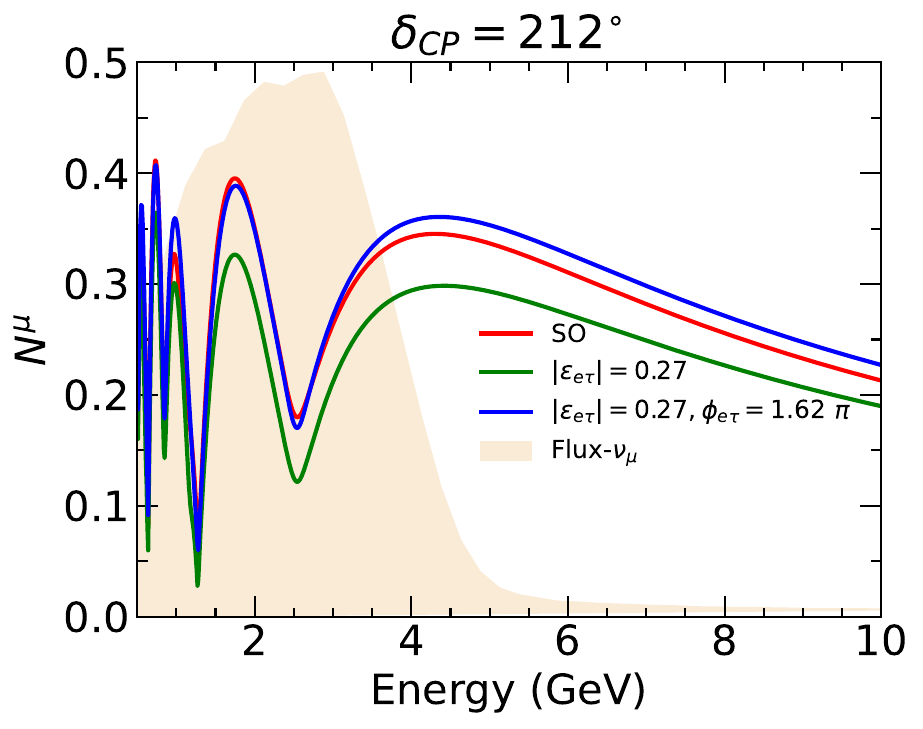}\\
    \includegraphics[height=50mm, width=54mm]{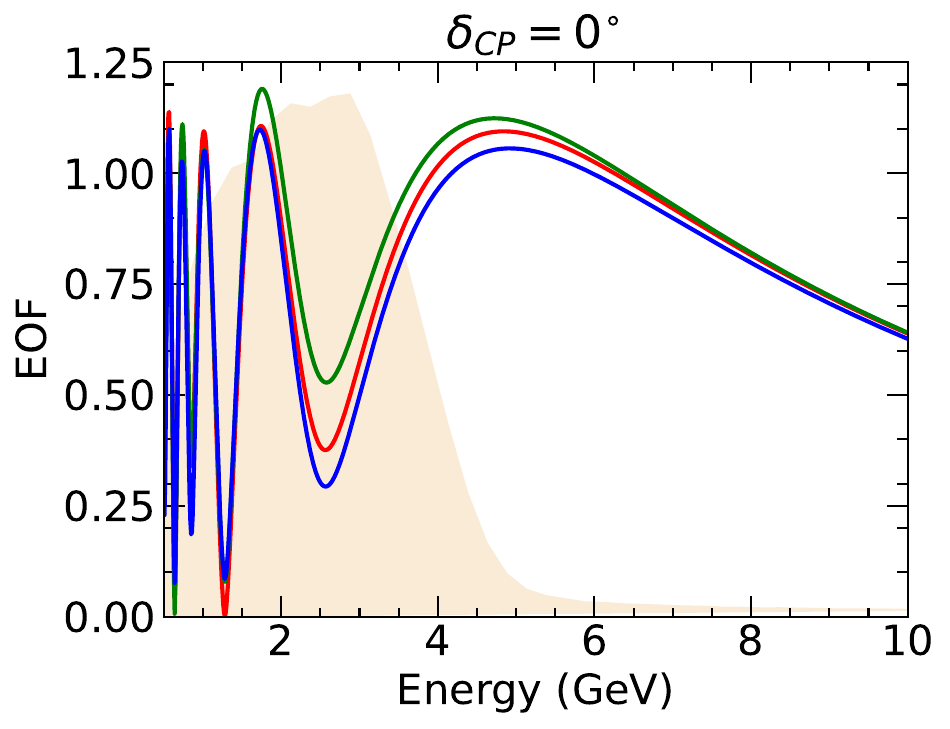}
    \includegraphics[height=50mm, width=54mm]{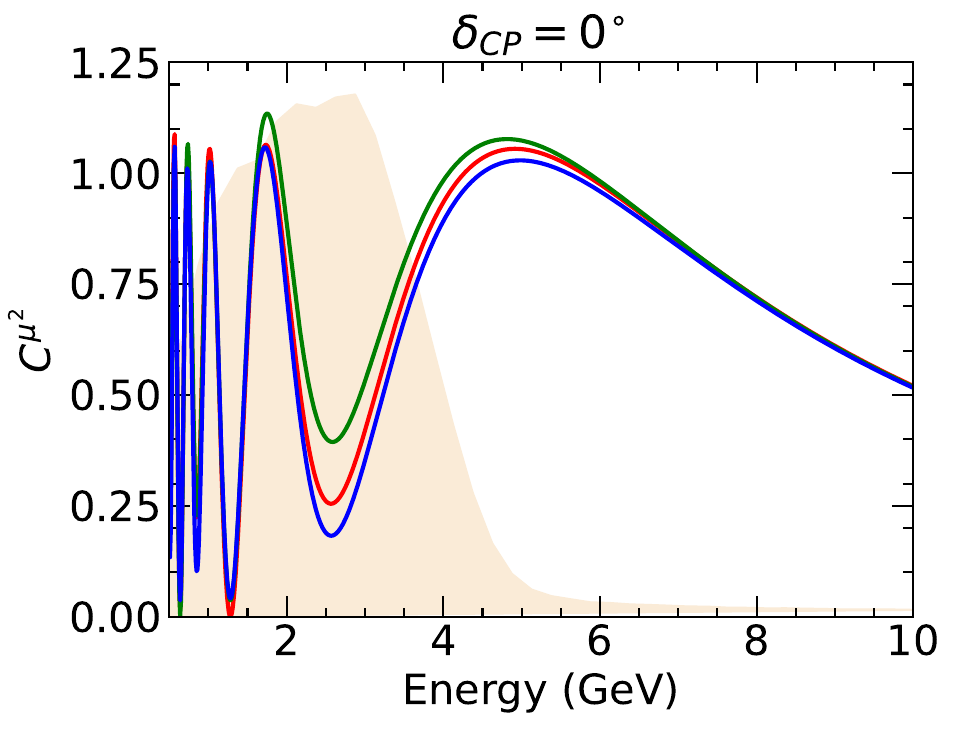}
    \includegraphics[height=50mm, width=54mm]{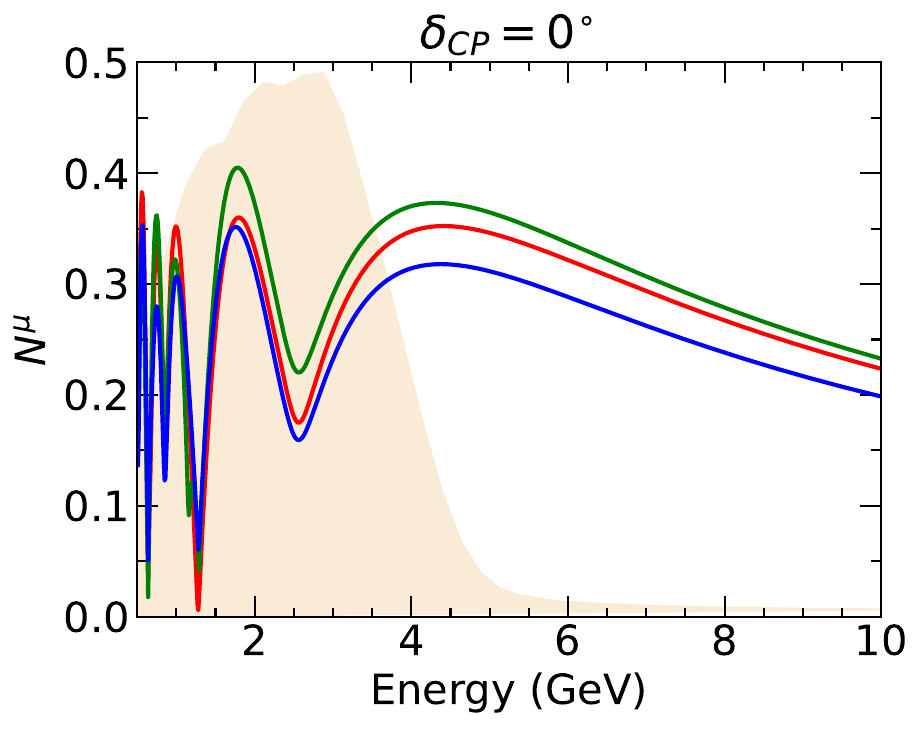}
    \caption{Upper (lower) row shows the NSI parameter ($\epsilon_{e\tau}$) dependency on EOF(left), Concurrence(middle) and Negativity(right) for DUNE experiment with $\delta_{CP} = 212^{\circ}~(0^{\circ})$.}
    \label{fig3}
\end{figure}
\begin{figure}[htb]
   \hspace{-0.2 cm}
   \includegraphics[height=50mm, width=54mm]{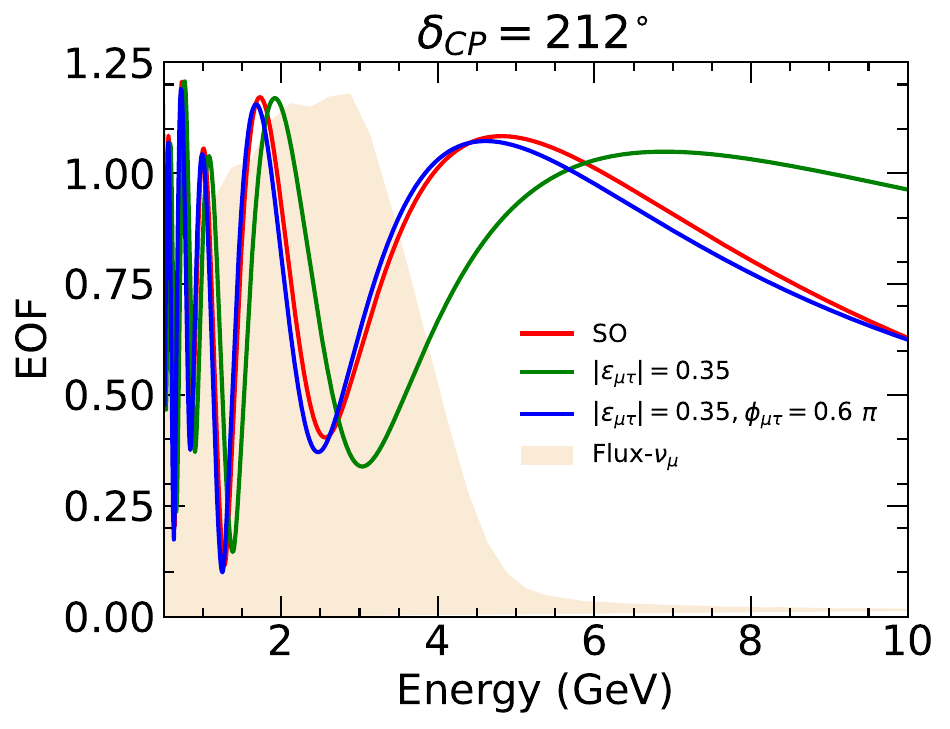}
    \includegraphics[height=50mm, width=54mm]{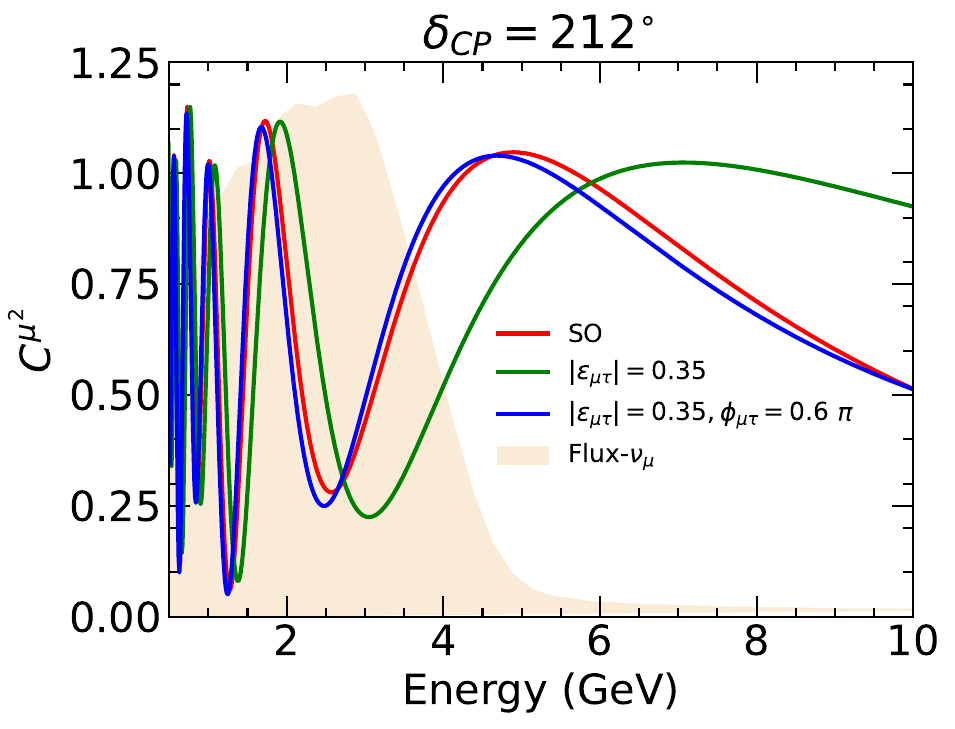}
    \includegraphics[height=50mm, width=54mm]{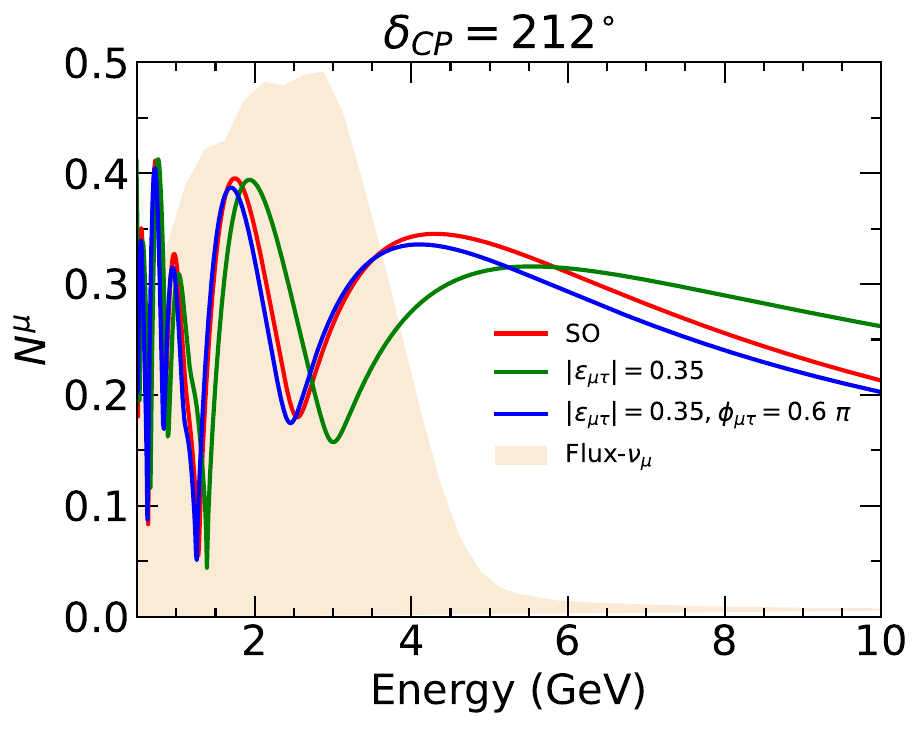}\\
    \includegraphics[height=50mm, width=54mm]{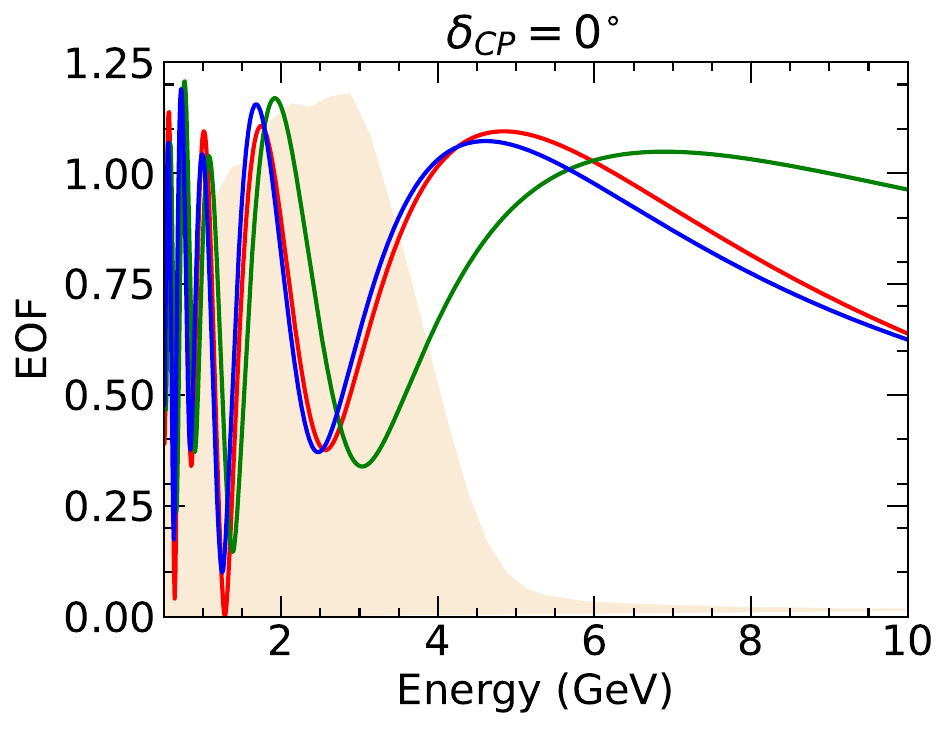}
    \includegraphics[height=50mm, width=54mm]{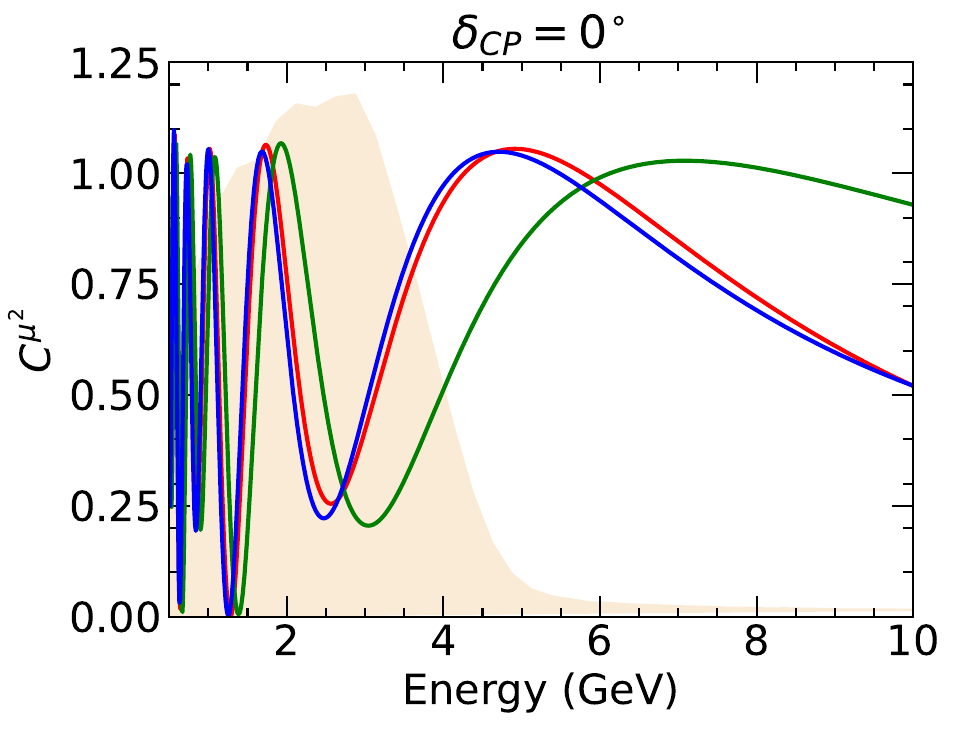}
    \includegraphics[height=50mm, width=54mm]{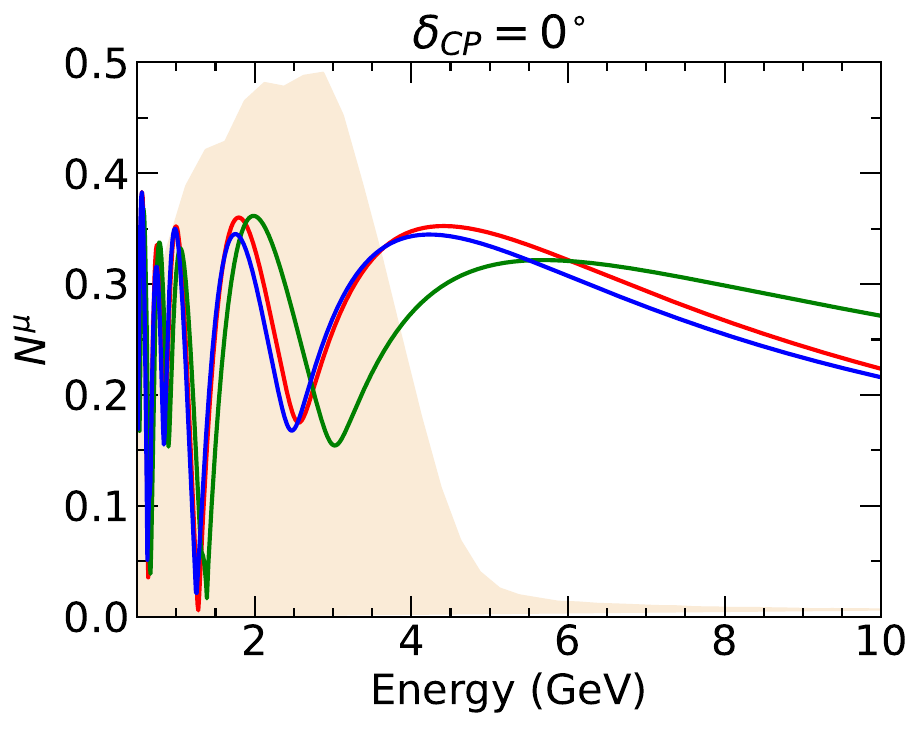}
    \caption{Upper (lower) row shows the NSI parameter ($\epsilon_{\mu \tau}$) dependency on EOF(left), Concurrence(middle) and Negativity(right) for DUNE experiment with $\delta_{CP} = 212^{\circ}~(0^{\circ})$.}
    \label{fig4}
\end{figure}
Similarly, the expression of Negativity in terms of $P_{\mu e}$ and $P_{\mu \mu}$ as:
\begin{equation}\label{NS}
    N^{\mu}=[\sqrt{P_{\mu e}}\sqrt{1-P_{\mu e}}\sqrt{1-P_{\mu \mu}}\sqrt{P_{\mu \mu}}\sqrt{P_{\mu e}+P_{\mu \mu}}\sqrt{1-P_{\mu e}- P_{\mu \mu}}]^{\frac{1}{3}}.
\end{equation}

If we examine the middle and right-most columns of Fig. \ref{fig2}, it becomes clear that both Concurrence and Negativity exhibit a similar pattern to that of the EOF in the presence of the $\epsilon_{e\mu}$ parameter. This similarity suggests that their behavior, like EOF, can be primarily attributed to the modifications in the appearance channel alone.
 
In conclusion, Fig. \ref{fig2} demonstrates that the presence of the off-diagonal NSI parameter $\epsilon_{e\mu}$ leads to a noticeable deviation in the behavior of all three entanglement measures compared to the standard scenario. This deviation becomes more pronounced when $\delta_{CP}$ corresponds to a CP-conserving value.

Figure \ref{fig3} illustrates the impact of the NSI parameter $\epsilon_{e \tau}$ on the three entanglement measures: EOF, Concurrence, and Negativity. The color scheme remains consistent with the previous figure; red denotes the standard oscillation scenario, green corresponds to a non-zero real value of $\epsilon_{e \tau}$ with phase $\phi_{e \tau} = 0$, and blue represents the case where both real and imaginary parts of $\epsilon_{e \tau}$, along with its associated phase $\phi_{e \tau}$, are considered. The shaded area corresponds to the DUNE $\nu_{\mu}$ neutrino flux in arbitrary units.

In the upper row with $\delta_{CP} = 212^\circ$, all three measures show that the green curve lies below both the red and blue curves at the oscillation peak, with suppressions of approximately 50\% for EOF, $\approx$ 62\% for Concurrence, and approx 33\% for Negativity. At higher neutrino energies, the three curves converge for EOF and Concurrence, indicating negligible differences. However, in the case of Negativity, the curves remain distinct even at higher energies. For energies above 4 GeV, the blue curve shows a modest enhancement of about 2\% relative to the red curve, whereas the green curve exhibits a significant suppression of approximately 14\% compared to the red curve. For $\delta_{CP} = 0^\circ$, shown in the lower row, the blue curve lies below the red and green curves at the oscillation maximum across all three entanglement measures. At this energy, relative to the red curve, the green curve exhibits an enhancement of about 36\% in EOF, $\approx$ 60\% in concurrence, and $\approx$ 30\% in negativity. In contrast, the blue curve shows a suppression of $\approx$ 22\% for EOF and approx 33\% for concurrence, while for negativity only a mild suppression is observed. As the energy increases, the curves tend to overlap for EOF and Concurrence, while they remain clearly separated for Negativity.
The overall behavior observed in Fig. \ref{fig3} can be attributed solely to the effect of $\epsilon_{e \tau}$ on the appearance probability, as the disappearance probability $P_{\mu\mu}$ is unaffected by this parameter.

\begin{figure*}[ht!]
\centering

\includegraphics[width=45mm]{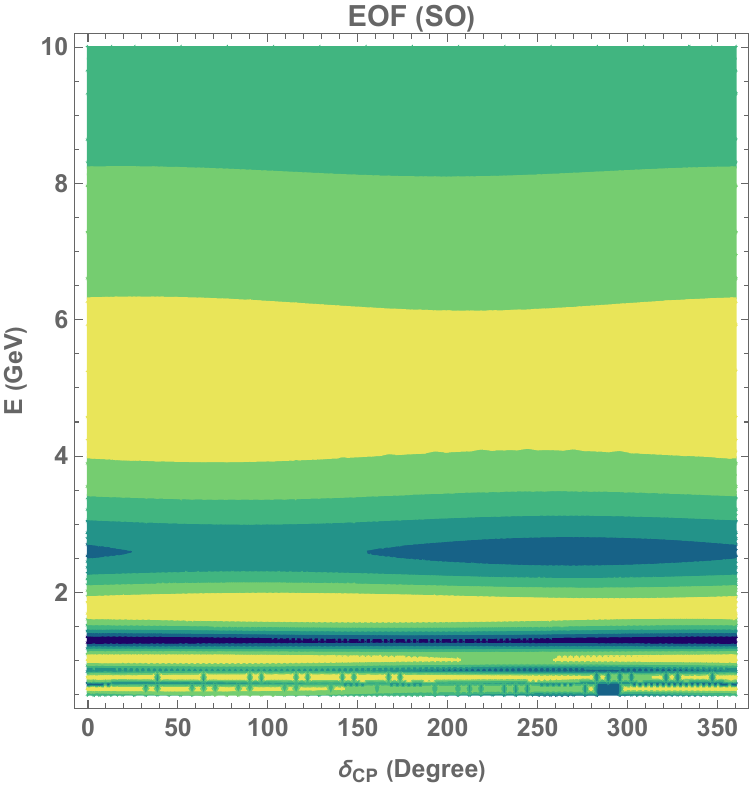}
\includegraphics[width=45mm]{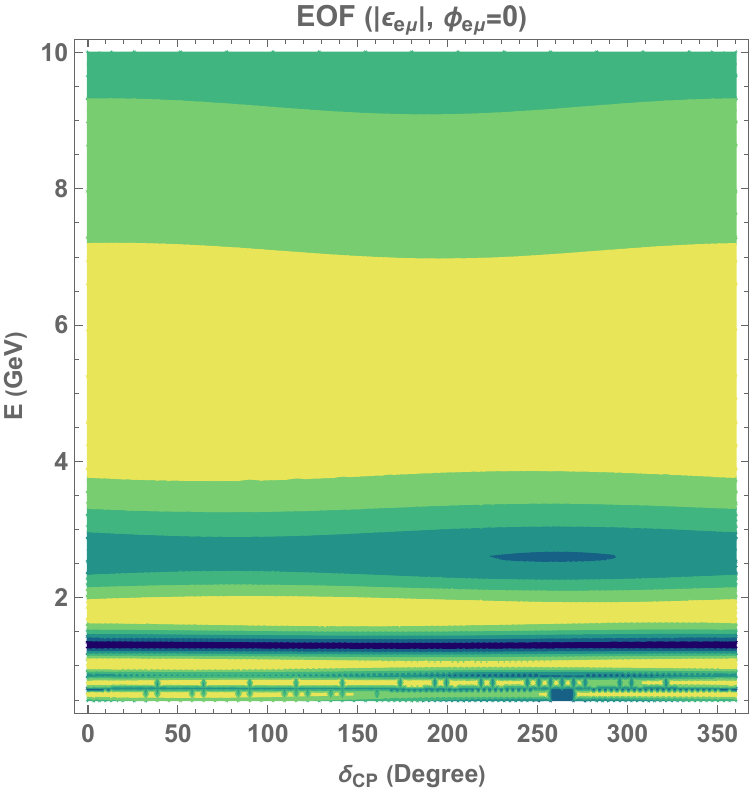}
\includegraphics[width=45mm]{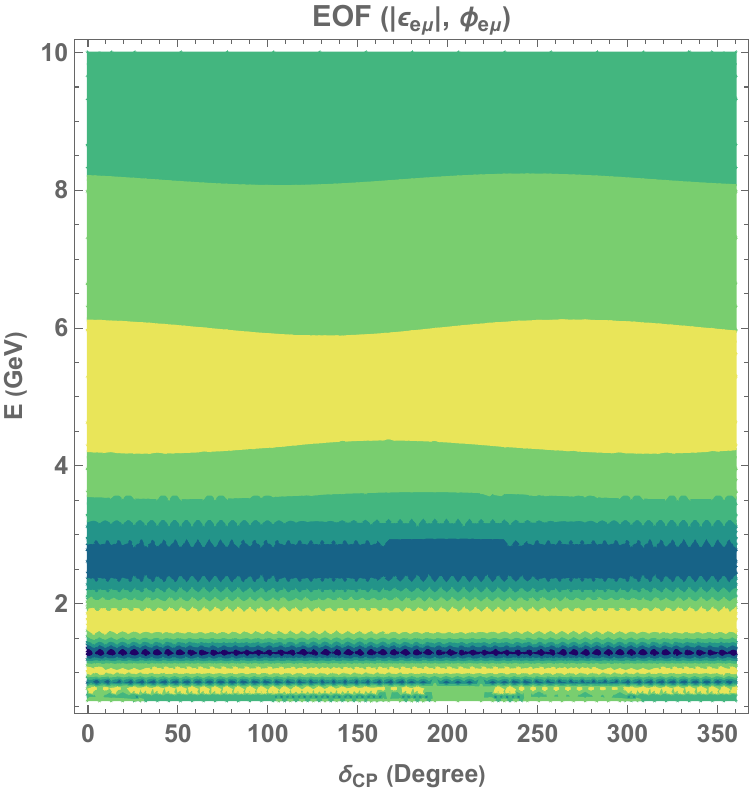}
\includegraphics[width=6.5mm]{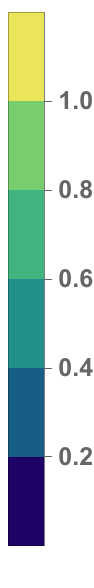}
\\
\includegraphics[width=45mm]{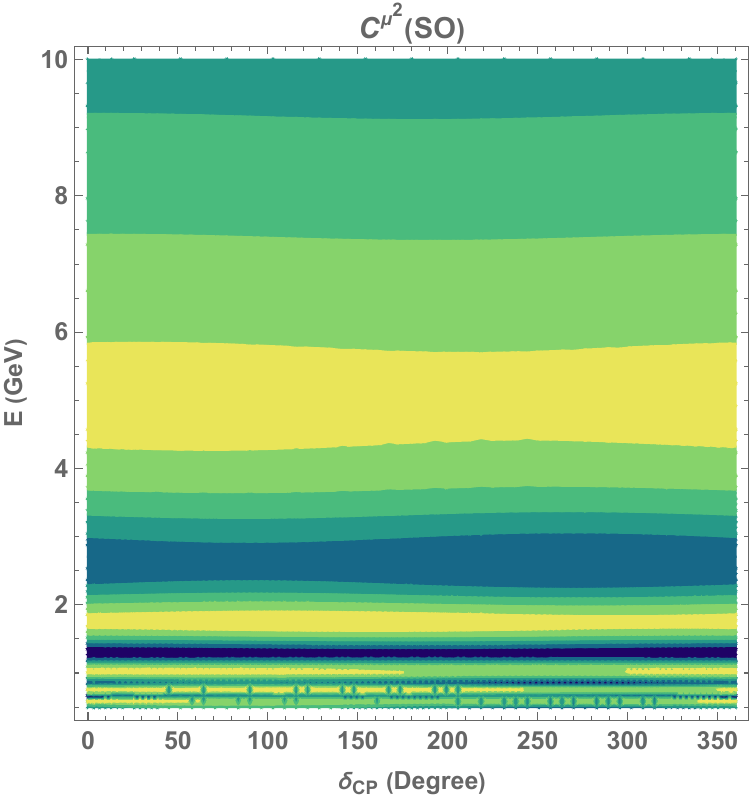}
\includegraphics[width=45mm]{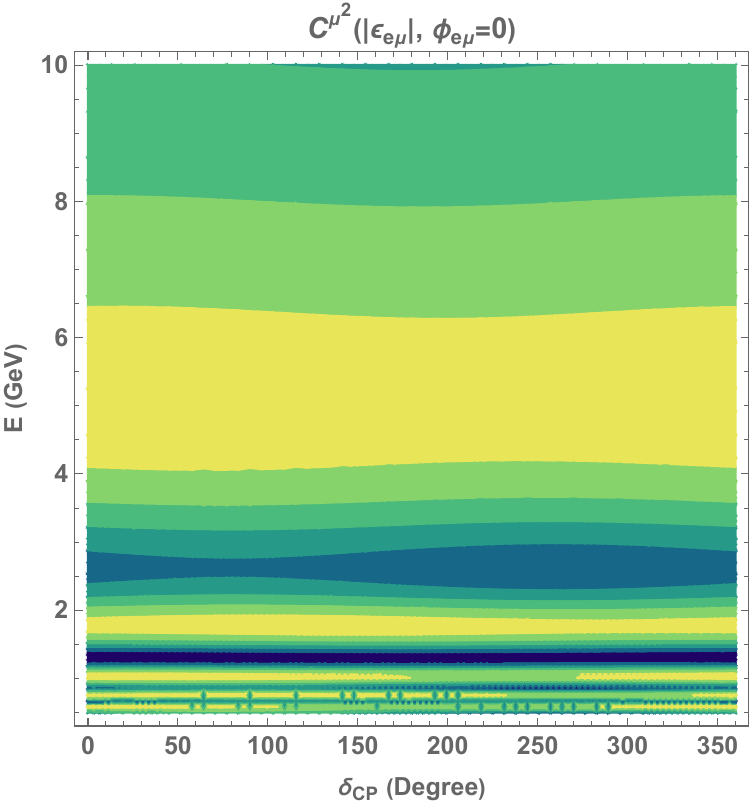}
\includegraphics[width=45mm]{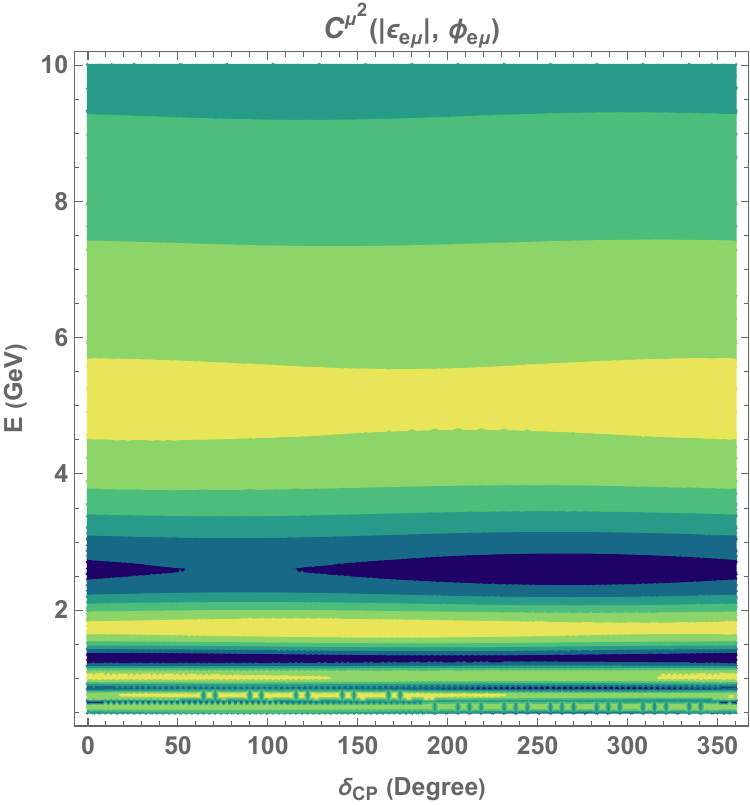}
\includegraphics[width=7.8mm]{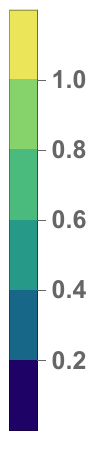}
\\
\includegraphics[width=45mm]{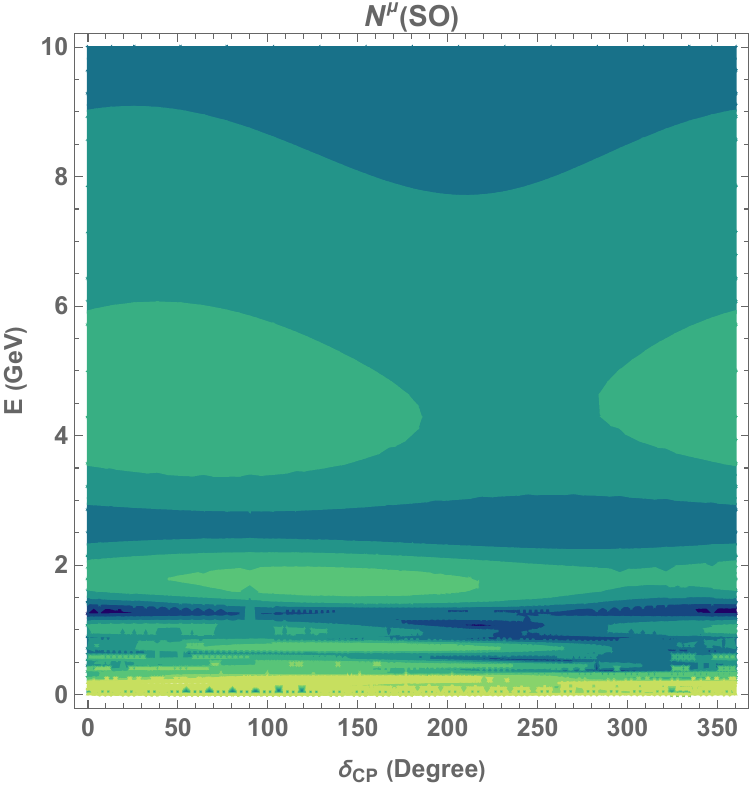}
\includegraphics[width=45mm]{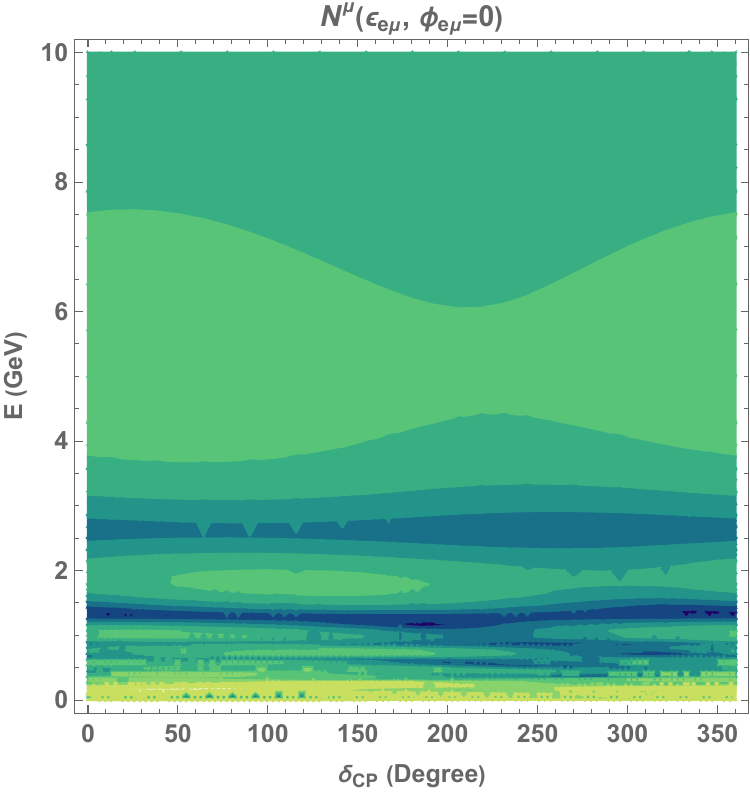}
\includegraphics[width=45mm]{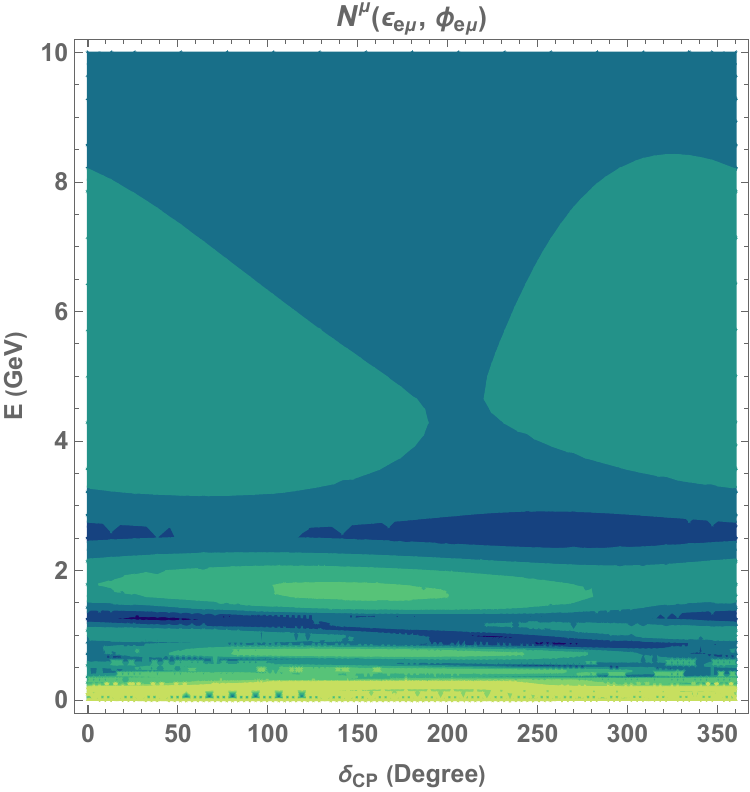}
\includegraphics[width=8.9mm]{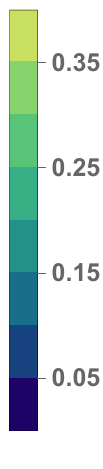}
\caption{The entanglement measures: EOF (top), Concurrence (middle), and Negativity (bottom) are shown in the plane of the neutrino energy E (in GeV) and the CP-violating phase $\delta_{CP}$ (in degree) for the SO (left), $\left |\epsilon_{e\mu} \right |, \phi_{e\mu}=0$ (middle)}, and $ \left |\epsilon_{e\mu} \right |, \phi_{e\mu}$ (right) scenarios for DUNE experimental set up.
\label{dfig1}
\end{figure*}
 
 In Fig. \ref{fig4}, all three entanglement measures; EOF, Concurrence, and Negativity exhibit a consistent pattern across the energy spectrum for the NSI scenario involving $\epsilon_{\mu\tau}$. In contrast to the cases with $\epsilon_{e\mu}$ and $\epsilon_{e\tau}$, a noticeable shift is observed in the curves for the scenario with $\phi_{\mu\tau} = 0$, particularly around the oscillation peak. This deviation highlights the significant role played by the phase $\phi_{\mu\tau}$ in shaping the entanglement behavior. As evident from Eqs. \ref{app} and \ref{dis}, the appearance probability $P_{\mu e}$ does not depend on $\epsilon_{\mu\tau}$ at leading order, whereas the disappearance probability $P_{\mu \mu}$ is strongly influenced by this parameter. Consequently, the behavior of EOF, Concurrence, and Negativity closely follows the trends shown in the rightmost panel of Fig. \ref{fig-pmumu}.

In summary, for long-baseline accelerator neutrino experiments, the influence of off-diagonal NSI parameters on entanglement measures is primarily governed by their impact on the appearance and disappearance channels. Specifically, the NSI parameters $\epsilon_{e\mu}$ and $\epsilon_{e\tau}$ predominantly affect the appearance probability $P_{\mu e}$, while $\epsilon_{\mu\tau}$ modifies the disappearance probability $P_{\mu \mu}$. 

Additionally, we have checked that similar patterns in the entanglement measures; EOF, Concurrence, and Negativity are observed in future long-baseline experiments such as P2SO when these off-diagonal NSI parameters are introduced. 

\begin{figure*}[ht!]
\centering
\includegraphics[width=45mm]{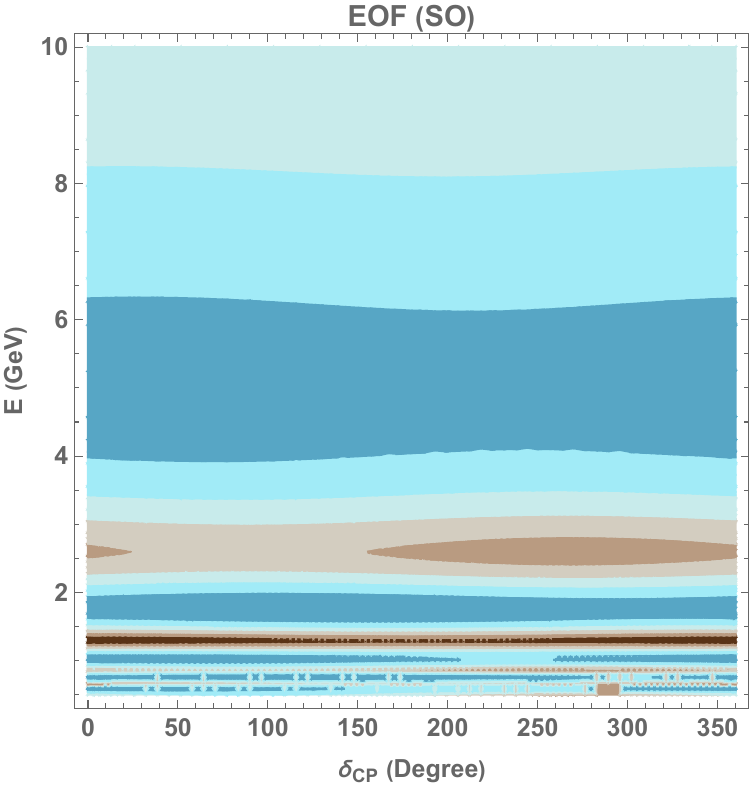}
\includegraphics[width=45mm]{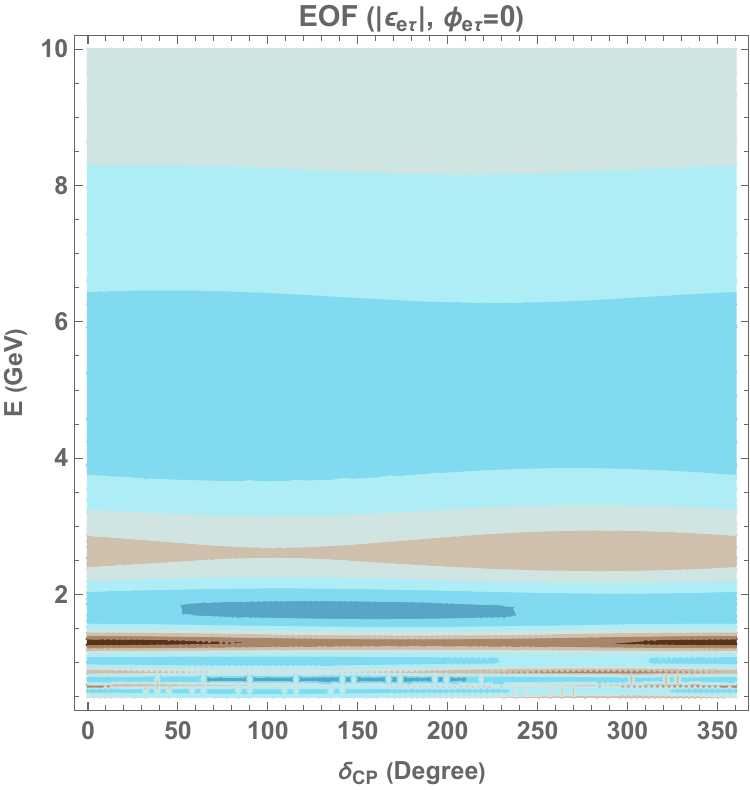}
\includegraphics[width=45mm]{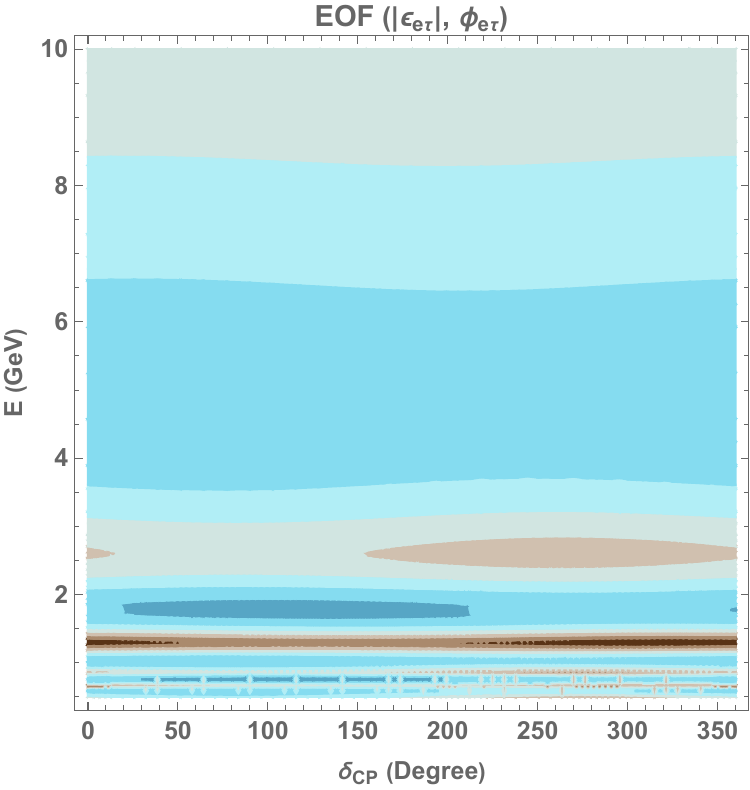}
\includegraphics[width=7.8mm]{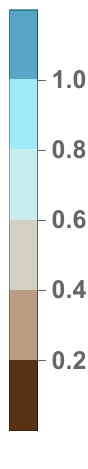}
\\
\includegraphics[width=45mm]{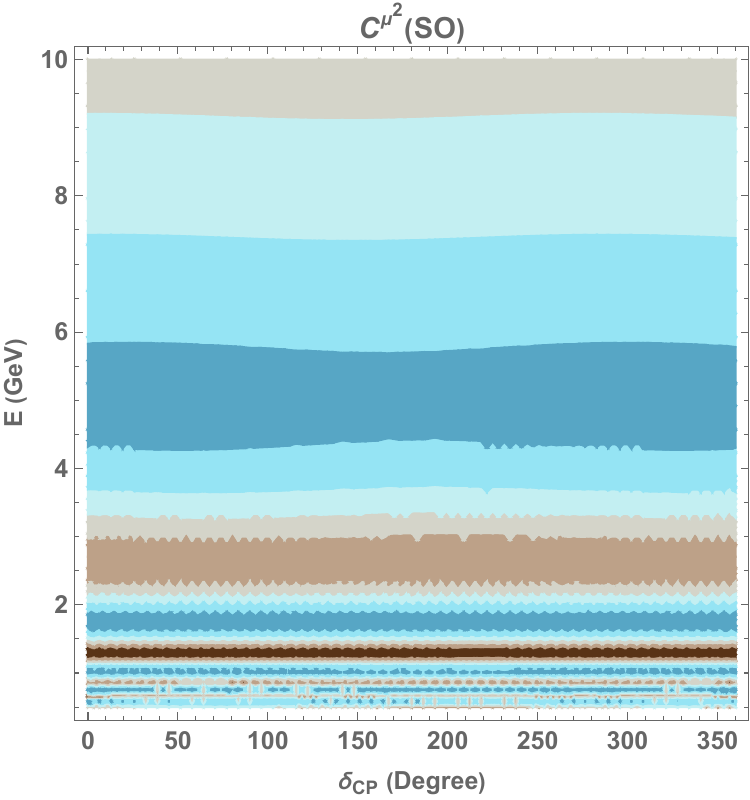}
\includegraphics[width=45mm]{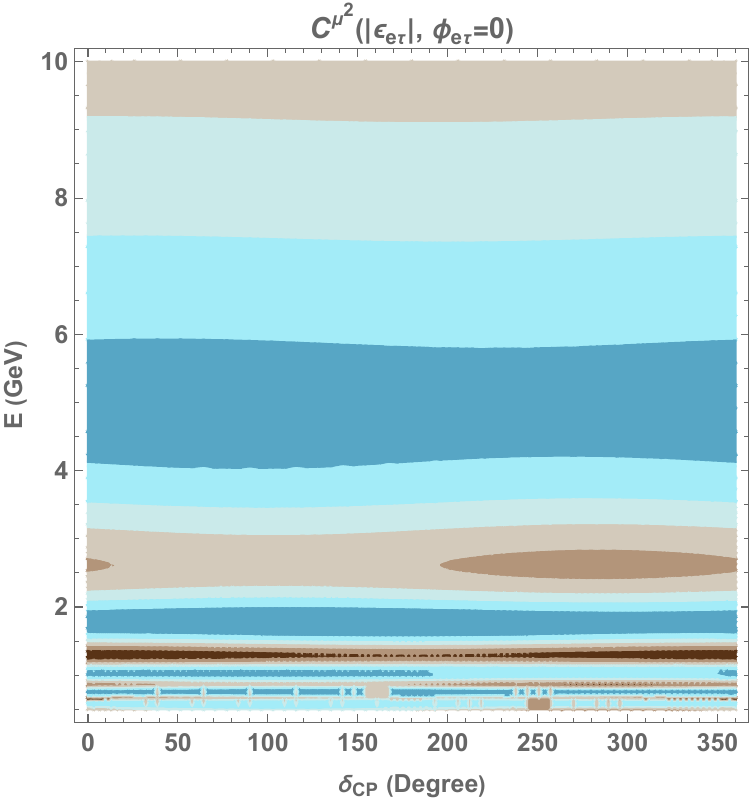}
\includegraphics[width=45mm]{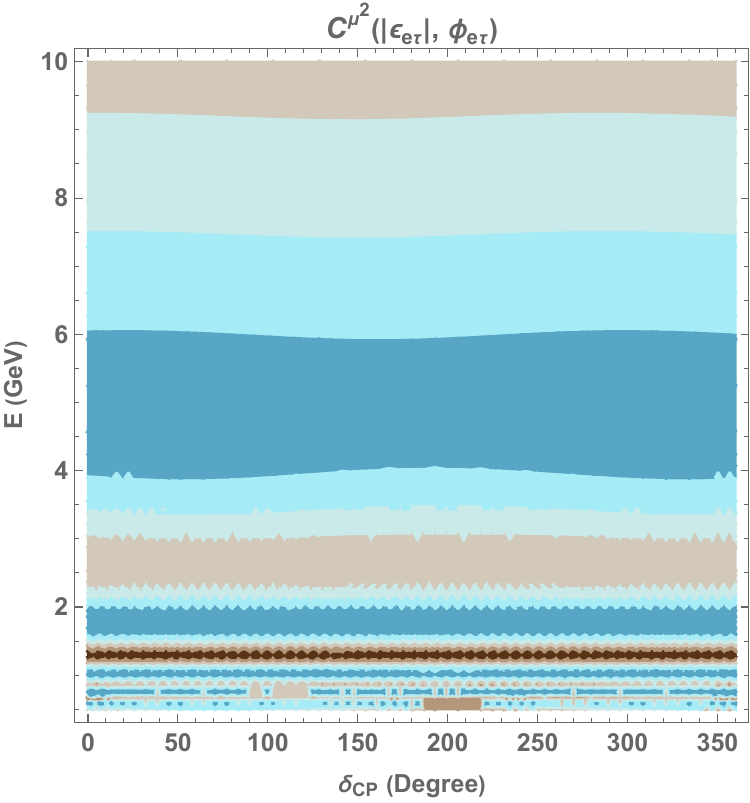}
\includegraphics[width=7.8mm]{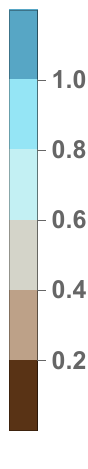}
\\
\includegraphics[width=45mm]{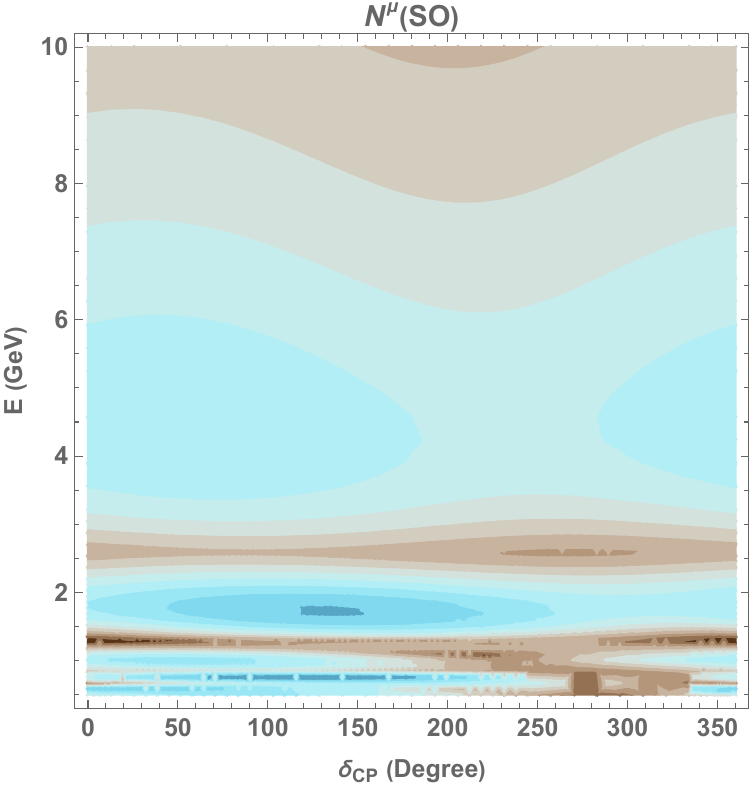}
\includegraphics[width=45mm]{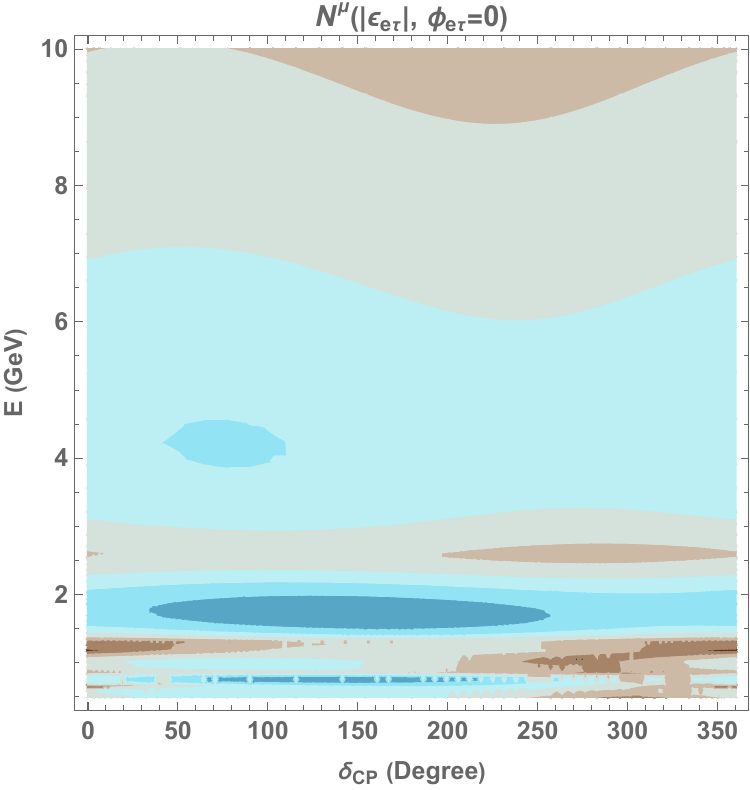}
\includegraphics[width=45mm]{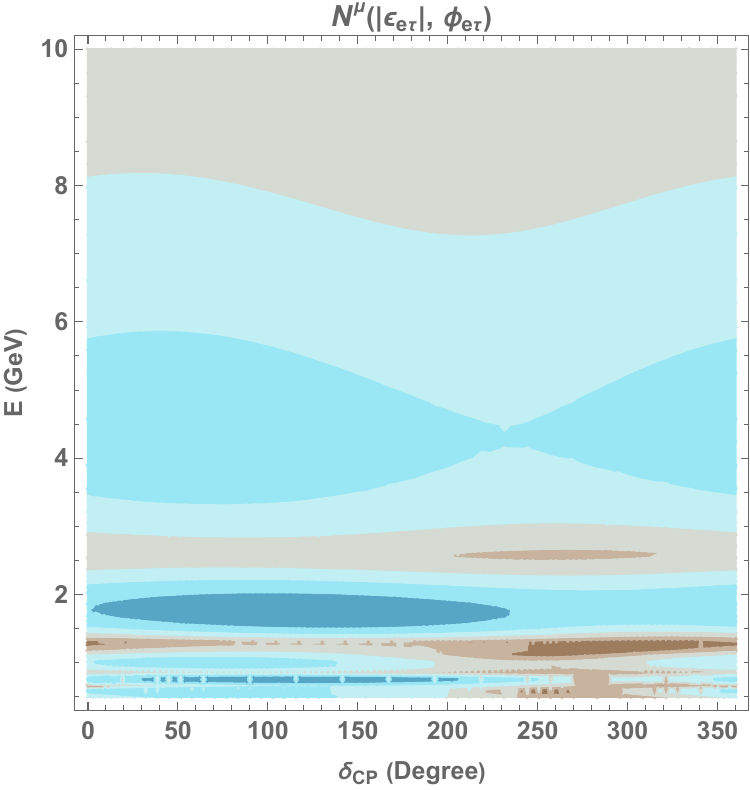}
\includegraphics[width=9mm]{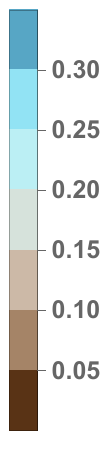}
\caption{ For the DUNE experiment, the entanglement measures: EOF (top), Concurrence (middle), and Negativity (bottom) are plotted in the (E - $\delta_{CP}$) plane for the SO (Left),  $\epsilon_{e\tau}, \phi_{e\tau}=0$ (middle) and, $\epsilon_{e\tau}, \phi_{e\tau}$ (right) scenarios.}
\label{dfig2}
\end{figure*}
\subsection{Dependency of $\delta_{CP}$ and energy on entanglement measures}
In this subsection, we mainly focus on the dependency of $\delta_{CP}$ and neutrino energy on EOF, Concurrence and Negativity in the absence and presence of NSI parameters. In Fig. \ref{dfig1}, we show the impact of off-diagonal NSI parameter $\left |\epsilon_{e \mu} \right |$ and its associated complex phase $\phi_{e\mu}$ on the entanglement measures as a function of neutrino energy and the CP-violating phase $\delta_{CP}$ for the DUNE experimental set-up. Each row in the figure corresponds to one entanglement measure, while the three columns represent different scenarios: the standard oscillation (SO), NSI with a real valued $\epsilon_{e\mu} $ $ (\phi_{e\mu}=0)$, and NSI with both magnitude and complex phase of $\epsilon_{e\mu}$ $ (\phi_{e\mu}\neq 0)$. 
It is observed that in the energy range of $[1.5 - 2]$ GeV, for all scenarios, EOF achieved the maximum value ($\sim 1.2$) for all the values of $\delta_{CP}$. Further in the energy range of around $[4 - 6]$ GeV, for SO and $\left |\epsilon_{e\mu } \right |$ with non-zero  $\phi_{e\mu}$ scenarios, the EOF shows maximum and for the scenario with $\phi_{e\mu}=0$, the energy range for maximum EOF value is around $[4 - 7]$ GeV for all values of $\delta_{CP}$ phase. At 2.5 GeV, which corresponds to the oscillation flux peak in DUNE, the EOF is observed to be relatively low. 
Similarly, Concurrence nearly mirrors the behavior of EOF, attaining a maximum value of about 1.2, at the energy range of $[0.5-2]$ GeV and $[4-6]$ GeV for SO and the condition when the magnitude and complex phase of NSI, both are present. The energy range for maximum Concurrence with the case of non-zero $\left |\epsilon_{e\mu} \right |$ and zero $\phi_{e\mu}$ is little broader, around $[4-6.5]$ GeV. At the oscillation peak of the DUNE setup, similar to EOF, the Concurrence value shows nearly zero for almost all the energy range of $\delta_{CP}$.

Unlike EOF and Concurrence, Negativity shows quite different behavior. Lower row of Fig. \ref{dfig1}, shows the variation of negativity with neutrino energy and CP-violating phase for SO (leftmost panel), presence of non-zero $\left |\epsilon_{e\mu} \right |$ and zero $\phi_{e\mu}$ (middle panel) and non-zero values of $\left |\epsilon_{e\mu} \right |$ and $\phi_{e \mu}$ (rightmost panel). For each panel, the maximum negativity can be seen only at the energy range of $[0-0.5]$ GeV with the full range of $\delta_{CP}$. For the SO scenario, the value of negativity reaches a maximum in the energy window of $[1.5 - 2]$ GeV, and this occurs over a distinct range of $50^{\circ} \leq \delta_{CP} \leq 225^{\circ}$ and in the case of the NSI $\epsilon_{e\mu}$ scenario without and with an associated phase $\phi_{e\mu}$, the maximum value of negativity observed within $50^{\circ} \leq \delta_{CP} \leq 200^{\circ}$ and $100^{\circ} \leq \delta_{CP} \leq 200^{\circ}$ respectively. In the energy range of $[4 - 6]$ GeV, the negativity shows clear differentiation between $\delta_{CP} = 0^\circ$ and $212^\circ$ in the SO scenario. For the NSI scenario with an associated complex phase $\phi_{e\mu}$, this distinctive behavior persists and even extends over a broader energy range of about $[4 - 8]$ GeV, indicating an enhanced phase dependent response in the presence of NSI effects.
At 2.5 GeV, negativity shows nearly zero value for all the three conditions. 

In Fig. \ref{dfig2}, three entanglement measures are shown in the plane of the neutrino energy $E$ and the CP-violating phase $\delta_{CP}$ for three scenarios:  the SO, the presence of an off-diagonal NSI parameter $\left |\epsilon_{e\tau} \right |$ with a vanishing complex phase $(\phi_{e\tau}=0)$, and a scenario where both the magnitude and complex phase of $\epsilon_{e\tau}$ are non-zero $(\phi_{e\tau}\neq 0)$. The figure is organized such that each row displays one of the entanglement measures, while the columns compare three scenarios: the first shows standard oscillation, the second includes NSI with a real-valued $\epsilon_{e\tau}$, and the third incorporates both the magnitude and complex phase of $\epsilon_{e\tau}$. In the SO case, the EOF reaches its maximum value of approx 1.2 within the energy ranges of $[1.5 - 2]$ GeV and $[4 - 6]$ GeV, consistently across all values of $\delta_{CP}$. In contrast, for the NSI scenario with a vanishing complex phase $(\phi_{e\tau}=0)$, the maximum EOF occurs within the energy range $[1.5 - 2]$ GeV and for $\delta_{CP}$ values around $[50^{\circ} - 250^{\circ}]$. Similarly, in the case of the NSI with non-vanishing complex phase scenario, the maximum EOF is also observed in the same energy range, but for $\delta_{CP}$ values approximately between 25$^\circ$ and 225$^\circ$. However, at this energy range, the EOF shows different values for $\delta_{CP}$, which is equal to 0$^\circ$ and 212$^\circ$, indicating a dependence on the CP-violating phase both in the presence and absence of the NSI phase. For the Concurrence, all the scenarios attain their maximum value, 1.2 in the energy ranges of $[1.5 - 2]$ GeV and $[4 - 6]$ GeV, consistently across the entire range of CP-violating phases, $\delta_{CP}$. However, negativity reaches its maximum value of 0.35 within the energy range of $[1.5-2]$ GeV for all scenarios, though the corresponding ranges of the CP-violating phase $\delta_{CP}$ differ. Specifically, the maximum occurs for $\delta_{CP}$ values in the ranges of $[120^{\circ} - 160^{\circ}]$ for the SO case, approx $[40^{\circ}-260^{\circ}]$ for the NSI scenario with vanishing complex phase, and $\approx[0^{\circ} - 240^{\circ}]$ for the NSI scenario with non-vanishing complex phase.

\begin{figure*}[ht!]
\centering
\includegraphics[width=50mm]{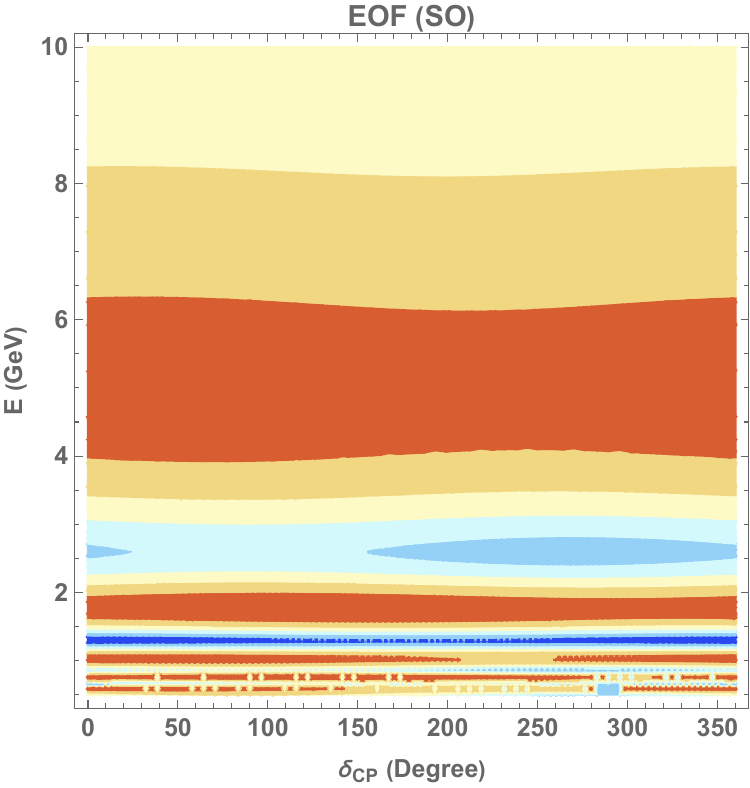}
\includegraphics[width=50mm]{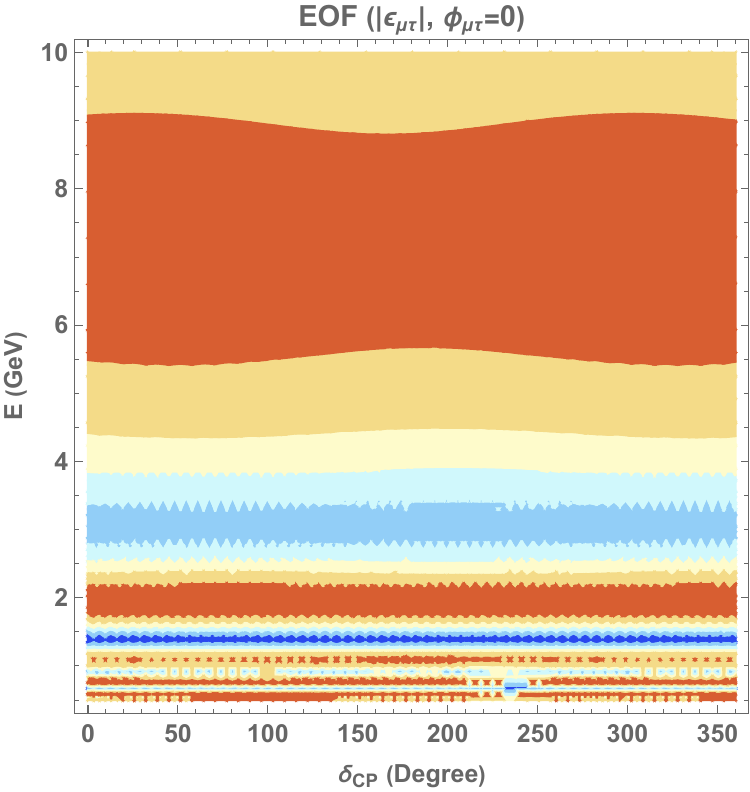}
\includegraphics[width=50mm]{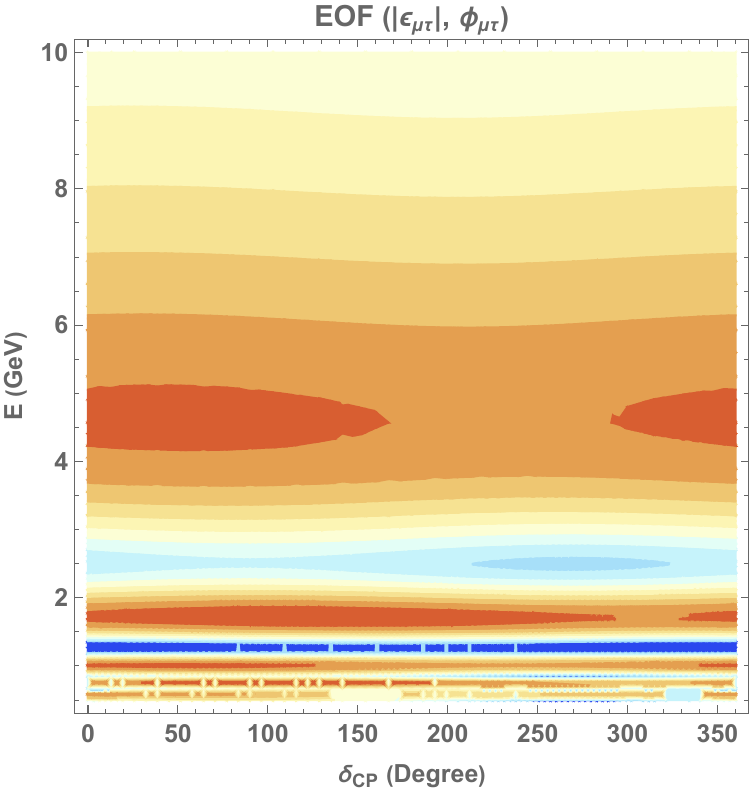}
\includegraphics[width=8mm]{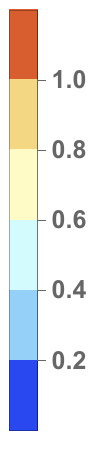}
\\
\includegraphics[width=50mm]{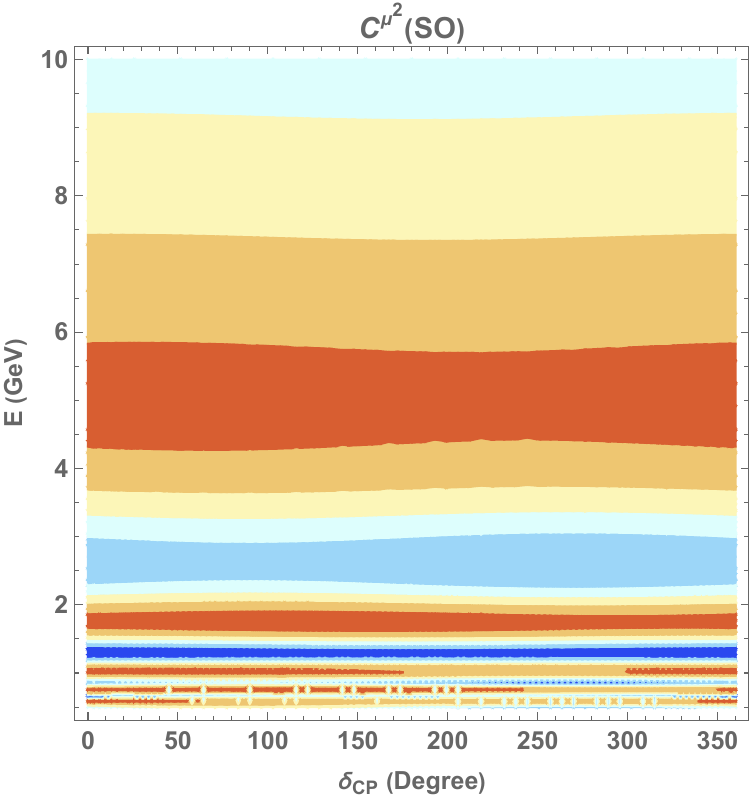}
\includegraphics[width=50mm]{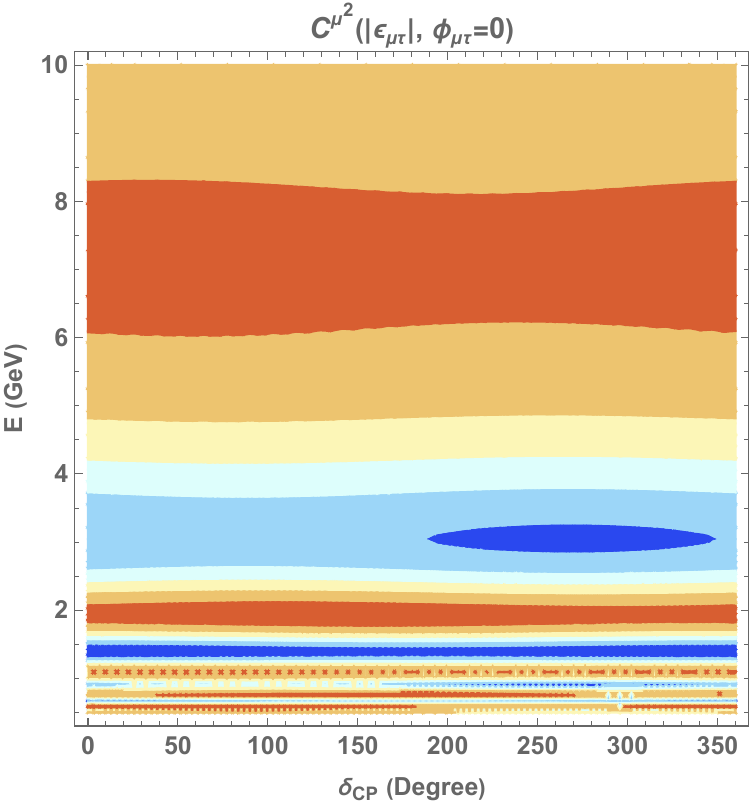}
\includegraphics[width=50mm]{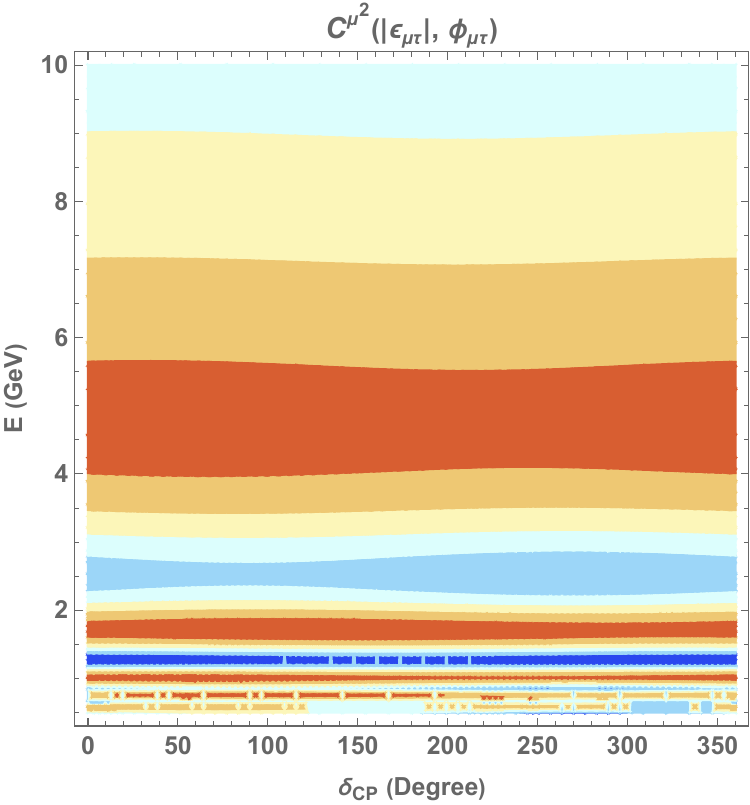}
\includegraphics[width=8mm]{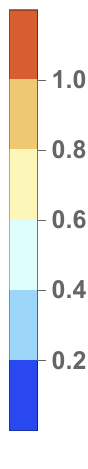}
\\
\includegraphics[width=50mm]{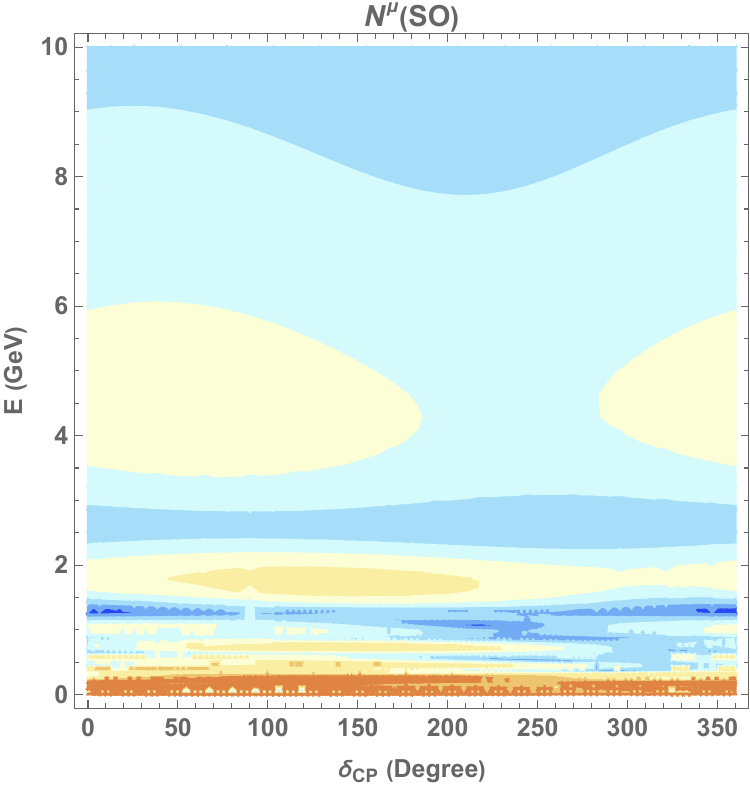}
\includegraphics[width=50mm]{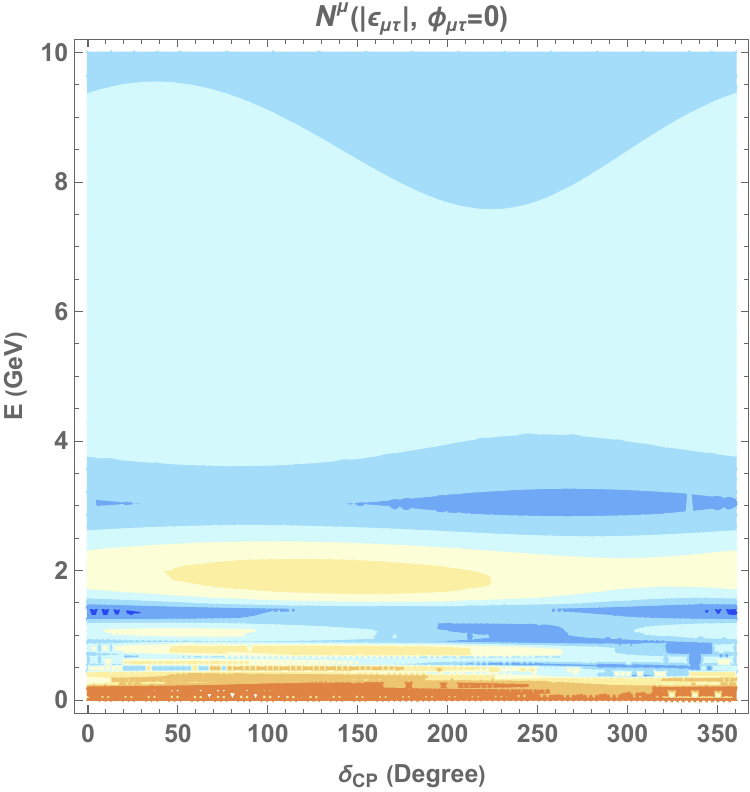}
\includegraphics[width=50mm]{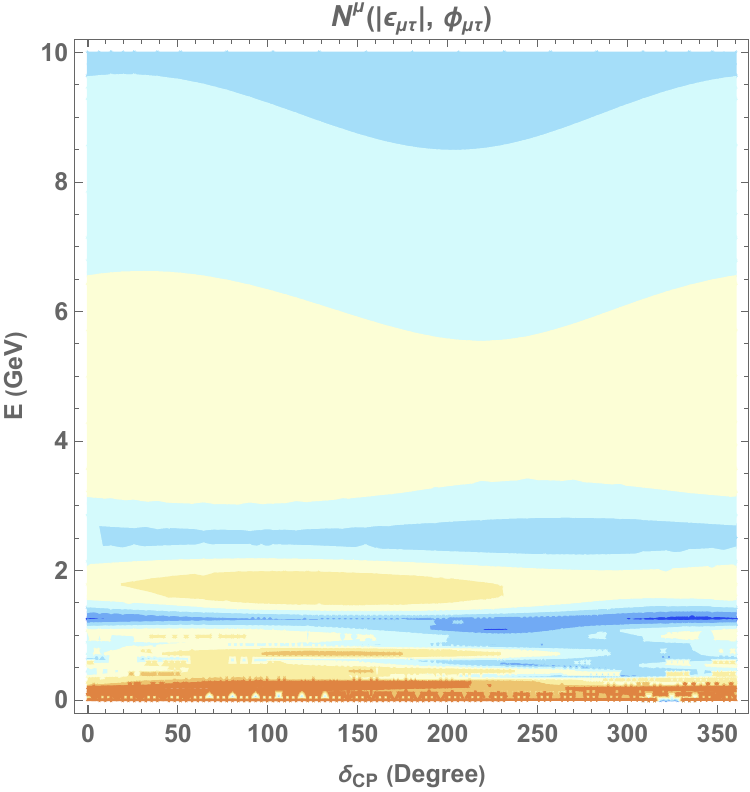}
\includegraphics[width=9.5mm]{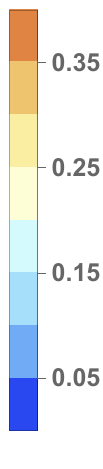}
\caption{ EOF (top), Concurrence (middle), and Negativity (bottom) are shown in the plane of the neutrino energy E (in GeV) and the CP-violating phase $\delta_{CP}$ (in degree) for the SO (Left),  $\epsilon_{\mu\tau}, \phi_{\mu\tau}=0$ (middle), and $\epsilon_{\mu\tau}, \phi_{\mu\tau}$ (right) scenarios for DUNE.}
\label{dfig3}
\end{figure*}
Figure \ref{dfig3} depicts the impact of NSI involving the $\epsilon_{\mu\tau}$ and their associated phase on entanglement measures, analyzed using three distinct measures plotted as functions of neutrino energy $E$ and the CP-violating phase $\delta_{CP}$. The analysis considers three separate scenarios:  the SO without NSI, the presence of a non-zero $\epsilon_{\mu\tau}$ with zero complex phase, and a situation with both the magnitude and phase of $\epsilon_{\mu\tau}$. The EOF exhibits a pattern for the SO scenario reaching its maximum within the energy intervals of $[1.5 - 2]$ GeV and $[4-6]$ GeV across all values of the CP-violating phase $\delta_{CP}$, whereas for the NSI case with a vanishing complex phase the maximum shifted into higher energy range, around $[5.5-9]$ GeV. However, in the NSI scenario with a non-vanishing complex phase, the peak of EOF is attained in the energy range $[4 - 5]$ GeV for the $\delta_{CP}$ values from 0$^\circ$ to 180$^\circ$ and 300$^\circ$ to 360$^\circ$. Notably, for this case, a distinction is observed between the EOF values at $\delta_{CP}$= 0$^\circ$ and 212$^\circ$, highlighting the phase sensitivity even in the absence of a complex NSI phase. A comparable trend is observed for Concurrence across all three scenarios, SO, NSI with vanishing phase, and NSI with non-vanishing phase, with the maximum occurring in the specific energy ranges, largely independent of $\delta_{CP}$. However, a distinct minimum region appears for Concurrence in the $\delta_{CP}$ range of 190$^\circ$ to 350$^\circ$ for energy nearly 3 GeV, with noticeable differences between the values at $\delta_{CP}$= 0$^\circ$ and 212$^\circ$ for the NSI with vanishing complex phase scenario. 
As for Negativity, the maximum value appears in the low energy for all scenarios. For the energy range $[1.5 - 2]$ GeV, the Negativity values at $\delta_{CP}$= 0$^\circ$ and 212$^\circ$, differ across all three scenarios, which are SO, NSI with vanishing phase, and NSI with non-vanishing phase.

\section{Conclusions}
\label{sec6}
\vspace{-0.3cm}
In this study, we investigated the impact of off-diagonal non-standard interaction (NSI) parameters on quantum entanglement within the three-flavor neutrino oscillation framework. By reformulating three commonly used entanglement measures; Entanglement of Formation (EOF), Concurrence, and Negativity in terms of oscillation probabilities, we analyzed how quantum correlations are modified by the presence of the off-diagonal NSI parameters $\left |\epsilon_{e\mu} \right |$, $\left |\epsilon_{e\tau} \right |$, and $\left |\epsilon_{\mu\tau} \right |$, including the influence of their complex phases. We illustrate our results for two representative choices of the $\delta_{CP}$: a CP-conserving value ($\delta_{CP}=0^\circ$) and the current best-fit value ($\delta_{CP}=212^\circ$) allowed by global analyses. However, depending on the true value of the $\delta_{CP}$, if it is determined precisely in future, one can conclude the NSI effect in different quantum measures. \\
Focusing on the DUNE experimental setup, we studied how both the magnitude and phase of these parameters affect the energy-dependent behavior of the entanglement measures. At the oscillation maximum, pronounced deviations are observed in all quantum correlation measures for all considered NSI parameters. Among the three measures, Negativity consistently shows the most pronounced sensitivity to NSI, clearly distinguishing scenarios with and without NSI non-zero phases from standard oscillations, at higher energies.
Moreover, we found that the entanglement behavior is predominantly governed by the oscillation (or survival) probabilities: the appearance channel plays the dominant role when $\epsilon_{e\mu}$ or $\epsilon_{e\tau}$ is considered individually, while the disappearance channel becomes central in the case of $\epsilon_{\mu\tau}$.\\
As event rates are derived from the oscillation probabilities by incorporating the source flux, interaction cross-sections, and detector effects, the overall structure and trends of the event rates closely mirror those of the underlying probabilities.\\
A detailed analysis of the entanglement measures in the plane of neutrino energy, $E$ and CP-violating phase, $\delta_{CP}$ highlights the distinct behavior of each measure under the influence of different off-diagonal NSI parameters. In all three NSI scenarios $\left |\epsilon_{e\mu} \right |$, $\left |\epsilon_{e\tau} \right |$, and $\left |\epsilon_{\mu\tau} \right |$ with their associated complex phases, Concurrence shows a broadly similar structure, with peaks in a specific energy ranges, and for all the value of the CP-violating phase $\delta_{CP}$. The behavior of EOF is different from that of Concurrence for all three NSI $\left |\epsilon_{e\mu} \right |$, $\left |\epsilon_{e\tau} \right |$, and $\left |\epsilon_{\mu\tau} \right |$ with their associated phase cases. The NSI $\epsilon_{e\mu}$ scenario for EOF is observed to be inconsistent across all CP violation phases. However, for the $\epsilon_{e\tau}$ and $\epsilon_{\mu\tau}$ NSI scenario, EOF displays a clear distinction between the SO and NSI with and without a complex phase $\phi_{e\tau}$, along with a noticeable dependence on $\delta_{CP}$.  Negativity demonstrates a stronger sensitivity to both energy and the CP-violating phase, exhibiting sharp variations in specific regions of the $E$ - $\delta_{CP}$ parameter space for all three NSI scenarios. \\
We also extended our analysis to the P2SO experiment and found that its results closely mirror those of DUNE, with slightly enhanced NSI effects attributable to the longer baseline of P2SO. This similarity indicates that both experiments exhibit comparable behavior in response to off-diagonal NSI parameters. Consequently, it can be inferred that other long-baseline neutrino experiments are likely to show similar trends in their entanglement behavior.

\section{Acknowledgements}
LK thanks the Ministry of Education (MoE) for financial support and the Indian Institute of Technology Jodhpur for providing necessary research facilities. PP wants to thank Prime Minister’s Research Fellows (PMRF) scheme for its financial support. We gratefully acknowledge the use of CMSD HPC facility of University of Hyderabad to carry out the computational works. 

\bibliography{main}

@article{Fukuda_1998,
   title={Evidence for Oscillation of Atmospheric Neutrinos},
   volume={81},
   ISSN={1079-7114},
   url={http://dx.doi.org/10.1103/PhysRevLett.81.1562},
   DOI={10.1103/physrevlett.81.1562},
   number={8},
   journal={Phys. Rev. Lett.},
   publisher={American Physical Society (APS)},
   author={Fukuda \textit{et al.} [Super-Kamiokande Collaboration], Y.} ,
   year={1998},
   month=aug, pages={1562–1567} }

@article{Ahmad_2002,
    author = "Ahmad \textit{et al.} [SNO Collaboration], Q. R.",
    title = "{Direct evidence for neutrino flavor transformation from neutral current interactions in the Sudbury Neutrino Observatory}",
    eprint = "nucl-ex/0204008",
    archivePrefix = "arXiv",
    doi = "10.1103/PhysRevLett.89.011301",
    journal = "Phys. Rev. Lett.",
    volume = "89",
    pages = "011301",
    year = "2002"
}

@article{Bennett:1996gf,
    author = "Bennett, Charles H. and DiVincenzo, David P. and Smolin, John A. and Wootters, William K.",
    title = "{Mixed state entanglement and quantum error correction}",
    eprint = "quant-ph/9604024",
    archivePrefix = "arXiv",
    doi = "10.1103/PhysRevA.54.3824",
    journal = "Phys. Rev. A",
    volume = "54",
    pages = "3824--3851",
    year = "1996"
}

@article{Guo2020,
  author       = {Yong Guo and Lin Zhang},
  title        = {Multipartite entanglement measure and complete monogamy relation},
  journal      = {Phys. Rev. A},
  volume       = {101},
  number       = {3},
  pages        = {032301},
  year         = {2020},
  doi          = {10.1103/PhysRevA.101.032301},
  publisher    = {American Physical Society}
}

@article{Peres:1996dw,
    author = "Peres, Asher",
    title = "{Separability criterion for density matrices}",
    eprint = "quant-ph/9604005",
    archivePrefix = "arXiv",
    doi = "10.1103/PhysRevLett.77.1413",
    journal = "Phys. Rev. Lett.",
    volume = "77",
    pages = "1413--1415",
    year = "1996"
}

@article{Vidal:2002zz,
    author = "Vidal, G. and Werner, R. F.",
    title = "{Computable measure of entanglement}",
    eprint = "quant-ph/0102117",
    archivePrefix = "arXiv",
    doi = "10.1103/PhysRevA.65.032314",
    journal = "Phys. Rev. A",
    volume = "65",
    pages = "032314",
    year = "2002"
}

@article{Sabin2008,
  author       = {C. Sab{\'i}n and G. Garc{\'i}a-Alcaine},
  title        = {A classification of entanglement in three-qubit systems},
  journal      = {Eur. Phys. J. C},
  volume       = {48},
  number       = {3},
  pages        = {435--442},
  year         = {2008},
  doi          = {10.1140/epjc/s10052-006-0041-1},
  publisher    = {Springer}
}

@article{Chatterjee:2020kkm,
    author = "Chatterjee, Sabya Sachi and Palazzo, Antonio",
    title = "{Nonstandard Neutrino Interactions as a Solution to the $NO\nu A$ and T2K Discrepancy}",
    eprint = "2008.04161",
    archivePrefix = "arXiv",
    primaryClass = "hep-ph",
    reportNumber = "IPPP/20/35",
    doi = "10.1103/PhysRevLett.126.051802",
    journal = "Phys. Rev. Lett.",
    volume = "126",
    number = "5",
    pages = "051802",
    year = "2021"
}

@article{Denton:2020uda,
    author = "Denton, Peter B. and Gehrlein, Julia and Pestes, Rebekah",
    title = "{$CP$ -Violating Neutrino Nonstandard Interactions in Long-Baseline-Accelerator Data}",
    eprint = "2008.01110",
    archivePrefix = "arXiv",
    primaryClass = "hep-ph",
    doi = "10.1103/PhysRevLett.126.051801",
    journal = "Phys. Rev. Lett.",
    volume = "126",
    number = "5",
    pages = "051801",
    year = "2021"
}

@article{Liao:2016orc,
    author = "Liao, Jiajun and Marfatia, Danny and Whisnant, Kerry",
    title = "{Nonstandard neutrino interactions at DUNE, T2HK and T2HKK}",
    eprint = "1612.01443",
    archivePrefix = "arXiv",
    primaryClass = "hep-ph",
    doi = "10.1007/JHEP01(2017)071",
    journal = "JHEP",
    volume = "01",
    pages = "071",
    year = "2017"
}

@article{Miranda:2015dra,
    author = "Miranda, O. G. and Nunokawa, H.",
    title = "{Non standard neutrino interactions: current status and future prospects}",
    eprint = "1505.06254",
    archivePrefix = "arXiv",
    primaryClass = "hep-ph",
    doi = "10.1088/1367-2630/17/9/095002",
    journal = "New J. Phys.",
    volume = "17",
    number = "9",
    pages = "095002",
    year = "2015"
}

@article{Ohlsson:2012kf,
    author = "Ohlsson, Tommy",
    title = "{Status of non-standard neutrino interactions}",
    eprint = "1209.2710",
    archivePrefix = "arXiv",
    primaryClass = "hep-ph",
    doi = "10.1088/0034-4885/76/4/044201",
    journal = "Rept. Prog. Phys.",
    volume = "76",
    pages = "044201",
    year = "2013"
}

@article{Guo2019,
  author       = {Yong Guo and Gilad Gour},
  title        = {Monogamy of the entanglement of formation},
  journal      = {Phys. Rev. A},
  volume       = {99},
  number       = {4},
  pages        = {042305},
  year         = {2019},
  doi          = {10.1103/PhysRevA.99.042305},
  publisher    = {American Physical Society}
}

@article{Wootters:1997id,
    author = "Wootters, William K.",
    title = "{Entanglement of formation of an arbitrary state of two qubits}",
    eprint = "quant-ph/9709029",
    archivePrefix = "arXiv",
    doi = "10.1103/PhysRevLett.80.2245",
    journal = "Phys. Rev. Lett.",
    volume = "80",
    pages = "2245--2248",
    year = "1998"
}

@article{Hill:1997pfa,
    author = "Hill, Scott and Wootters, William K.",
    title = "{Entanglement of a pair of quantum bits}",
    eprint = "quant-ph/9703041",
    archivePrefix = "arXiv",
    doi = "10.1103/PhysRevLett.78.5022",
    journal = "Phys. Rev. Lett.",
    volume = "78",
    pages = "5022--5025",
    year = "1997"
}

@article{Blasone_2008,
    author = "Blasone, M. and Dell'Anno, F. and De Siena, S. and Di Mauro, M. and Illuminati, F.",
    title = "{Multipartite entangled states in particle mixing}",
    eprint = "0711.2268",
    archivePrefix = "arXiv",
    primaryClass = "quant-ph",
    doi = "10.1103/PhysRevD.77.096002",
    journal = "Phys. Rev. D",
    volume = "77",
    pages = "096002",
    year = "2008"
}

@article{Blasone_2009Entanglement,
   title={Entanglement in neutrino oscillations},
   volume={85},
   ISSN={1286-4854},
   url={http://dx.doi.org/10.1209/0295-5075/85/50002},
   DOI={10.1209/0295-5075/85/50002},
   number={5},
   journal={EPL},
   publisher={IOP Publishing},
   author={Blasone, M. and Dell’Anno, F. and De Siena, S. and Illuminati, F.},
   year={2009},
   month=mar, pages={50002} }

@article{Horodecki_2009,
   title={Quantum entanglement},
   volume={81},
   ISSN={1539-0756},
   url={http://dx.doi.org/10.1103/RevModPhys.81.865},
   DOI={10.1103/revmodphys.81.865},
   number={2},
   journal={Rev. Mod. Phys.},
   publisher={American Physical Society (APS)},
   author={Horodecki, Ryszard and Horodecki, Paweł and Horodecki, Michał and Horodecki, Karol},
   year={2009},
   month=jun, pages={865–942} }

@article{Abi:2020wmh,
  author = {Abi \textit{et al.} [DUNE Collaboration], B.},
  title = {{Deep Underground Neutrino Experiment (DUNE), Far Detector Technical Design Report, Volume I Introduction to DUNE}},
  journal = {JINST},
  volume = {15},
  year = {2020},
  number = {08},
  pages = {T08008},
  doi = {10.1088/1748-0221/15/08/T08008},
  eprint = {2002.02967},
  archivePrefix = {arXiv},
  primaryClass = {physics.ins-det}
}

@article{Abe_2023,
    author = "Abe \textit{et al.} [T2K Collaboration], K.",
    title = "{Measurements of neutrino oscillation parameters from the T2K experiment using $3.6\times 10^{21}$ protons on target}",
    eprint = "2303.03222",
    archivePrefix = "arXiv",
    primaryClass = "hep-ex",
    doi = "10.1140/epjc/s10052-023-11819-x",
    journal = "Eur. Phys. J. C",
    volume = "83",
    number = "9",
    pages = "782",
    year = "2023"
}

@article{Abi_2021,
    author = "Abi \textit{et al.} [DUNE Collaboration], B.",
    title = "{Prospects for beyond the Standard Model physics searches at the Deep Underground Neutrino Experiment}",
    eprint = "2008.12769",
    archivePrefix = "arXiv",
    primaryClass = "hep-ex",
    reportNumber = "FERMILAB-PUB-20-459-LBNF-ND, FERMILAB-PUB-20-459-LBNF-ND",
    doi = "10.1140/epjc/s10052-021-09007-w",
    journal = "Eur. Phys. J. C",
    volume = "81",
    number = "4",
    pages = "322",
    year = "2021"
}

@article{Biggio_2009,
   title={General bounds on non-standard neutrino interactions},
   volume={2009},
   ISSN={1029-8479},
   url={http://dx.doi.org/10.1088/1126-6708/2009/08/090},
   DOI={10.1088/1126-6708/2009/08/090},
   number={08},
   journal={JHEP},
   publisher={Springer Science and Business Media LLC},
   author={Biggio, Carla and Blennow, Mattias and Fernández-Martínez, Enrique},
   year={2009},
   month=aug, pages={090–090} }

@article{farzan2018,
    author = "Farzan, Y. and Tortola, M.",
    title = "{Neutrino oscillations and Non-Standard Interactions}",
    eprint = "1710.09360",
    archivePrefix = "arXiv",
    primaryClass = "hep-ph",
    doi = "10.3389/fphy.2018.00010",
    journal = "Front. in Phys.",
    volume = "6",
    pages = "10",
    year = "2018"
}

@article{KumarJha:2020pke,
    author = "K. Jha, Abhishek and Mukherjee, Supratik and Bambah, Bindu A.",
    title = "{Tri-Partite entanglement in Neutrino Oscillations}",
    eprint = "2004.14853",
    archivePrefix = "arXiv",
    primaryClass = "hep-ph",
    doi = "10.1142/S0217732321500565",
    journal = "Mod. Phys. Lett. A",
    volume = "36",
    number = "09",
    pages = "2150056",
    year = "2021"
}

@article{Bouri:2024kcl,
    author = "Bouri, Subhadip and Jha, Abhishek Kumar and Banerjee, Subhashish",
    title = "{Probing CP violation and mass ordering in neutrino oscillations in matter through quantum speed limits}",
    eprint = "2405.13114",
    archivePrefix = "arXiv",
    primaryClass = "hep-ph",
    doi = "10.1103/w6rq-l3ql",
    journal = "Phys. Rev. D",
    volume = "112",
    number = "3",
    pages = "036007",
    year = "2025"
}

@article{Konwar_2024,
   title={NSI effects on tripartite entanglement in neutrino oscillations},
   volume={1002},
   ISSN={0550-3213},
   url={http://dx.doi.org/10.1016/j.nuclphysb.2024.116544},
   DOI={10.1016/j.nuclphysb.2024.116544},
   journal={Nucl. Phys. B},
   publisher={Elsevier BV},
   author={Konwar, Lekhashri and Yadav, Bhavna},
   year={2024},
   month=may, pages={116544} }

@article{Blasone_2021,
    author = "Blasone, Massimo and De Siena, Silvio and Matrella, Cristina",
    title = "{Wave packet approach to quantum correlations in neutrino oscillations}",
    eprint = "2104.03166",
    archivePrefix = "arXiv",
    primaryClass = "quant-ph",
    doi = "10.1140/epjc/s10052-021-09471-4",
    journal = "Eur. Phys. J. C",
    volume = "81",
    number = "7",
    pages = "660",
    year = "2021"
}

@article{Blasone_2010,
   title={On entanglement in neutrino mixing and oscillations},
   volume={237},
   ISSN={1742-6596},
   url={http://dx.doi.org/10.1088/1742-6596/237/1/012007},
   DOI={10.1088/1742-6596/237/1/012007},
  journal={Journal of Physics: Conference Series},
   publisher={IOP Publishing},
   author={Blasone, Massimo and Dell’Anno, Fabio and Siena, Silvio De and Illuminati, Fabrizio},
   year={2010},
   month=jun, pages={012007} }

@article{li2021characterizing,
    author = "Li, Li-Juan and Ming, Fei and Song, Xue-Ke and Ye, Liu and Wang, Dong",
    title = "{Characterizing entanglement and measurement{\textquoteright}s uncertainty in neutrino oscillations}",
    doi = "10.1140/epjc/s10052-021-09503-z",
    journal = "Eur. Phys. J. C",
    volume = "81",
    number = "8",
    pages = "728",
    year = "2021"
}

@article{deepthi2015,
    author = "Deepthi, K. N. and C, Soumya and Mohanta, R.",
    title = "{Revisiting the sensitivity studies for leptonic CP-violation and mass hierarchy with T2K, NOnuA and LBNE experiments}",
    eprint = "1409.2343",
    archivePrefix = "arXiv",
    primaryClass = "hep-ph",
    doi = "10.1088/1367-2630/17/2/023035",
    journal = "New J. Phys.",
    volume = "17",
    number = "2",
    pages = "023035",
    year = "2015"
}

@article{fukasawa2017,
    author = "Fukasawa, Shinya and Ghosh, Monojit and Yasuda, Osamu",
    title = "{Complementarity Between Hyperkamiokande and DUNE in Determining Neutrino Oscillation Parameters}",
    eprint = "1607.03758",
    archivePrefix = "arXiv",
    primaryClass = "hep-ph",
    doi = "10.1016/j.nuclphysb.2017.02.008",
    journal = "Nucl. Phys. B",
    volume = "918",
    pages = "337--357",
    year = "2017"
}

@article{masud2016probing,
    author = "Masud, Mehedi and Chatterjee, Animesh and Mehta, Poonam",
    title = "{Probing CP violation signal at DUNE in presence of non-standard neutrino interactions}",
    eprint = "1510.08261",
    archivePrefix = "arXiv",
    primaryClass = "hep-ph",
    doi = "10.1088/0954-3899/43/9/095005/meta",
    journal = "J. Phys. G",
    volume = "43",
    number = "9",
    pages = "095005",
    year = "2016"
}

@article{de2016non,
    author = "de Gouv{\^e}a, Andr{\'e} and Kelly, Kevin J.",
    title = "{Non-standard neutrino interactions at DUNE}",
    eprint = "1511.05562",
    archivePrefix = "arXiv",
    primaryClass = "hep-ph",
    reportNumber = "NUHEP-TH-15-10",
    doi = "10.1016/j.nuclphysb.2016.03.013",
    journal = "Nucl. Phys. B",
    volume = "908",
    pages = "318--335",
    year = "2016"
}

@article{Deepthi_2018,
   title={Challenges posed by non-standard neutrino interactions in the determination of δ at DUNE},
   volume={936},
   ISSN={0550-3213},
   url={http://dx.doi.org/10.1016/j.nuclphysb.2018.09.004},
   DOI={10.1016/j.nuclphysb.2018.09.004},
   journal={Nucl. Phys. B},
   publisher={Elsevier BV},
   author={Deepthi, K.N. and Goswami, Srubabati and Nath, Newton},
   year={2018},
   month=nov, pages={91–105} }

@article{Blennow_2017,
    author = "Blennow, Mattias and Coloma, Pilar and Fernandez-Martinez, Enrique and Hernandez-Garcia, Josu and Lopez-Pavon, Jacobo",
    title = "{Non-Unitarity, sterile neutrinos, and Non-Standard neutrino Interactions}",
    eprint = "1609.08637",
    archivePrefix = "arXiv",
    primaryClass = "hep-ph",
    reportNumber = "IFT-UAM-CSIC-16-090, FTUAM-16-35, FERMILAB-PUB-16-400-T",
    doi = "10.1007/JHEP04(2017)153",
    journal = "JHEP",
    volume = "04",
    pages = "153",
    year = "2017"
}

@article{Agarwalla_2016,
   title={Degeneracy between θ23 octant and neutrino non-standard interactions at DUNE},
   volume={762},
   ISSN={0370-2693},
   url={http://dx.doi.org/10.1016/j.physletb.2016.09.020},
   DOI={10.1016/j.physletb.2016.09.020},
   journal={Phys. Lett. B},
   publisher={Elsevier BV},
   author={Agarwalla, Sanjib Kumar and Chatterjee, Sabya Sachi and Palazzo, Antonio},
   year={2016},
   month=nov, pages={64–71} }

@article{Masud_2016,
    author = "Masud, Mehedi and Mehta, Poonam",
    title = "{Nonstandard interactions and resolving the ordering of neutrino masses at DUNE and other long baseline experiments}",
    eprint = "1606.05662",
    archivePrefix = "arXiv",
    primaryClass = "hep-ph",
    doi = "10.1103/PhysRevD.94.053007",
    journal = "Phys. Rev. D",
    volume = "94",
    number = "5",
    pages = "053007",
    year = "2016"
}

@article{Coloma_2017,
    author = "Coloma, Pilar and Schwetz, Thomas",
    title = "{Generalized mass ordering degeneracy in neutrino oscillation experiments}",
    eprint = "1604.05772",
    archivePrefix = "arXiv",
    primaryClass = "hep-ph",
    reportNumber = "FERMILAB-PUB-16-115-T",
    doi = "10.1103/PhysRevD.94.055005",
    journal = "Phys. Rev. D",
    volume = "94",
    number = "5",
    pages = "055005",
    year = "2016"
}

@article{Liao_2016,
    author = "Liao, Jiajun and Marfatia, Danny and Whisnant, Kerry",
    title = "{Degeneracies in long-baseline neutrino experiments from nonstandard interactions}",
    eprint = "1601.00927",
    archivePrefix = "arXiv",
    primaryClass = "hep-ph",
    doi = "10.1103/PhysRevD.93.093016",
    journal = "Phys. Rev. D",
    volume = "93",
    number = "9",
    pages = "093016",
    year = "2016"
}

@article{Coloma_2016,
    author = "Coloma, Pilar",
    title = "{Non-Standard Interactions in propagation at the Deep Underground Neutrino Experiment}",
    eprint = "1511.06357",
    archivePrefix = "arXiv",
    primaryClass = "hep-ph",
    reportNumber = "FERMILAB-PUB-15-501-T",
    doi = "10.1007/JHEP03(2016)016",
    journal = "JHEP",
    volume = "03",
    pages = "016",
    year = "2016"
}

@article{banerjee2024analysis,
      title={Analysis of neutrino oscillation parameters in the light on quantum entanglement}, 
      author={Rajrupa Banerjee and Papia Panda and Rukmani Mohanta and Sudhanwa Patra},
      year={2024},
      eprint={2410.05727},
      archivePrefix={arXiv},
      primaryClass={hep-ph},
      url={https://arxiv.org/abs/2410.05727}, 
}

@article{Alok_2016q,
   title={Quantum correlations in terms of neutrino oscillation probabilities},
   volume={909},
   ISSN={0550-3213},
   url={http://dx.doi.org/10.1016/j.nuclphysb.2016.05.001},
   DOI={10.1016/j.nuclphysb.2016.05.001},
   journal={Nucl. Phys. B},
   publisher={Elsevier BV},
   author={Alok, Ashutosh Kumar and Banerjee, Subhashish and Uma Sankar, S.},
   year={2016},
   month=aug, pages={65–72} }

@article{banerjee2015quantum,
  title={A quantum-information theoretic analysis of three-flavor neutrino oscillations: Quantum entanglement, nonlocal and nonclassical features of neutrinos},
  author={Banerjee, Subhashish and Alok, Ashutosh Kumar and Srikanth, R and Hiesmayr, Beatrix C},
  journal={ Eur. Phys. J. C},
  volume={75},
  number={10},
  pages={487},
  year={2015},
  publisher={Springer},
  url={http://dx.doi.org/10.1140/epjc/s10052-015-3717-x},
   DOI={10.1140/epjc/s10052-015-3717-x}
}

@article{Li:2022mus,
    author = "Li, Yu-Wen and Li, Li-Juan and Song, Xue-Ke and Wang, Dong and Ye, Liu",
    title = "{Geuine tripartite entanglement in three-flavor neutrino oscillations}",
    eprint = "2205.11058",
    archivePrefix = "arXiv",
    primaryClass = "quant-ph",
    doi = "10.1140/epjc/s10052-022-10759-2",
    journal = "Eur. Phys. J. C",
    volume = "82",
    number = "9",
    pages = "799",
    year = "2022"
}

@article{fu2017testing,
    author = "Fu, Qiang and Chen, Xurong",
    title = "{Testing violation of the Leggett{\textendash}Garg-type inequality in neutrino oscillations of the Daya Bay experiment}",
    eprint = "1705.08601",
    archivePrefix = "arXiv",
    primaryClass = "hep-ph",
    doi = "10.1140/epjc/s10052-017-5371-y",
    journal = "Eur. Phys. J. C",
    volume = "77",
    number = "11",
    pages = "775",
    year = "2017"
}

@article{formaggio2016violation,
    author = "Formaggio, J. A. and Kaiser, D. I. and Murskyj, M. M. and Weiss, T. E.",
    title = "{Violation of the Leggett-Garg Inequality in Neutrino Oscillations}",
    eprint = "1602.00041",
    archivePrefix = "arXiv",
    primaryClass = "quant-ph",
    doi = "10.1103/PhysRevLett.117.050402",
    journal = "Phys. Rev. Lett.",
    volume = "117",
    number = "5",
    pages = "050402",
    year = "2016"
}

@article{naikoo2020quantum,
   title={A quantum information theoretic quantity sensitive to the neutrino mass-hierarchy},
   volume={951},
   ISSN={0550-3213},
   url={http://dx.doi.org/10.1016/j.nuclphysb.2019.114872},
   DOI={10.1016/j.nuclphysb.2019.114872},
   journal={Nucl. Phys. B},
   publisher={Elsevier BV},
   author={Naikoo, Javid and Alok, Ashutosh Kumar and Banerjee, Subhashish and Uma Sankar, S. and Guarnieri, Giacomo and Schultze, Christiane and Hiesmayr, Beatrix C.},
   year={2020},
   month=feb, pages={114872} }

@article{naikoo2019leggett,
    author = "Naikoo, Javid and Kumar Alok, Ashutosh and Banerjee, Subhashish and Uma Sankar, S.",
    title = "{Leggett-Garg inequality in the context of three flavour neutrino oscillation}",
    eprint = "1901.10859",
    archivePrefix = "arXiv",
    primaryClass = "hep-ph",
    doi = "10.1103/PhysRevD.99.095001",
    journal = "Phys. Rev. D",
    volume = "99",
    number = "9",
    pages = "095001",
    year = "2019"
}

@article{shafaq2021enhanced,
   title={Enhanced violation of Leggett–Garg inequality in three flavour neutrino oscillations via non-standard interactions},
   volume={48},
   ISSN={1361-6471},
   url={http://dx.doi.org/10.1088/1361-6471/abff0d},
   DOI={10.1088/1361-6471/abff0d},
   number={8},
   journal={J. Phys. G},
   publisher={IOP Publishing},
   author={Shafaq, Sheeba and Mehta, Poonam},
   year={2021},
   month=jun, pages={085002} }

@article{sarkar2021effects,
    author = "Sarkar, Trisha and Dixit, Khushboo",
    title = "{Effects of nonstandard interaction on temporal and spatial correlations in neutrino oscillations}",
    eprint = "2010.02175",
    archivePrefix = "arXiv",
    primaryClass = "hep-ph",
    doi = "10.1140/epjc/s10052-021-08874-7",
    journal = "Eur. Phys. J. C",
    volume = "81",
    number = "1",
    pages = "88",
    year = "2021"
}

@article{blasone2023leggett,
    author = "Blasone, Massimo and Illuminati, Fabrizio and Petruzziello, Luciano and Smaldone, Luca",
    title = "{Leggett-Garg inequalities in the quantum field theory of neutrino oscillations}",
    eprint = "2111.09979",
    archivePrefix = "arXiv",
    primaryClass = "quant-ph",
    doi = "10.1103/PhysRevA.108.032210",
    journal = "Phys. Rev. A",
    volume = "108",
    number = "3",
    pages = "032210",
    year = "2023"
}

@article{chattopadhyay2023quantum,
    author = "Chattopadhyay, Dibya S. and Dighe, Amol",
    title = "{Quantum mismatch: A powerful measure of quantumness in neutrino oscillations}",
    eprint = "2304.02475",
    archivePrefix = "arXiv",
    primaryClass = "hep-ph",
    reportNumber = "TIFR/TH/23-2",
    doi = "10.1103/PhysRevD.108.112013",
    journal = "Phys. Rev. D",
    volume = "108",
    number = "11",
    pages = "112013",
    year = "2023"
}

@article{konwar2024violation,
    author = "Konwar, Lekhashri and Vardani, Juhi and Yadav, Bhavna",
    title = "{Violation of LGtI inequalities in the light of NO$\nu $A and T2K anomaly}",
    eprint = "2401.02886",
    archivePrefix = "arXiv",
    primaryClass = "hep-ph",
    doi = "10.1140/epjc/s10052-024-13370-9",
    journal = "Eur. Phys. J. C",
    volume = "84",
    number = "10",
    pages = "1103",
    year = "2024"
}

@article{dixit2024quantum,
    author = "Dixit, Khushboo and Haque, S. Shajidul and Razzaque, Soebur",
    title = "{Quantum spread complexity in neutrino oscillations}",
    eprint = "2305.17025",
    archivePrefix = "arXiv",
    primaryClass = "hep-ph",
    doi = "10.1140/epjc/s10052-024-12620-0",
    journal = "Eur. Phys. J. C",
    volume = "84",
    number = "3",
    pages = "260",
    year = "2024"
}

@article{konwar2025steering,
    author = "Konwar, Lekhashri and Yadav, Bhavna",
    title = "{Steering in neutrino oscillations with non-standard interaction}",
    eprint = "2411.14234",
    archivePrefix = "arXiv",
    primaryClass = "hep-ph",
    doi = "10.1088/1361-6471/adbfb0",
    journal = "J. Phys. G",
    volume = "52",
    number = "4",
    pages = "045001",
    year = "2025"
}

@article{konwar2025neutrino,
    author = "Konwar, Lekhashri",
    title = "{Neutrino mass ordering and CP violation via quantum steering with NSI}",
    doi = "10.1016/j.jspc.2025.100065",
    journal = "J. Subatomic Part. Cosmol.",
    volume = "3",
    pages = "100065",
    year = "2025"
}

@article{Acero_2022,
    author = "Acero \textit{et al.} [NOvA Collaboration], M. A.",
    title = "{Improved measurement of neutrino oscillation parameters by the NOvA experiment}",
    eprint = "2108.08219",
    archivePrefix = "arXiv",
    primaryClass = "hep-ex",
    reportNumber = "FERMILAB-PUB-21-373-ND",
    doi = "10.1103/PhysRevD.106.032004",
    journal = "Phys. Rev. D",
    volume = "106",
    number = "3",
    pages = "032004",
    year = "2022"
}

@article{JUNO:2021vlw,
    author = "Abusleme \textit{et al.} [JUNO Collaboration], A.",
    title = "{JUNO physics and detector}",
    eprint = "2104.02565",
    archivePrefix = "arXiv",
    primaryClass = "hep-ex",
    doi = "10.1016/j.ppnp.2021.103927",
    journal = "Prog. Part. Nucl. Phys.",
    volume = "123",
    pages = "103927",
    year = "2022"
}

@article{Esteban_2024,
    author = "Esteban, Ivan and Gonzalez-Garcia, M. C. and Maltoni, Michele and Martinez-Soler, Ivan and Pinheiro, Jo{\~a}o Paulo and Schwetz, Thomas",
    title = "{NuFit-6.0: updated global analysis of three-flavor neutrino oscillations}",
    eprint = "2410.05380",
    archivePrefix = "arXiv",
    primaryClass = "hep-ph",
    reportNumber = "IFT-UAM/CSIC-24-140, YITP-SB-2024-24, IPPP/24/64, IPPP/24/64, IFT-UAM/CSIC-24-140, YITP-SB-2024-24",
    doi = "10.1007/JHEP12(2024)216",
    journal = "JHEP",
    volume = "12",
    pages = "216",
    year = "2024"
}

@article{DUNE:2021cuw,
    author = "Abi, B. and others",
    collaboration = "DUNE",
    title = "{Experiment Simulation Configurations Approximating DUNE TDR}",
    eprint = "2103.04797",
    archivePrefix = "arXiv",
    primaryClass = "hep-ex",
    reportNumber = "FERMILAB-FN-1125-ND",
    month = "3",
    year = "2021"
}

@article{Huber:2004ka,
    author = "Huber, Patrick and Lindner, M. and Winter, W.",
    title = "{Simulation of long-baseline neutrino oscillation experiments with GLoBES (General Long Baseline Experiment Simulator)}",
    eprint = "hep-ph/0407333",
    archivePrefix = "arXiv",
    reportNumber = "TUM-HEP-553-04",
    doi = "10.1016/j.cpc.2005.01.003",
    journal = "Comput. Phys. Commun.",
    volume = "167",
    pages = "195",
    year = "2005"
}

@article{Huber:2007ji,
    author = "Huber, Patrick and Kopp, Joachim and Lindner, Manfred and Rolinec, Mark and Winter, Walter",
    title = "{New features in the simulation of neutrino oscillation experiments with GLoBES 3.0: General Long Baseline Experiment Simulator}",
    eprint = "hep-ph/0701187",
    archivePrefix = "arXiv",
    reportNumber = "TUM-HEP-656-07",
    doi = "10.1016/j.cpc.2007.05.004",
    journal = "Comput. Phys. Commun.",
    volume = "177",
    pages = "432--438",
    year = "2007"
}

@article{Panda:2024avc,
    author = "Panda, Papia and Mohanta, Rukmani",
    title = "{Probing neutrino mass ordering with supernova neutrinos at NO{\ensuremath{\nu}}A including the effect of sterile neutrinos}",
    eprint = "2412.05213",
    archivePrefix = "arXiv",
    primaryClass = "hep-ph",
    doi = "10.1016/j.nuclphysb.2025.117002",
    journal = "Nucl. Phys. B",
    volume = "1018",
    pages = "117002",
    year = "2025"
}

@article{Panda:2024qsh,
    author = "Panda, Papia and Singha, Dinesh Kumar and Ghosh, Monojit and Mohanta, Rukmani",
    title = "{Effect of torsion in long-baseline neutrino oscillation experiments}",
    eprint = "2403.09105",
    archivePrefix = "arXiv",
    primaryClass = "hep-ph",
    doi = "10.1140/epjc/s10052-025-13771-4",
    journal = "Eur. Phys. J. C",
    volume = "85",
    number = "1",
    pages = "67",
    year = "2025"
}

@inproceedings{mohanta:2025,
    author = "Mohanta, Rukmani",
    title = "{Exploring Physics beyond the Standard Model with Neutrinos}",
    booktitle = "{Particle Physics and Cosmology in the Himalayas}",
    eprint = "2503.12985",
    archivePrefix = "arXiv",
    primaryClass = "hep-ph",
    month = "3",
    year = "2025"
}

@article{bera2025,
    author = "Bera, Chinmay and Deepthi, K. N. and Mohanta, Rukmani",
    title = "{The effect of non-standard interactions and environmental decoherence at DUNE}",
    eprint = "2501.14383",
    archivePrefix = "arXiv",
    primaryClass = "hep-ph",
    doi = "10.1007/JHEP06(2025)179",
    journal = "JHEP",
    volume = "06",
    pages = "179",
    year = "2025"
}

@article{pusty2024,
    author = "Pusty, Sambit Kumar and Majhi, Rudra and Singha, Dinesh Kumar and Ghosh, Monojit and Mohanta, Rukmani",
    title = "{Impact of scalar NSI with off-diagonal parameters at DUNE and P2SO}",
    eprint = "2410.23014",
    archivePrefix = "arXiv",
    primaryClass = "hep-ph",
    month = "10",
    year = "2024"
}

@article{Majhi_2023,
    author = "Majhi, Rudra and Singha, Dinesh Kumar and Ghosh, Monojit and Mohanta, Rukmani",
    title = "{Distinguishing nonstandard interaction and Lorentz invariance violation at the Protvino to super-ORCA experiment}",
    eprint = "2212.07244",
    archivePrefix = "arXiv",
    primaryClass = "hep-ph",
    doi = "10.1103/PhysRevD.107.075036",
    journal = "Phys. Rev. D",
    volume = "107",
    number = "7",
    pages = "075036",
    year = "2023"
}

@article{singha2024,
    author = "Singha, Dinesh Kumar and Majhi, Rudra and Panda, Lipsarani and Ghosh, Monojit and Mohanta, Rukmani",
    title = "{Study of scalar nonstandard interaction at the Protvino to super-ORCA experiment}",
    eprint = "2308.10789",
    archivePrefix = "arXiv",
    primaryClass = "hep-ph",
    doi = "10.1103/PhysRevD.109.095038",
    journal = "Phys. Rev. D",
    volume = "109",
    number = "9",
    pages = "095038",
    year = "2024"
}

@article{Denton:2022pxt,
    author = "Denton, Peter B. and Giarnetti, Alessio and Meloni, Davide",
    title = "{How to identify different new neutrino oscillation physics scenarios at DUNE}",
    eprint = "2210.00109",
    archivePrefix = "arXiv",
    primaryClass = "hep-ph",
    doi = "10.1007/JHEP02(2023)210",
    journal = "JHEP",
    volume = "02",
    pages = "210",
    year = "2023"
}

@book{Giunti:2007ry,
    author = "Giunti, Carlo and Kim, Chung W.",
    title = "{Fundamentals of Neutrino Physics and Astrophysics}",
    doi = "10.1093/acprof:oso/9780198508717.001.0001",
    isbn = "978-0-19-850871-7",
    year = "2007"
}

@article{blasone2014field,
  title={A field-theoretical approach to entanglement in neutrino mixing and oscillations},
  author={Blasone, Massimo and Dell'Anno, Fabio and De Siena, Silvio and Illuminati, Fabrizio},
  journal={EPL},
  volume={106},
  number={3},
  pages={30002},
  year={2014},
  publisher={IOP Publishing},
 url={http://dx.doi.org/10.1209/0295-5075/106/30002},
   DOI={10.1209/0295-5075/106/30002}
}

@article{Liao:2016hsa,
    author = "Liao, Jiajun and Marfatia, Danny and Whisnant, Kerry",
    title = "{Degeneracies in long-baseline neutrino experiments from nonstandard interactions}",
    eprint = "1601.00927",
    archivePrefix = "arXiv",
    primaryClass = "hep-ph",
    doi = "10.1103/PhysRevD.93.093016",
    journal = "Phys. Rev. D",
    volume = "93",
    number = "9",
    pages = "093016",
    year = "2016"
}

@article{Kopp:2007ne,
    author = "Kopp, Joachim and Lindner, Manfred and Ota, Toshihiko and Sato, Joe",
    title = "{Non-standard neutrino interactions in reactor and superbeam experiments}",
    eprint = "0708.0152",
    archivePrefix = "arXiv",
    primaryClass = "hep-ph",
    reportNumber = "STUPP-07-192",
    doi = "10.1103/PhysRevD.77.013007",
    journal = "Phys. Rev. D",
    volume = "77",
    pages = "013007",
    year = "2008"
}

@article{Kikuchi:2008vq,
    author = "Kikuchi, Takashi and Minakata, Hisakazu and Uchinami, Shoichi",
    title = "{Perturbation Theory of Neutrino Oscillation with Nonstandard Neutrino Interactions}",
    eprint = "0809.3312",
    archivePrefix = "arXiv",
    primaryClass = "hep-ph",
    doi = "10.1088/1126-6708/2009/03/114",
    journal = "JHEP",
    volume = "03",
    pages = "114",
    year = "2009"
}

\newpage
\appendix

\section{ Entanglement measures in Neutrino Oscillations}
\label{sec:analytic}
\subsection{Entanglement of Formation in terms of Neutrino Oscillation Probabilities}
From the definition of EOF in Eq. \ref{EOF3}, the density matrix $\rho_{ABC}$ of the tripartite system with A, B, and C as the subsystems is defined as
\begin{equation} \label{A1}
   EOF(\rho _{ABC}(t))=\frac{1}{2}[S(\rho _{A})+S(\rho _{B})+S(\rho _{C})],
\end{equation}
where $S(\rho _{A})$, $ S(\rho _{B})$ and $ S(\rho _{C})$ are von Neumann entropies defined as $S(\rho _{A})=-Tr(\rho _{A}\log\rho _{A})$ and same with $S(\rho _{B})$ and $S(\rho _{C})$. $\rho_{A}$, $\rho_{B}$ and $\rho_{C}$ are reduced density matrices which have the expressions, $\rho _{A}=Tr_{BC}\left (\rho _{ABC}(t)\right )$, $\rho _{B}=Tr_{AC}\left (\rho _{ABC}(t)\right )$ and $\rho _{C}=Tr_{AB}\left (\rho _{ABC}(t)\right )$. \\

\textbf{Density matrix for neutrino system:}
 For a three-qubit system, the total Hilbert space has $2^{3}=8$ basis states, which are
 $$\left\{\ket{000}, \ket{001}, \ket{010}, \ket{011}, \ket{100}, \ket{101}, \ket{110}, \ket{111}\right\}$$
 The time-evolved neutrino state in the flavor basis is given by:\\
$\ket{\nu _{\alpha }\left ( t \right )}=\Bar{U}_{\alpha e}^{f}\left ( t \right )\ket{100}_{e}+\Bar{U}_{\alpha \mu}^{f}\left ( t \right )\ket{010}_{\mu} +\Bar{U}_{\alpha \tau}^{f}\left ( t \right ) \ket{001}_{\tau}.$ \\ 
Here $\Bar{U}_{\alpha e}^{f}\left ( t \right ),\Bar{U}_{\alpha \mu}^{f}\left ( t \right )$, and $  \Bar{U}_{\alpha \tau}^{f}\left ( t \right )$ are the time-evolution operator of the respective flavor states with $\Bar U^f_{\alpha \beta}(t)= U e^{-\iota H_{tot}t}U^{-1}$, where $H_{tot}$ is defined in Eq. \eqref{nsi2}. 
\\
The corresponding conjugate state is: 
$\bra{\nu _{\alpha }\left ( t \right )}=\bar{U}^{f*}_{\alpha e}\left ( t \right )\bra{100}_{e}+\Bar{U}^{f*}_{\alpha \mu}\left ( t \right )\bra{010}_{\mu} +\Bar{U}^{f*}_{\alpha \tau}\left ( t \right ) \bra{001}_{\tau}.$\\
Since for the state $\ket{\nu _{\alpha }\left ( t \right )}$, only $\ket{100}, \ket{010}, \ket{001}$  have nonzero coefficients, we represent this as an 8-dimensional column vector of the order 
$\left (  \ket{000}, \ket{001}, \ket{010}, \ket{011}, \ket{100}, \ket{101}, \ket{110}, \ket{111} \right )$ as:
\begin{equation}
    \ket{\nu _{\alpha }\left ( t \right )}= \begin{bmatrix}
0 \\ 
 \Bar{U}_{\alpha \tau}^{f}\left ( t \right )\\\Bar{U}_{\alpha \mu}^{f}\left ( t \right )
 \\
 0\\
 \Bar{U}_{\alpha e}^{f}\left ( t \right )\\
 0\\
 0\\0
\end{bmatrix}
\end{equation}

Similarly, $\bra{\nu _{\alpha }\left ( t \right )}$ is written as:
\begin{equation}
    \bra{\nu _{\alpha }\left ( t \right )}= \begin{bmatrix}
0 & \Bar{U}^{f*}_{\alpha \tau}\left ( t \right ) &  \Bar{U}^{f*}_{\alpha \mu}\left ( t \right )& 0 & \Bar{U}^{f*}_{\alpha e}\left ( t \right ) & 0& 0 & 0 \\
\end{bmatrix}
\end{equation}
The density matrix is given by:
\begin{equation}
    \rho _{e\mu\tau}^{\alpha}(t) = \ket{\nu _{\alpha }(t)}\bra{\nu _{\alpha }(t)}
\end{equation}
Expanding this,
\begin{equation}
    \rho _{e\mu \tau}^{\alpha}(t)= \begin{bmatrix}
0 \\ 
 \Bar{U}_{\alpha \tau}^{f}\left ( t \right )\\\Bar{U}_{\alpha \mu}^{f}\left ( t \right )
 \\
 0\\
 \Bar{U}_{\alpha e}^{f}\left ( t \right )\\
 0\\
 0\\0
\end{bmatrix} \begin{bmatrix}
0 & \Bar{U}^{f*}_{\alpha \tau}\left ( t \right ) &  \Bar{U}^{f*}_{\alpha \mu}\left ( t \right )& 0 & \Bar{U}^{f*}_{\alpha e}\left ( t \right ) & 0& 0 & 0 \\
\end{bmatrix}.
\end{equation}
Performing the outer product:
\begin{equation} 
\rho _{e\mu\tau}^{\alpha}(t)= \begin{pmatrix}
0 & 0 & 0 & 0 & 0 & 0 & 0 & 0\\ 
0 & \left |\Bar{U}_{\alpha \tau }^{f}(t) \right |^{2} & \Bar{U}_{\alpha \tau }^{f}(t)\Bar{U}_{\alpha \mu }^{f \ast }(t) & 0 & \Bar{U}_{\alpha \tau }(t)\Bar{U}_{\alpha e }^{f \ast }(t) & 0 & 0 & 0\\ 
0 & \Bar{U}_{\alpha \mu }^{f}(t)\Bar{U}_{\alpha \tau }^{f \ast}(t) & \left |\Bar{U}_{\alpha \mu }^{f}(t) \right |^{2} & 0 & \Bar{U}_{\alpha \mu }^{f}(t)\Bar{U}_{\alpha e }^{f \ast }(t) & 0 & 0 & 0\\ 
0 & 0 & 0 & 0 & 0 & 0 & 0 & 0\\ 
0 & \Bar{U}_{\alpha e }^{f}(t)\Bar{U}_{\alpha \tau }^{f \ast }(t) & \Bar{U}_{\alpha e }^{f}(t)\Bar{U}_{\alpha \mu }^{f \ast }(t) & 0 & \left |\Bar{U}_{\alpha e }^{f}(t) \right |^{2}& 0 & 0 & 0\\ 
0 & 0 & 0 & 0 & 0 &0  & 0 & 0\\ 
0 & 0 & 0 & 0 & 0 &0  & 0 & 0\\
0 & 0 & 0 & 0 & 0 &0  & 0 & 0 
\end{pmatrix}
\end{equation}

\begin{equation}\label{A7}
    =\begin{pmatrix}
0 & 0 & 0 & 0 & 0 & 0 & 0 & 0\\ 
0 & \rho _{22}^{\alpha } & \rho _{23}^{\alpha } & 0 & \rho _{25}^{\alpha } & 0 & 0 & 0\\ 
0 & \rho _{32}^{\alpha } & \rho _{33}^{\alpha } & 0 & \rho _{35}^{\alpha } & 0 & 0 & 0\\ 
0 & 0 & 0 & 0 & 0 & 0 & 0 & 0\\ 
0 & \rho _{52}^{\alpha } & \rho _{53}^{\alpha } & 0 & \rho _{55}^{\alpha } & 0 & 0 & 0\\ 
0 & 0 & 0 & 0 & 0 &0  & 0 & 0\\ 
0 & 0 & 0 & 0 & 0 &0  & 0 & 0\\
0 & 0 & 0 & 0 & 0 &0  & 0 & 0 
\end{pmatrix},
\end{equation}
where the elements of this matrix can be represented as in Eq. \ref{rhop}. \\
For the neutrino, the density matrix is represented by $\rho_{e\mu\tau}^{\alpha}$, as defined in Eq. \ref{A7}. The corresponding entropies, $S(\rho_{e}^{\alpha})$, $S(\rho_{\mu}^{\alpha})$, and $S(\rho_{\tau}^{\alpha})$, are von Neumann entropies, with
\begin{equation}\label{A2}
S(\rho_{e}^{\alpha})=-\text{Tr}(\rho_{e}^{\alpha}\log \rho_{e}^{\alpha}),
\end{equation}
and similarly for $S(\rho_{\mu})$ and $S(\rho_{\tau})$ with 
\begin{equation}\label{A3}
\rho_{e}^{\alpha}=Tr _{\mu\tau}\left [\rho _{e\mu\tau}^{\alpha}(t)  \right ], \quad
\rho_{\mu}^{\alpha}=Tr _{e\tau}\left [\rho _{e\mu\tau}^{\alpha}(t)  \right ], \quad
\rho_{\tau}^{\alpha}=Tr _{e\mu}\left [\rho _{e\mu\tau}^{\alpha}(t)  \right ].
\end{equation}
The reduced density matrices for neutrino $ \rho _{e\mu \tau}^{\alpha}(t)$ are
 \begin{eqnarray}
     \rho _{e\mu}^{\alpha}=Tr _{\tau}\left [\rho _{e\mu\tau}^{\alpha}  \right ] =\begin{pmatrix}
\rho _{22}^{\alpha} & 0 & 0 & 0 \\
0 & \rho _{33}^{\alpha} & \rho _{35}^{\alpha} & 0 \\
0 & \rho _{53}^{\alpha} & \rho _{55}^{\alpha} & 0 \\
0 & 0 & 0 & 0 \\
\end{pmatrix},
 \end{eqnarray}
 \begin{eqnarray}
     \rho _{e\tau}^{\alpha}=Tr _{\mu}\left [\rho _{e\mu\tau}^{\alpha}  \right ] =\begin{pmatrix}
\rho _{33}^{\alpha} & 0 & 0 & 0 \\
0 & \rho _{22}^{\alpha} & \rho _{25}^{\alpha} & 0 \\
0 & \rho _{52}^{\alpha} & \rho _{55}^{\alpha} & 0 \\
0 & 0 & 0 & 0 \\
\end{pmatrix},
 \end{eqnarray}
 \begin{eqnarray}
     \rho _{\mu\tau}^{\alpha}=Tr _{e}\left [\rho _{e\mu\tau}^{\alpha}  \right ] =\begin{pmatrix}
\rho _{55}^{\alpha} & 0 & 0 & 0 \\
0 & \rho _{22}^{\alpha} & \rho _{23}^{\alpha} & 0 \\
0 & \rho _{32}^{\alpha} & \rho _{33}^{\alpha} & 0 \\
0 & 0 & 0 & 0 \\
\end{pmatrix}.
 \end{eqnarray}
 
Again the reduced density matrix $\rho_{e}$, $\rho_{\mu}$ and $\rho_{\tau}$ is obtained from the full density matrix $\rho_{e\mu\tau}^{\alpha}$ by tracing over the matrices $\rho_{\mu\tau}$, $\rho_{e\tau}$ and $\rho_{e\mu}$ respectively. And are defined as,
\begin{eqnarray}\label{rhoe}
    \rho _{e}^{\alpha}=Tr _{\mu\tau}\left [\rho _{e\mu\tau}^{\alpha}  \right ] =\begin{pmatrix}
\rho _{22}^{\alpha}+ \rho _{33}^{\alpha} & 0 \\
0 &  \rho _{55}^{\alpha}\\
\end{pmatrix},
 \end{eqnarray}
 
 \begin{eqnarray}\label{rhomu}
    \rho _{\mu}^{\alpha}= Tr _{e\tau}\left [\rho _{e\mu\tau}^{\alpha}  \right ] =\begin{pmatrix}
\rho _{22}^{\alpha}+ \rho _{55}^{\alpha} & 0 \\
0 &  \rho _{33}^{\alpha}\\
\end{pmatrix},
 \end{eqnarray}
 
 \begin{eqnarray}\label{rhotau}
     \rho _{\tau}^{\alpha}=Tr _{e\mu}\left [\rho _{e\mu\tau}^{\alpha}  \right ] =\begin{pmatrix}
\rho _{33}^{\alpha}+ \rho _{55}^{\alpha} & 0 \\
0 &  \rho _{22}^{\alpha}\\
\end{pmatrix}.
 \end{eqnarray}

The entropy $S(\rho_{e}^{\alpha})$ can be evaluated as
\begin{equation}\label{A11}
\begin{aligned}
S(\rho_{e}^{\alpha}) &= -\text{Tr}(\rho_{e}\log \rho_{e}) \\ 
&= -\text{Tr}\left [ 
\begin{pmatrix}
\rho _{22}^{\alpha}+ \rho _{33}^{\alpha} & 0 \\
0 &  \rho _{55}^{\alpha}
\end{pmatrix}
\begin{pmatrix}
\log\!\left ( \rho _{22}^{\alpha}+ \rho _{33}^{\alpha}  \right )& 0 \\
0 & \log\!\left (  \rho _{55}^{\alpha} \right )
\end{pmatrix} \right ] \\
&= -(\rho _{22}^{\alpha}+ \rho _{33}^{\alpha})\log\!\left ( \rho _{22}^{\alpha}+ \rho _{33}^{\alpha}  \right )
   - \rho _{55}^{\alpha}\log\!\left (  \rho _{55}^{\alpha} \right ).
\end{aligned}
\end{equation}
Similarly we can write for the $S(\rho_{\mu}^{\alpha})$ and $S(\rho_{\tau}^{\alpha})$ as,
\begin{equation}\label{A12}
   S(\rho_{\mu}^{\alpha}) = -(\rho _{22}^{\alpha}+ \rho _{55}^{\alpha})\log\!\left ( \rho _{22}^{\alpha}+ \rho _{55}^{\alpha}  \right )
   - \rho _{33}^{\alpha}\log\!\left (  \rho _{33}^{\alpha} \right ),
\end{equation}
and 
\begin{equation}\label{A13}
   S(\rho_{\tau}^{\alpha}) = -(\rho _{33}^{\alpha}+ \rho _{55}^{\alpha})\log\!\left ( \rho _{33}^{\alpha}+ \rho _{55}^{\alpha}  \right )
   - \rho _{22}^{\alpha}\log\!\left (  \rho _{22}^{\alpha} \right ).
\end{equation}
As $\rho _{55}^{\alpha}= P_{\alpha e}(t)= \left |\Bar{U}^{f}_{\alpha e }(t) \right |^{2}$,
 $\rho _{33}^{\alpha}= P_{\alpha \mu}(t)= \left |\Bar{U}^{f}_{\alpha \mu }(t) \right |^{2}$, and 
 $\rho _{22}^{\alpha}= P_{\alpha \tau}(t)= \left |\Bar{U}^{f}_{\alpha \tau }(t) \right |^{2}$ are the associated probabilities, therefore, \eqref{A11}, \eqref{A12}, \eqref{A13} can be expressed in terms of neutrino oscillation probabilities for $\mu$ as initial flavor as
 \begin{equation}\label{}
   S(\rho_{e}^{\mu}) = -(P _{\mu\tau}+ P _{\mu e})\log\!\left ( P _{\mu\tau}+ P _{\mu e}  \right )
   - P _{\mu\mu}\log\!\left (  P _{\mu\mu}\right ),
\end{equation}

\begin{equation}\label{}
   S(\rho_{\mu}^{\mu}) = -(P _{\mu\mu}+ P _{\mu \tau})\log\!\left ( P _{\mu\mu}+ P _{\mu \tau}  \right )
   - P _{\mu e}\log\!\left (  P _{\mu e}\right ),
\end{equation}

\begin{equation}\label{}
   S(\rho_{\tau}^{\mu}) = -(P _{\mu\mu}+ P _{\mu e})\log\!\left ( P _{\mu\mu}+ P _{\mu e}  \right )
   - P _{\mu\tau}\log\!\left (  P _{\mu\tau}\right ).
\end{equation}
Therefore, the \eqref{A1} can be expressed in terms of neutrino oscillations probabilities as follows:
\begin{equation} \label{A17}
\begin{aligned}
EOF^{\mu} = \tfrac{1}{2}\Big[ 
  &-(P _{\mu\tau}+ P _{\mu e})\log\!\left ( P _{\mu\tau}+ P _{\mu e}  \right )
   - P _{\mu\mu}\log\!\left (  P _{\mu\mu}\right ) \\
  &-(P _{\mu\mu}+ P _{\mu \tau})\log\!\left ( P _{\mu\mu}+ P _{\mu \tau}  \right )
   - P _{\mu e}\log\!\left (  P _{\mu e}\right ) \\
  &-(P _{\mu\mu}+ P _{\mu e})\log\!\left ( P _{\mu\mu}+ P _{\mu e}  \right )
   - P _{\mu\tau}\log\!\left (  P _{\mu\tau}\right ) \Big].
\end{aligned}
\end{equation}

\subsection{Concurrence in terms of Neutrino Oscillation Probabilities}
The Concurrence in Eq. \eqref{con} for the tripartite density matrix $\rho_{ABC}$ is defined as,
\begin{equation}
    C(\rho _{ABC})=[3-Tr(\rho _{A})^{2}-Tr(\rho _{B})^{2}-Tr(\rho _{C})^{2}]^{\frac{1}{2}},
\end{equation}
where $\rho _{A}$, $\rho _{B}$, and  $\rho _{C}$ are the reduced density matrices.
For the neutrino case, these reduced density matrices are $\rho_{e}$, $\rho_{\mu}$, and $\rho_{\tau}$ are defined in Eqs. \eqref{rhoe}, \eqref{rhomu} and \eqref{rhotau}, and thus the above expression provides the Concurrence for the three  flavor neutrino system, as
\begin{equation}
    C(\rho _{e\mu \tau})= [3- \rho _{22}^{\alpha 2}-(\rho _{33}^{\alpha }+\rho _{55}^{\alpha })^{ 2}- \rho _{33}^{\alpha 2}-(\rho _{22}^\alpha + \rho _{55}^\alpha )^{2}-\rho _{55}^{\alpha2}-(\rho _{22}^{\alpha} + \rho _{33}^{\alpha})^{2}]^{\frac{1}{2}}.
\end{equation}
In terms of neutrino oscillation probabilities, Concurrence for the neutrino can be written as,
\begin{equation}\label{Con} 
 C^{\mu}=\sqrt{3-3(P_{\mu e}^{2}+P_{\mu \mu}^{2}+P_{\mu \tau}^{2})-2P_{\mu \mu}P_{\mu \tau}-2P_{\mu e}(P_{\mu \mu}+P_{\mu \tau})}.
\end{equation}

\subsection{Negativity in terms of Neutrino Oscillation Probabilities}
The tripartite Negativity of a state $\rho_{ABC}$ is defined as \cite{Sabin2008}
$$N=(N_{A-BC}N_{B-CA}N_{C-AB})^{\frac{1}{3}},$$
with $N_{I-JK} = -2 \sum_{i} \sigma_{i} (\rho_{ABC}^{T_{I}})$ being the negative eigenvalues of $\rho^{T_{I}}$, the partial transpose of $\rho_{ABC}$ with respect to subsystem $I$, with $I = A, B, C,$ and $JK = BC, CA, AB$, respectively.
\\
For the neutrino, the density matrix is defined by \eqref{rho1}, and we can find out the partial transpose of the density matrix.\\
The partial transpose, $\rho _{e\mu\tau}^{\alpha T_{e}}$ of the neutrino density matrix $\rho_{e\mu\tau}^{\alpha}$ with respect to subsystem $e$ is,
\begin{equation}
    \rho _{e\mu\tau}^{\alpha T_{e}}(t)=\begin{pmatrix}
0 & 0 & 0 & 0 & 0 & \rho _{52}^{\alpha } & \rho _{53}^{\alpha } & 0\\ 
0 & \rho _{22}^{\alpha } & \rho _{23}^{\alpha } & 0 & 0 & 0 & 0 & 0\\ 
0 & \rho _{32}^{\alpha } & \rho _{33}^{\alpha } & 0 & 0 & 0 & 0 & 0\\ 
0 & 0 & 0 & 0 & 0 & 0 & 0 & 0\\ 
0 & 0 & 0 & 0 & \rho _{55}^{\alpha } & 0 & 0 & 0\\ 
\rho _{25}^{\alpha } & 0 & 0 & 0 & 0 &0  & 0 & 0\\ 
\rho _{35}^{\alpha } & 0 & 0 & 0 & 0 &0  & 0 & 0\\
0 & 0 & 0 & 0 & 0 &0  & 0 & 0 
\end{pmatrix},
\end{equation}
and the negative eigenvalue is
\begin{equation}
    \sqrt{\rho _{22}^{\alpha }\rho _{55}^{\alpha }+\rho _{33}^{\alpha }\rho _{55}^{\alpha }}.
\end{equation}
Therefore,
\begin{equation}\label{n1}
    N_{e-\mu\tau}^{\alpha }= \sqrt{\rho _{22}^{\alpha }\rho _{55}^{\alpha }+\rho _{33}^{\alpha }\rho _{55}^{\alpha }}= \sqrt{\rho _{55}^{\alpha }}\sqrt{\rho _{22}^{\alpha }+\rho _{33}^{\alpha }}.
\end{equation}

The partial transpose, $\rho _{e\mu\tau}^{\alpha T_{\mu}}$ of the neutrino density matrix $\rho_{e\mu\tau}^{\alpha}$ with respect to subsystem $\mu$ is,
\begin{equation}
    \rho _{e\mu\tau}^{\alpha T_{\mu}}(t)=\begin{pmatrix}
0 & 0 & 0 & \rho _{32}^{\alpha } & 0 & 0 & \rho _{35}^{\alpha } & 0\\ 
0 & \rho _{22}^{\alpha } & 0 & 0 & \rho _{25}^{\alpha } & 0 & 0 & 0\\ 
0 & 0 & \rho _{33}^{\alpha } & 0 & 0 & 0 & 0 & 0\\ 
\rho _{23}^{\alpha }& 0 & 0 & 0 & 0 & 0 & 0 & 0\\ 
0 & \rho _{52}^{\alpha } & 0 & 0 & \rho _{55}^{\alpha } & 0 & 0 & 0\\ 
0& 0 & 0 & 0 & 0 &0  & 0 & 0\\ 
\rho _{53}^{\alpha } & 0 & 0 & 0 & 0 &0  & 0 & 0\\
0 & 0 & 0 & 0 & 0 &0  & 0 & 0 
\end{pmatrix},
\end{equation}
and the negative eigenvalue is 
\begin{equation}
    \sqrt{\rho _{22}^{\alpha }\rho _{33}^{\alpha }+\rho _{33}^{\alpha }\rho _{55}^{\alpha }}.
\end{equation}
Therefore,
\begin{equation}\label{n2}
    N_{\mu-e\tau}^{\alpha }= \sqrt{\rho _{22}^{\alpha }\rho _{33}^{\alpha }+\rho _{33}^{\alpha }\rho _{55}^{\alpha }}= \sqrt{\rho _{33}^{\alpha }} \sqrt{\rho _{22}^{\alpha }+\rho _{55}^{\alpha }}.
\end{equation}

The partial transpose, $\rho _{e\mu\tau}^{\alpha T_{\tau}}$ of the neutrino density matrix $\rho_{e\mu\tau}^{\alpha}$ with respect to subsystem $\tau$ is,
\begin{equation}
    \rho _{e\mu\tau}^{\alpha T_{\tau}}(t)=\begin{pmatrix}
0 & 0 & 0 & \rho _{23}^{\alpha } & 0 & \rho _{25}^{\alpha } & 0 & 0\\ 
0 & \rho _{22}^{\alpha } & 0 & 0 & 0 & 0 & 0 & 0\\ 
0 & 0 & \rho _{33}^{\alpha } & 0 & \rho _{35}^{\alpha } & 0 & 0 & 0\\ 
\rho _{32}^{\alpha }& 0 & 0 & 0 & 0 & 0 & 0 & 0\\ 
0 & 0 & \rho _{53}^{\alpha } & 0 & \rho _{55}^{\alpha } & 0 & 0 & 0\\ 
\rho _{53}^{\alpha }& 0 & 0 & 0 & 0 &0  & 0 & 0\\ 
0 & 0 & 0 & 0 & 0 &0  & 0 & 0\\
0 & 0 & 0 & 0 & 0 &0  & 0 & 0 
\end{pmatrix},
\end{equation}
and the negative eigenvalue is
\begin{equation}
    \sqrt{\rho _{22}^{\alpha }\rho _{33}^{\alpha }+\rho _{22}^{\alpha }\rho _{55}^{\alpha }}.
\end{equation}
Therefore,
\begin{equation}\label{n3}
    N_{\tau-\mu e}^{\alpha }= \sqrt{\rho _{22}^{\alpha }\rho _{33}^{\alpha }+\rho _{22}^{\alpha }\rho _{55}^{\alpha }}= \sqrt{\rho _{22}^{\alpha }} \sqrt{\rho _{33}^{\alpha }+\rho _{55}^{\alpha }}.
\end{equation}
The Eq. \eqref{n1}, \eqref{n2} and \eqref{n3} in terms of neutrino oscillation probabilities for initial muon flavor can be written as,
\begin{equation}
    N_{e-\mu\tau}^{\mu }= \sqrt{P_{\mu e}}\sqrt{P_{\mu \mu}+P_{\mu \tau}},
\end{equation}

\begin{equation}
    N_{\mu-e\tau}^{\mu }= \sqrt{P_{\mu \mu}} \sqrt{P_{\mu e}+P_{\mu \tau}},
\end{equation}

\begin{equation}
    N_{\tau-\mu e}^{\mu }= \sqrt{P_{\mu\tau}} \sqrt{P_{\mu\mu}+P_{\mu e}}.
\end{equation}

In terms of survival and oscillation probabilities of neutrino oscillation, Negativity for $\mu$ as initial flavor, is given as :
\begin{equation}\label{N}
    N^{\mu}=[\sqrt{P_{\mu e}}\sqrt{P_{\mu \mu}+P_{\mu \tau}}\sqrt{P_{\mu e}+P_{\mu \tau}}\sqrt{P_{\mu \mu}}\sqrt{P_{\mu e}+P_{\mu \mu}}\sqrt{P_{\mu \tau}}]^{\frac{1}{3}}.
\end{equation}

\end{document}